# *COSMIC EVOLUTION THROUGH UV SURVEYS (CETUS)*

# FINAL REPORT

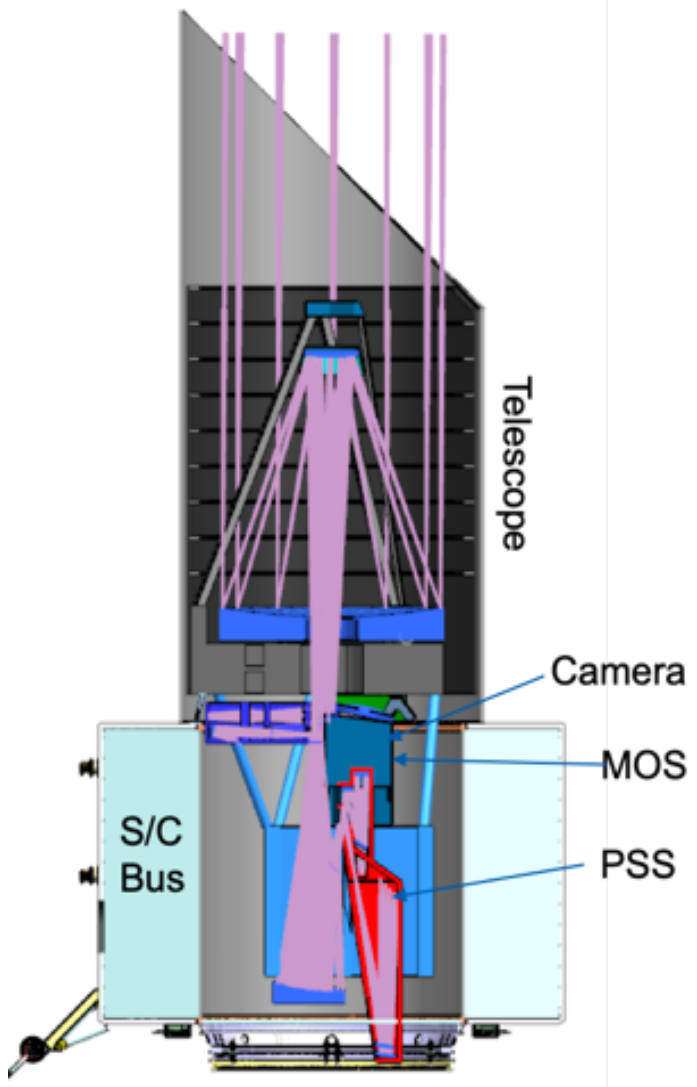

**Thematic Activity: Project (probe mission concept)**
**Program: Electromagnetic observations from space**

**Authors of Final Report:**
Jonathan Arenberg, Northrop Grumman Corporation
Sally Heap, Univ. of Maryland, sara.heap@gmail.com
Tony Hull, Univ. of New Mexico
Steve Kendrick, Kendrick Aerospace Consulting LLC
Bob Woodruff, Woodruff Consulting

**Scientific Contributors:** Maarten Baes, Rachel Bezanson, Luciana Bianchi, David Bowen, Brad Cenko, Yi-Kuan Chiang, Rachel Cochrane, Mike Corcoran, Paul Crowther, Simon Driver, Bill Danchi, Eli Dwek, Brian Fleming, Kevin France, Pradip Gatkine, Suvi Gezari, Lea Hagen, Chris Hayward, Matthew Hayes, Tim Heckman, Edmund Hodges-Kluck, Alexander Kutyrev, Thierry Lanz, John MacKenty, Steve McCandliss, Harvey Moseley, Coralie Neiner, Goren Östlin, Camilla Pacifici, Marc Rafelski, Bernie Rauscher, Jane Rigby, Ian Roederer, David Spergel, Dan Stark, Alexander Szalay, Bryan Terrazas, Jonathan Trump, Arjun van der Wel, Sylvain Veilleux, Kate Whitaker, Isak Wold, Rosemary Wyse

**Technical Contributors**: Jim Burge, Kelly Dodson, Chip Eckles, Brian Fleming, Jamie Kennea, Gerry Lemson, John MacKenty, Steve McCandliss, Greg Mehle, Shouleh Nikzad, Trent Newswander, Lloyd Purves, Manuel Quijada, Ossy Siegmund, Dave Sheikh, Phil Stahl, Ani Thakar, John Vallerga, Marty Valente, the Goddard IDC/MDL.

# Cosmic Evolution Through UV Surveys (CETUS)

## TABLE OF CONTENTS

# 1. INTRODUCTION TO CETUS

We introduce the CETUS probe-class mission concept by comparing it with a mission we already know: the Hubble Space Telescope. Hubble is a UV-optical-IR telescope. By design, it does things that only a telescope in space can do: it obtains exquisite images of astronomical sources unmarred by atmospheric seeing effects; and from above Earth's atmosphere, it observes UV radiation from astronomical sources. CETUS is an all-UV space mission concept, and it does things that only CETUS can do. The four main capabilities of CETUS that even Hubble doesn't have are: (1) wide-field (17.4'x17.4') imaging and spectroscopy of astronomical sources with ≤0.5" resolution; (2) spectral sensitivity to UV radiation at wavelengths as short as 1000 Å; (3) near-UV multi-object slit spectroscopy; and (4) rapid-response UV spectroscopy and deep imaging of transients like GW 170817; and (5) 23 times higher sensitivity to extended sources. Table 1-1 gives a full list of CETUS parameters and capabilities.

To UV observers on Hubble, CETUS is obviously an outgrowth of Hubble:

*The CETUS FUV+NUV*[1] *camera* is similar to Hubble's WFC3 working in the NUV and Hubble's ACS solar-blind channel whose forté is in the FUV. The CETUS camera is not only a consolidation of the two instruments but the result of new, added capabilities. One new capability is an enormous field of view of 17.4'x17.4', which is 1,000 times larger than that of ACS and nearly 40 times larger than that of WFC3. The other is a fast f-ratio, f/5, of the CETUS telescope that enables the CETUS camera to detect emission from extended sources 23 times fainter than can be detected by any Hubble camera.

*The CETUS FUV+NUV point/slit spectrograph (PSS)* is a near-copy of COS' far-UV spectrograph and STIS's near-UV echelle spectrograph. Again, the CETUS spectrograph comes with added capabilities. One new capability is sensitivity to UV radiation at wavelengths as short as 1,000 Å, whereas STIS or COS are insensitive below 1,150 Å. This extra 150 Å is important as it contains the only spectral diagnostic of warm-hot plasma as well as numerous $H_2$ lines. In addition, the FUV PSS has a 6'-long slit for imaging spectroscopy - 12 times longer than STIS's long slit, while COS has no imaging capabilities at all. Like STIS, the NUV PSS has a R~40,000 spectrograph, which is essential for research on stars like the extremely metal-poor stars whose NUV spectra bear the imprint of explosive ejecta from the first stars.

*The CETUS NUV Multi-Object Spectrograph (MOS)* has no analogue on Hubble. With its microshutter array (MSA), its closest cousin is the MOS on JWST's NIRspec instrument. The CETUS MSA has over 72,000 shutters in a 380x190 array. Each shutter is commandable to be open or closed. The MSA shutters can be configured to contain a set of 17.4'-long slits - >40 times longer than the long slit of STIS.

CETUS' strategic objective is to serve the astronomical community as a worthy successor to Hubble. Like Hubble, CETUS will make a reconnaissance of the UV universe to address a broad suite of scientific questions, including 9 of the 20 "Key Science Questions" posed by Astro2010. Although it is a probe-class mission concept, its science program includes all Hubble programs being carried out under the Hubble UV Initiative except those requiring extremely high angular resolution.

> "In considering a new [UV-optical] facility, we should look not only at how to best provide answers to longstanding theoretical questions. Rather, we should also ponder the choice of instruments that will provide novel kinds of observations that could raise new incisive questions." Harwit (2003)

The new capabilities of CETUS will enable new types of observations, so the potential for discovery is high. The lifetime of CETUS is arbitrarily long, limited only by the amount of propellent. Hence, its science program is bound to evolve as CETUS or ground- or space-based telescopes make new findings.

Unlike Hubble, CETUS is a survey telescope. As CETUS observations may be used by multiple astronomers for different purposes, it makes sense for CETUS to have no proprietary time. All data in the CETUS archive will be immediately available to all. The archival data will be science-ready. Archival data

[1] Throughout this report, we use the abbreviation "FUV" for far ultraviolet, and "NUV" for near ultraviolet

will include not only raw and processed data typical of Hubble processed data but also measurements of the processed data.

CETUS will bring benefits to the astronomy community similar to those of large missions like Hubble. According to a committee convened by NASA to consider large, strategic missions (*Powering Science*, NAP 2016), a large, strategic mission:

1. Captures science data that cannot be obtained in any other way;
2. Answers many of the most compelling scientific questions facing the scientific field;
3. Opens new windows of scientific inquiry;
4. Provides high-quality (precise and with stable absolute calibration) observations sustained over an extended period of time;
5. Supports the workforce, the industrial base, and technology development;
6. Maintains U.S. leadership in space & maintains U.S. scientific leadership;
7. Produces scientific results and discoveries that capture the public's imagination;
8. Receives a high degree of external visibility;
9. Provides greater opportunities for international participation, cooperation, and collaboration.

CETUS will provide many of these benefits. The main area where CETUS falls short is item 5: supporting the workforce. As a cost-capped Probe mission concept, the full life-cycle cost of CETUS cannot exceed $1.0 billion. We are therefore trying to ensure that CETUS yields science-ready data, so that astronomers - both observers and theoreticians - get maximum benefits. We are also considering a shortened primary mission and undergoing review for a funded extended mission.

The CETUS instruments have effective apertures that will continue and extend the discovery space of major past (FUSE) and present (COS, STIS) UV instruments and will provide operation in future years when Hubble is no longer functional. **Figures 1-2** and **1-3** show comparisons of effective aperture and wavelength range with other instruments.

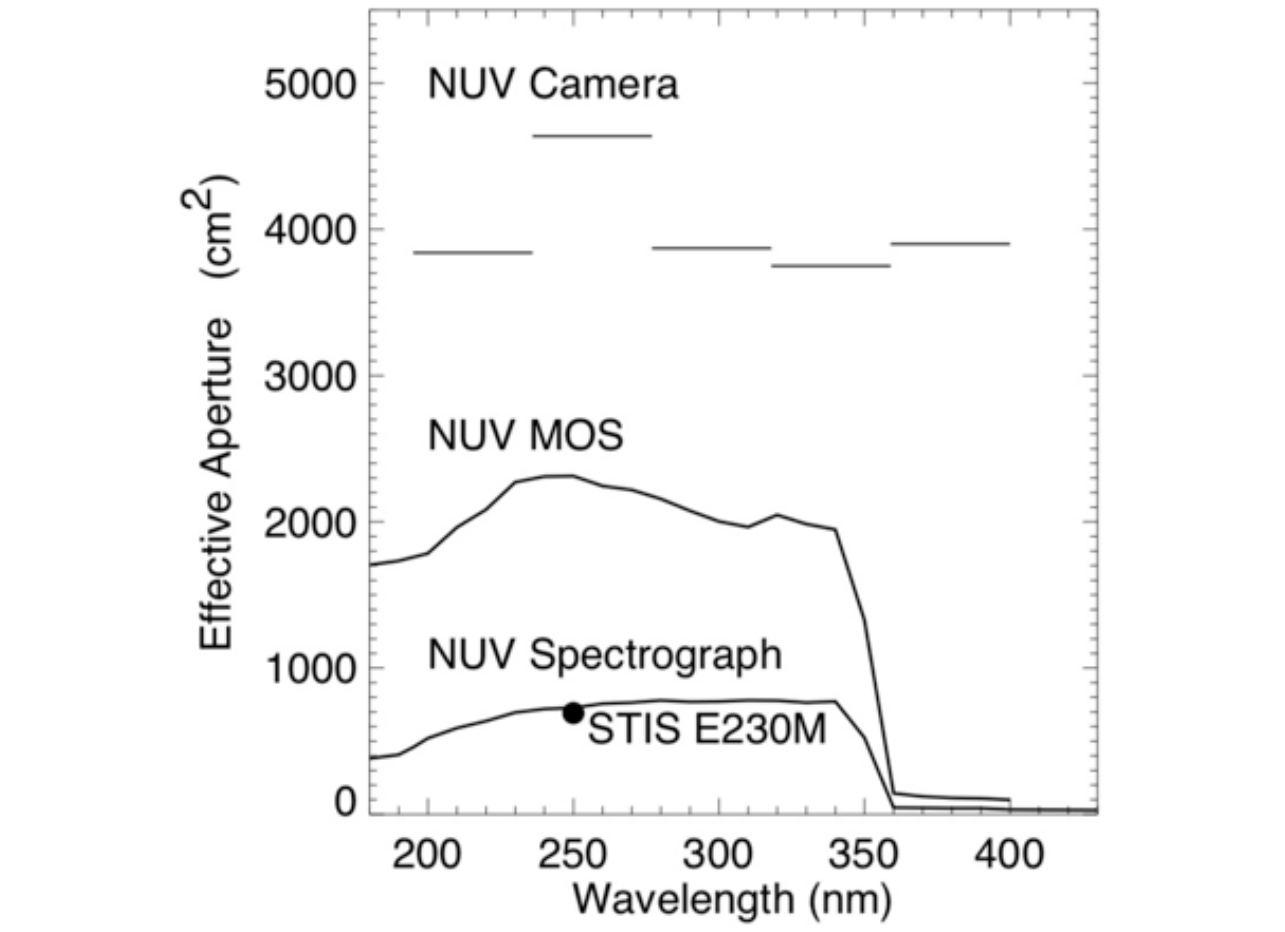

**Figure 1-2. Effective apertures for CETUS NUV Camera, MOS, and Spectrograph compared to STIS**

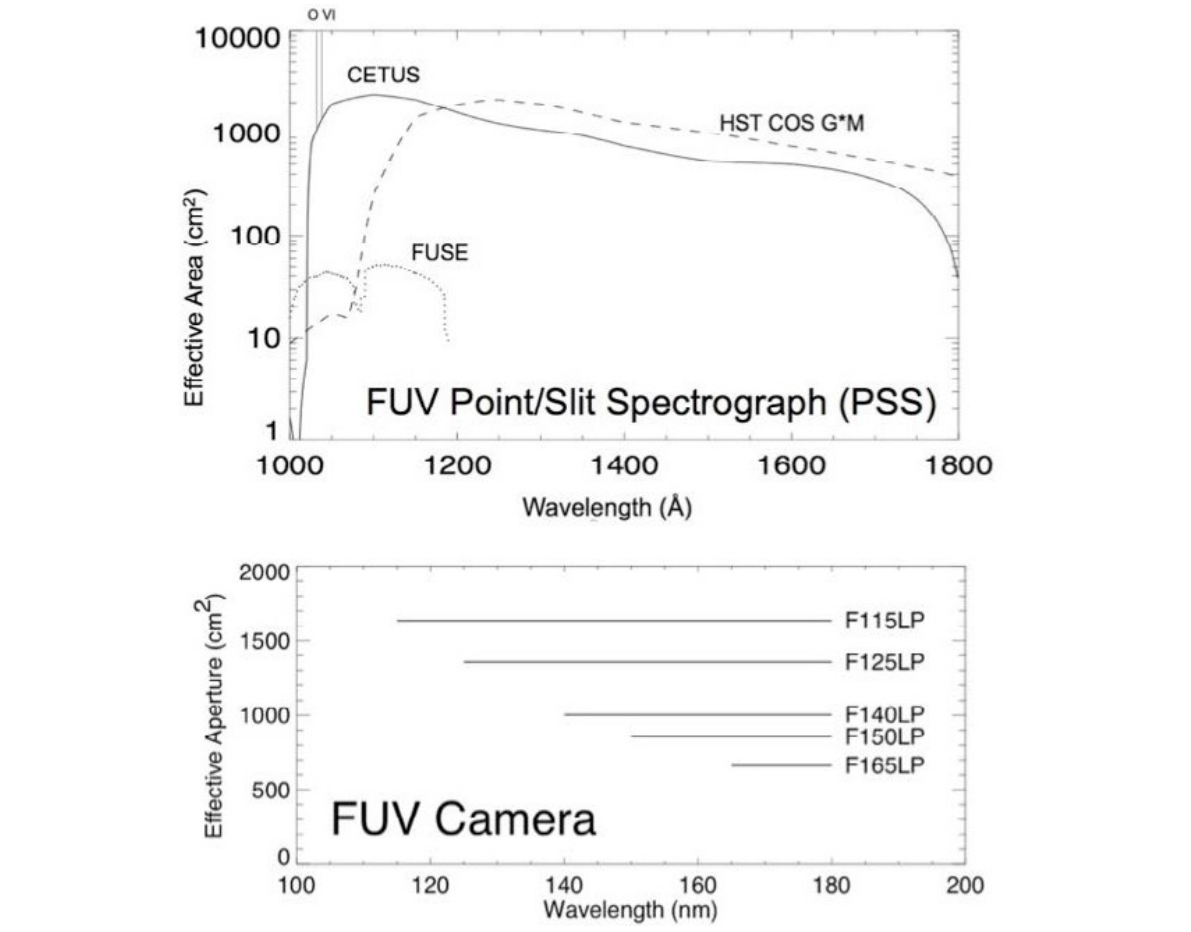

**Figure 1-3. (Top) Comparison of CETUS effective area with COS and FUSE. (Bottom) Effective areas of FUV camera filters.**

**Table 1-1** highlights the properties of the telescope and the three scientific instruments. CETUS' NUV MOS and FUV spectrograph, in particular, are instruments that will make new kinds of observations that are sure to provoke incisive questions.

**Table 1-1. CETUS has a 1.5-m telescope feeding 3 instruments operating in the NUV and FUV.**
***[Unique capabilities are written in blue font]***

| Instrument | Properties and Capabilities |
|---|---|
| Telescope | • 1.5-m diameter, telescope<br>• f/5 f-ratio yielding wide field of view accommodating all 3 science instruments operating in parallel<br>• f/5 f-ratio yielding 23 times greater sensitivity than Hubble to low surface-brightness sources<br>• Far-UV sensitivity down to 100 nm<br>• Heritage: Copernicus (OAO-3 telescope with 0.8-m primary), Kepler (1.4-m telescope) |
| Near-UV multi-object spectrograph (MOS) With microshutter array | • Near-UV spectra of up to 100 sources simultaneously<br>• Field of view: 17.4’ x 17.4’<br>• MSA Shutter size: 2.75”x5.50”; 8x16-pixel imaging within shutter<br>• Long-slit option: 2.75”x17.4’ (via 1 column of open MSA shutters)<br>• Subsampling and dithering enabled by internal mechanism<br>• Wavelength range:180-350 nm<br>• Spectral resolving power: R~1,000 at 250 nm<br>• Limiting sensitivity: $F_\lambda$ = $4\times10^{-18}$ erg/s/cm$^2$/Å<br>• Heritage: MOS in JWST’s NIRspec |
| Far-UV and near-UV camera (CAM) | • Field of view: 17.4’ x 17.4’<br>• Subsampling and dithering enabled by internal mechanism<br>FUV / NUV<br>• Angular resolution: 0.55” / 0.33”<br>• Wavelength range: 125-180 nm / 180-400 nm<br>• Filters: 5 long-pass / 5 with <FWHM>~40nm<br>• Sensitivity (1 hour): $m_{AB}$=27 / $m_{AB}$=26<br>• Heritage: HST/STIS, ACS / HST/WFC3 |
| Far-UV and near-UV point/slit spectrograph (PSS) | FUV / NUV<br>• Wavelength range: 100-172 nm / 180-350 nm<br>• Entrance slit: 2”x360”, 0.2”x3” / 2”x360”<br>• Spectral RP: 12,000, 20,000 / 40,000<br>• Effective area (cm$^2$): 2,000 (max) / 600-1000<br>• Heritage: HST/COS / HST/GHRS, STIS |
| | |

CETUS’ unique capabilities enable it to conduct four major science programs:

1. Understanding the universe at low redshift by surveying a large variety of nearby galaxies and their circumgalactic gas and dust both in absorption and emission;
2. Understanding the universe at mid-life by surveying the rest far-UV spectra of z~1 galaxies and Lyman-alpha emitters;
3. Understanding transients like LIGO sources, tidal disruption events, etc. by responding promptly to major transient events and monitoring their UV emission down to faint magnitudes, e.g. AB=27;
4. Making new discoveries based on CETUS UV surveys combined with those made by multi-wavelength surveys that will be carried out in the 2020’s and 2030’s.

These science programs are described in the next chapter. It should be understood that they are only a sampling of what might done. For example, programs to observe solar-system bodies are not included although we have been assured by Clarke (SciWP #90) that “Much of what we do in planetary science does not require the highest angular resolution. It depends more on time series and having a telescope when you need it (i.e. S/L 9 at Jupiter)”. Also, the science program does not and cannot account for new uses of CETUS instruments devised by clever astronomers.

# 2. SCIENCE CASE FOR THE CETUS PROBE MISSION

## 2.1 THE LOW-Z UNIVERSE

### Dust in Galaxies Hagen #593[2], Gordon #441

Knowledge of the properties of interstellar dust is essential for understanding the universe at all scales from the stellar properties of a star-forming region in a nearby galaxy to the cosmic spectral energy distribution (SED). Nevertheless, the properties of dust, particularly its attenuation at UV wavelengths are not well constrained. Part of the problem is that the attenuation curves vary from galaxy to galaxy and even in within a single galaxy (Figure 2-1). What we do know is that there is a clear relation between the UV slope, $\beta_{uv}$ and the IRX=FFIR/FUV for isolated galaxies at z~0 (Meurer et al., 1999), but with deviations from that relationship in merging/recently merged galaxies (Safarzadah et al. 2017). Studies of UV attenuation are needed to interpret the UV-IR SED of unresolved star-forming regions.

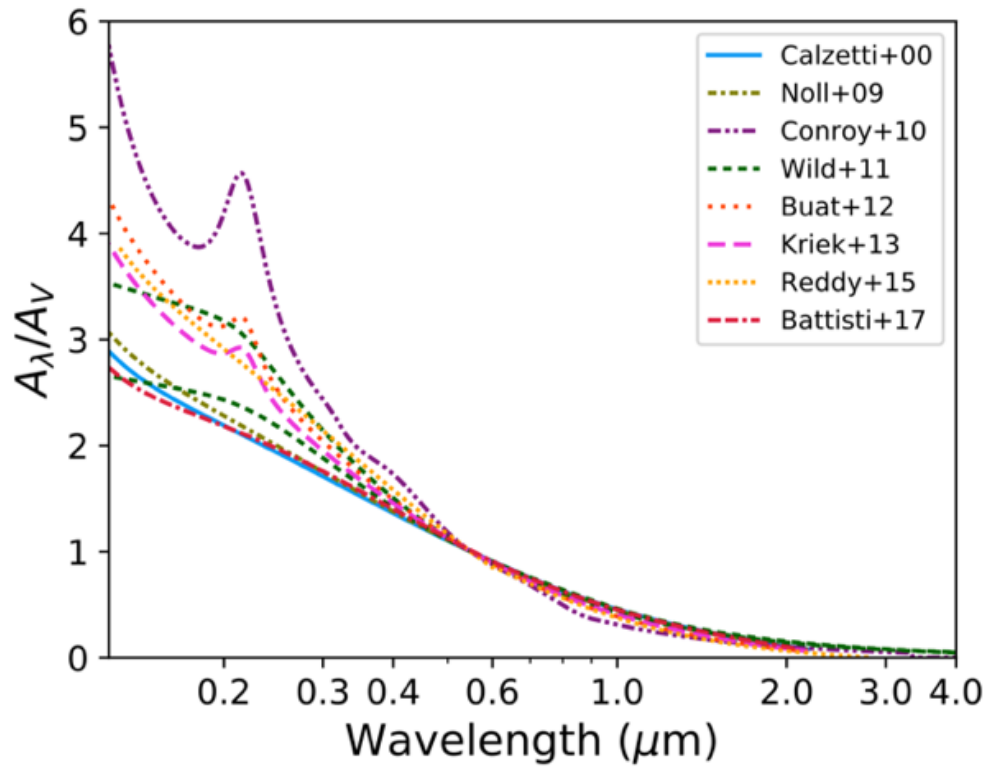

**Figure 2-1. A collection of measured dust attenuation curves in galaxies illustrates the range in the 2175-Å bump and the UV slope, $\beta_{uv}$. From Hagen #593**

Observations to determine the resolved UV attenuation properties of dust in galaxies outside the Local Group has already started with Decleir et al.'s study of M 74 (NGC 628) using UV photometry (1900-2700 Å) from the SWIFT UVOT telescope (Figure 2-2). However, these studies are limited by the small light-gathering power of the UVOT telescope and lack of FUV coverage. The CETUS camera has the same wide field of view as UVOT (17'x17'), FUV coverage down to ~1200 Å, sub-arcsec resolution (0.41" in NUV, 0.55" in FUV), and 25 X larger light-gathering power than UVOT. All these capabilities make CETUS the ideal telescope to detect and measure the 2175 Å UV bump, UV slope, and their variations within a galaxy. In combination with telescopes like ALMA, CETUS will search for correlations between these UV attenuation properties and far-IR continuum luminosity, galaxy type and environment with the ultimate goal of predicting the attenuation properties of unresolved galaxies and evaluating the uncertainty in properties of their (unreddened) stellar populations.

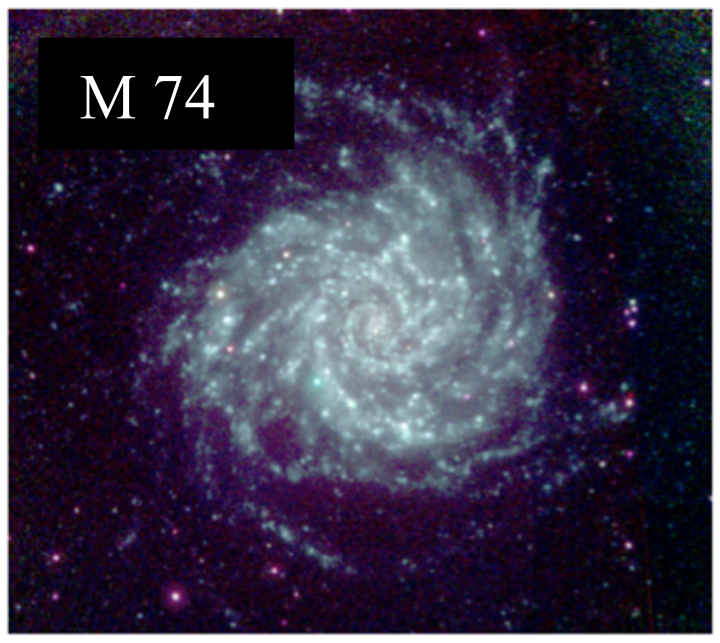

**Figure 2-2. UVOT UV images completely cover M74. CETUS has the same field of view, a set of NUV & FUV filters, and 25 times light-gathering power of UVOT.**

The CETUS attenuation parameters for galaxies of different morphological types, stellar masses, and environment will lead to improved values of the theoretical UV slope, $\beta_{uv}$ and IRX=FFIR/FUV for comparison with observation. Knowledge of dust properties will inform dust radiative transfer codes like SKIRT 3-D (Baes 2011) to predict wide-baseline spectral energy distributions and wavelength-dependent morphologies of simulated galaxies.

Understanding star formation and its effects on the surroundings is a fundamental goal. Star formation is a complex process involving multiple stages: molecular cloud formation, collapse, ionization of the surrounding ISM by massive newly born stars, dust dispersal, and cloud disruption. Dust plays a role in promoting star formation by providing a site for molecules to form and in sustaining star formation through shielding molecules from hard UV radiation from newly born stars. There are currently major campaigns to observe the process of star formation on scales much finer than the 1-kpc scale of the Kennicutt-Schmidt law. Different telescopes sensitive in different spectral regions will focus on different stages of star formation. **Figure 2-3** from Bianchi (2011) gives a hint of what we can expect from future multi-wavelength campaigns. It shows a region of the Andromeda galaxy including NGC 206. The NUV image from GALEX has a resolution of 5” or 27 pc at the galaxy. The quad-picture shows that NUV emission is prominent where hot stars have dispersed the dust, and IR emission from heated dust is prominent in embedded star-forming sites where the UV flux of newly born hot stars has been completely absorbed.

[2] References to science white papers submitted to Astro2020 give the name of the principal author and ID # of the paper. Full information is given in the References section.

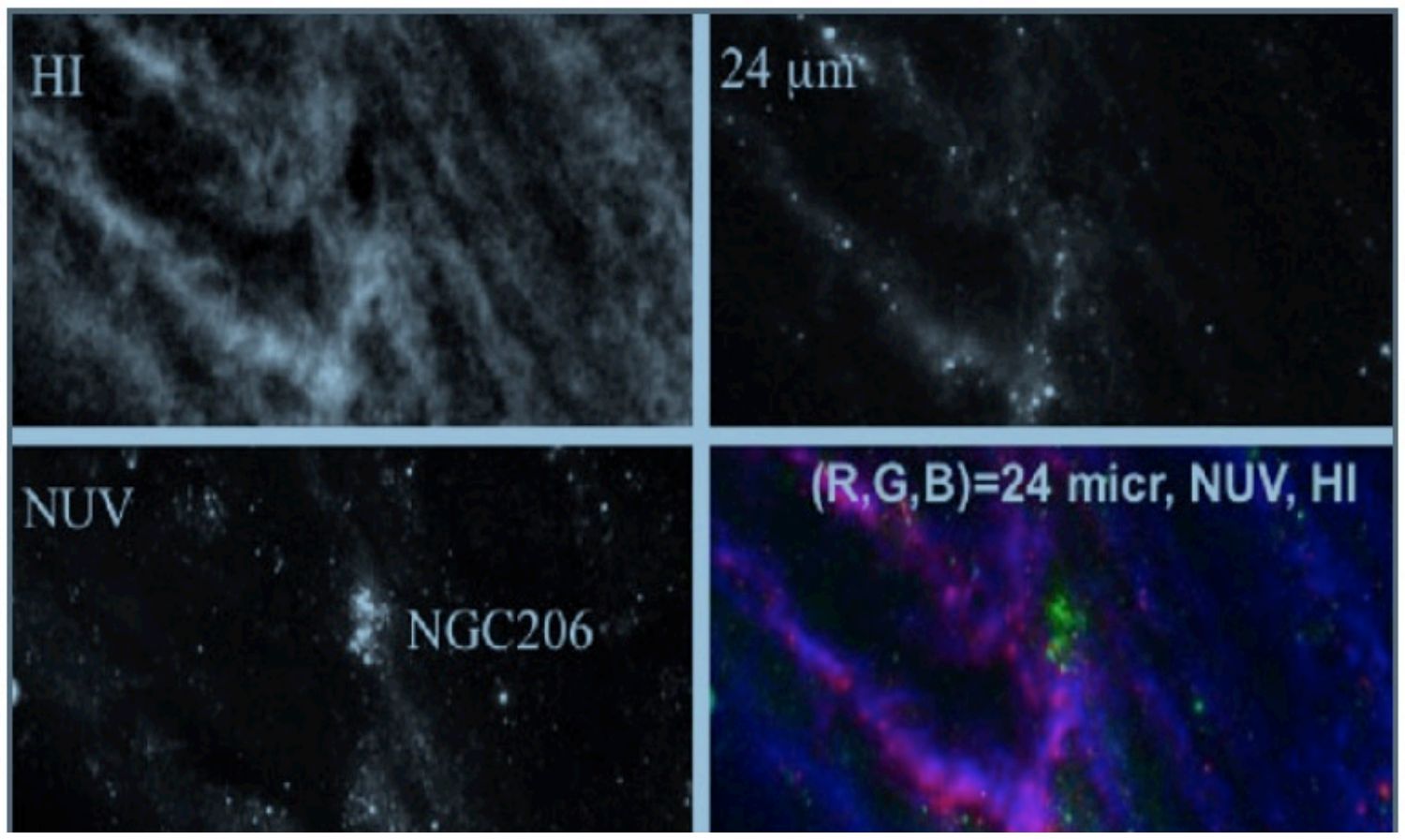

**Figure 2-3. The combination of UV, optical, IR, HI images puts powerful constraints on the star-formation process. CETUS will obtain FUV/NUV imagery of galaxies in the PHANGS sample. In a single pointing, CETUS will observe the entirety of each galaxy with a median resolution of 30-40 pc. Figure credit: Bianchi (2011)**

The PHANGS survey (Physics at High Angular resolution in Nearby GalaxieS) is using ALMA to observe CO emission arising from 300,000 star-formation sites in 74 spiral galaxies. Hubble is providing UV-optical imagery, and VLT MUSE is providing complementary Hα images. There are already H I images on hand (Hibbard, Lockman, van Gorkom, priv. comm.). CETUS will join in providing FUV and NUV images with a median resolution of 30-40 pc. These observations should reveal the physical interactions involved in star formation – the sequence of events, time scales at each stage, any overlaps in time of different stages and at different scales.

CETUS will also observe galaxies not in the PHANGS sample such as the newly identified “Little Blue Spheroids” (LBS’s) found in the recent GAMA survey (Moffett et al. 2016). In addition, CETUS will obtain NUV R~1,000 spectra covering 1,800 – 3,500 A of up to 100 star-forming regions within a galaxy with its multi-object spectrograph (MOS) **(Figure 2-4)** as recommended by Gordon (SciWP #441) . Figure 2-4. The CETUS MOS will obtain NUV spectra of up to 100 star-forming regions (circled in red) in a galaxy.

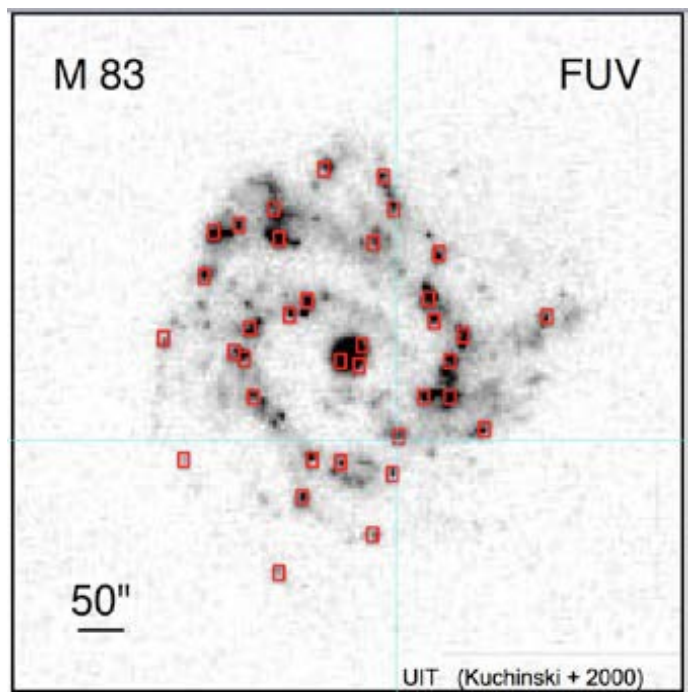

**Figure 2-4. The CETUS MOS will obtain NUV spectra of up to 100 star-forming regions (circled in red) in a galaxy.**

CETUS will provide more and better information on UV attenuation to further our understanding of the cosmic SED. The cosmic SED is the total wavelength-dependent emissivity of all galaxies within a given volume at a given redshift. It is a complex function of both the volume density of different galaxy types and the different processes [particularly UV dust attenuation and IR dust re-emission] that shape the SED’s of individual galaxies. A comparison by Baes et al. (2019) of the cosmic SED derived from an EAGLE simulation to the observed (z=0) cosmic SED derived from the GAMA survey (Driver et al. 2016) shows obvious discrepancies in the UV, which Baes et al. ascribe to an underestimate of the UV attenuation in the post-processing of the EAGLE simulation. CETUS will help solve such discrepancies by providing more and better data on UV attenuation by galaxies of different types, stellar mass, and environments.

## The Outskirts of Galaxies

*Low-z Starburst Galaxies* **Tim Heckman**

Galactic winds driven by the energy and momentum supplied by massive stars and supernovae play a crucial role in the evolution of galaxies and the inter-galactic medium. Most of the current data on these outflows measure the properties of interstellar absorption-lines in the rest-UV. While highly useful in measuring outflow velocities, these data provide no direct information on the spatial extent or structure of the outflow. Without this, we cannot reliably measure the outflow rates of mass, metals, momentum, and kinetic energy nor can we test competing models for the acceleration of the outflowing gas. This can be best addressed by imaging spectroscopy of these same UV resonance lines in emission. Therefore, CETUS would be a game-changer for understanding low-z starburst galaxies that are driving winds. These can serve as local laboratories for better understanding the winds seen in high-redshift galaxies.

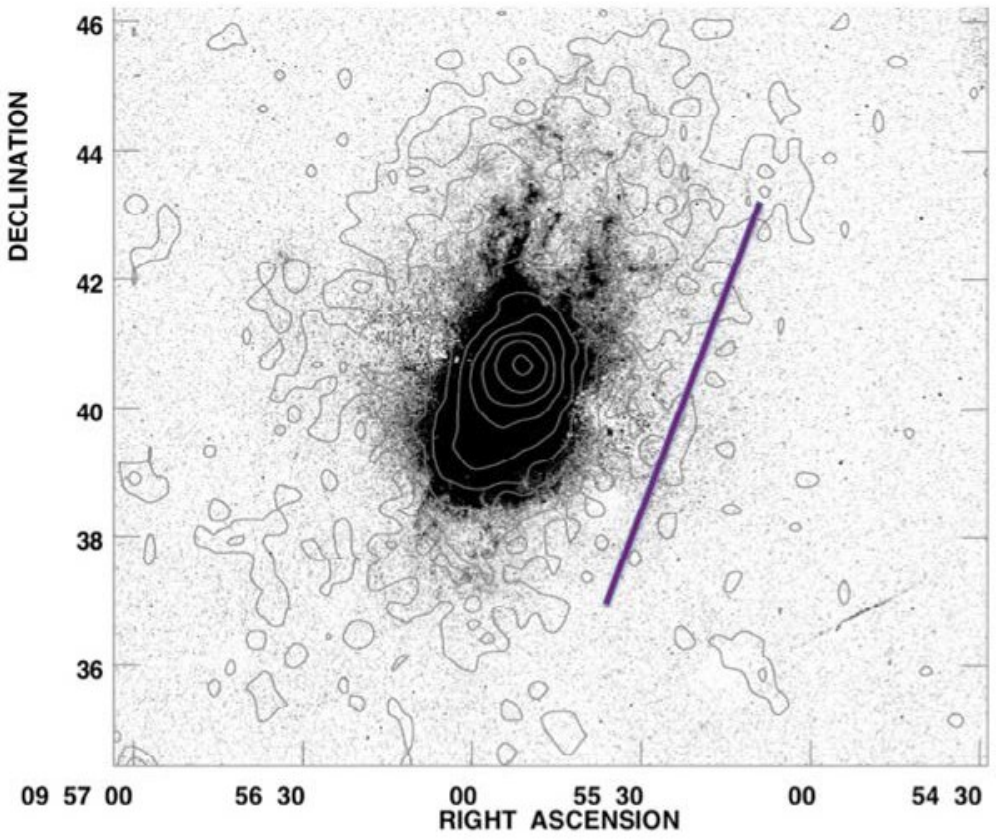

**Figure 2-5. The 2"x6'-long spectroscopic slit will scan across M82 producing a spectrogram covering 1,000-1,800 A in 2 grating settings.**

As an example, consider the prototypical starburst galaxy M82 and its famous wind. At a distance of only 3.6 Mpc, 1" subtends ~17 pc at the galaxy. The galaxy is seen nearly edge-on, while above and below the galactic plane, the outflow is traced by a filamentary structure of dust, gas, and X-ray emission. **Figure 2-5** shows a composite of Hα (grayscale) and X-ray emission (contours) of M82 (Lehnert, Heckman, & Weaver 1999). Note that the surface brightness of the Hα and X-ray emission depends on the line-of-sight integral of the square of the gas density, while the surface brightness of resonantly scattered UV line emission depends on the integral of the density. Thus, these UV lines are better probes of the less dense gas known to dominate the outflow energetics, and they can be directly connected to the gas seen in absorption.

The figure also shows the CETUS 6'-long slit poised to sample the FUV spectrum of the galaxy and its ejecta in the G120M mode as it is moved push-broom style across M82 parallel to the galaxy disk. G120M spectra span the wavelength range from 1000 Å to 1400 Å, including O VI 1032Å, 1038Å, Si III 1206.5Å, N V 1238Å, 1242Å, Si II 1260Å, O I 1302Å, and C II 1334.5Å, Si IV 1393 Å. The G160M mode covers the longer wavelengths. Although the CETUS optics produce a spectral resolving power, $\lambda/\Delta\lambda$=R~20,000, the long slit is 2" wide (TBR) so that for extended sources, the resolving power is reduced to R~2,000. This resolving power is well-matched to the line widths measured in Hα (a few hundred km/s).

The low redshift of M 82 precludes observations of Lyα, because it lies in the damping wings of the Milky Way absorption-line. However, there are many other bright starburst winds in galaxies at redshifts that provide access to Lyα (e.g. NGC 1482, NGC 3079, NGC 6240, and Arp 220). While these more distant systems have smaller angular sizes than the M82 wind, we expect Lyα emission to be detectable to larger radii than the metal lines due to its greater relative brightness.

*The Dusty Outskirts of Galaxies* **Edmund Hodges-Kluck, SWP #276**

Star-forming galaxies are surrounded by reflection nebulae produced by halo dust scattering some of the starlight from the disk into our line of sight. Such reflection nebulae are best seen around edge-on galaxies, like M82, in the UV continuum because the sky is dark and the scattering cross-section of dust is high. The UV/optical image of M82 from the *Swift* UVOT (**Figure 2-7**), shows a highly structured reflection nebula in blue. As dust can only form in dense regions, the dust must have originated in the galaxy. The dust lifetime in the halo outskirts can exceed the Hubble time, so the total amount of dust traces the prior history of galactic outflows. Recent studies (e.g. Ménard et al. 2010) indicate that there is at least as much dust outside of galaxies as within them.

The dust halo may also carry a large amount of metals, since refractory metals are depleted on dust in the ISM. Detection of variations in nebular absorption lines within the halo as well as differences in metallicity between the dust in the halo and that in the disk, which can be measured from the dust SED, will provide constraints on dust properties.

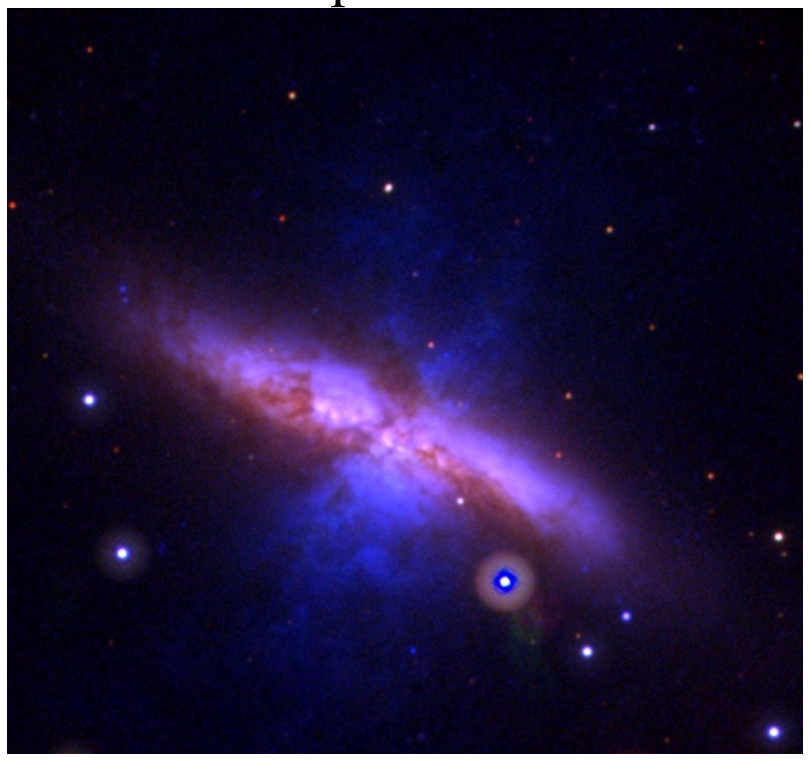

**Figure 2-6. The dusty outflow of M82 is visible in the near-UV (blue).**

CETUS will make a large population study of dusty galactic outflows in nearby galaxies of all types. With its wide field of view and 10 filters covering both the NUV and FUV, the CETUS camera will obtain a coarse SED to be compared with that of the galactic disk. The FUV is particularly important for measuring the UV slope parameter, □UV, which in turn constrains the dust composition. In addition, the CETUS R~20,000 FUV spectrograph with 6'-long-slit, CETUS will measure the depletion ratios in halo gas back-lit by a quasar or the UV afterglow of a GRB through measurements of absorption lines of refractory elements like O, Si and Fe.

*Emission-Line Halos Around Low-z Galaxies of All Masses* **Matthew Hayes**

**Martin # 565 (z=1); Chen #361; Burchett #591 (ICM); McCandliss #592 (LyC)**

Far-UV long-pass filters on Hubble's Advanced Camera for Surveys (ACS) have been used to isolate Lyα emission halos. The technique devised by Hayes et al. (2013) and Östlin et al. (2014) is to construct virtual narrow-band filters from subtraction of adjacent long-pass filters as shown in the right panel of **Figure 2-7**. The shorter-wavelength virtual filter registers Lyman-α, and the longer-wavelength filter gives a measure of the far-UV continuum flux.

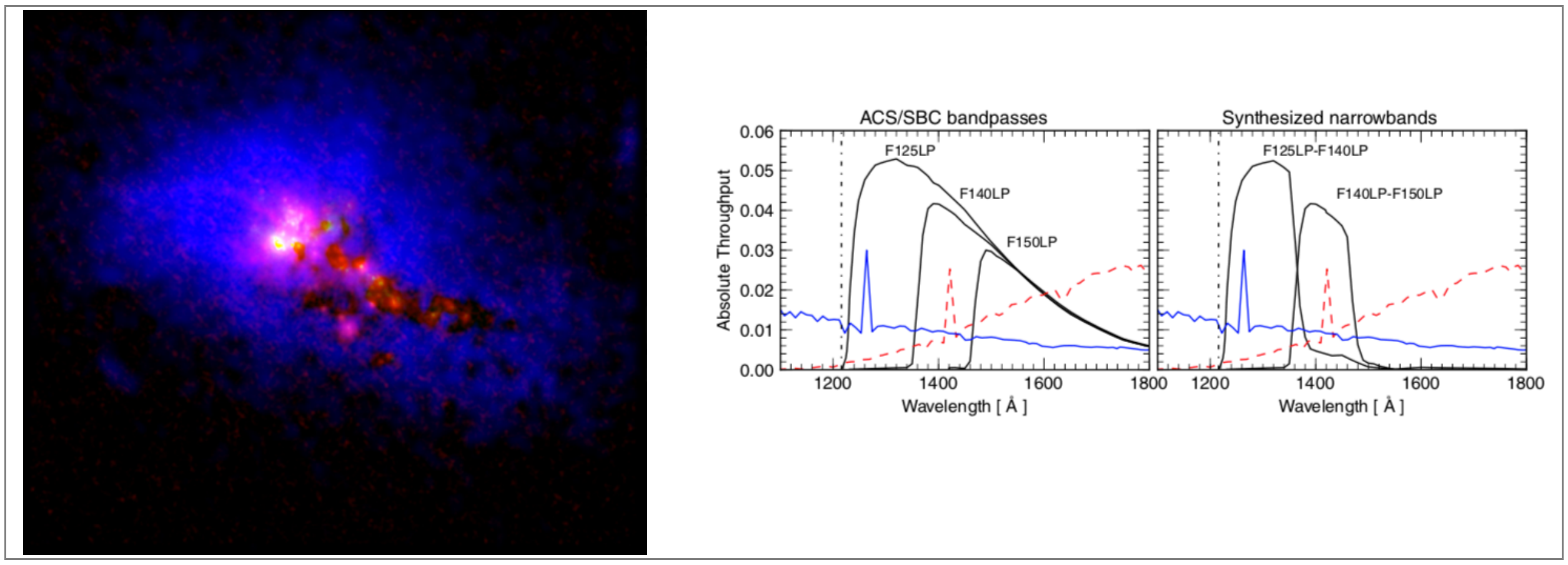

**Figure 2-7.** ***Left:*** **low-z starburst galaxy Mrk 259. Lya is shown by blue light, and far-UV continuum radiation in red, as captured by the long-pass filters of Hubble's ACS/SBC camera, shown in the *right* panel. (Hayes et al. 2013)**

The CETUS Far-UV Camera will make use of the same set of long-pass filters as Hubble's ACS. And with its 17'x17' field size and 23-fold increase in sensitivity to extended sources compared to HST, CETUS will trace Lyα radiation to surface-brightness levels more than an order of magnitude fainter than possible with HST. Such faint emission will be found at galactocentric radii several tens of kpc out into the circumgalactic medium, at radial distances where currently only chance QSO sightlines can probe the halo gas using absorption techniques.

The dramatic increase in both sensitivity and field size of CETUS opens unique possibilities:

- Determine exactly how large galaxies are in Lyα, and examine the transition between ISM and circumgalactic gas;
- Study galaxies of significantly larger angular size, e.g. SDSS IV (MANGA) systems, or halo gas in galaxy groups, e.g. Hickson Compact Groups;
- Discover {25,35,30} new galaxies at <z>={0.05, 0.15,0.27} in 1-hr exposures using filters {F125LP, F140LP,F150LP} (based on Lyα luminosity function of Wold et al., 2017);
- Image metal-line emitting regions in deeper exposures such as OVI 1031, 1037 emission in LAE's identified in F150LP (z~0.25), and CIV (1548,1551) in LAE's identified in F125LP (z~0.05) (Hayes et al. 2016). In Hubble program, GO 15655, Johnson et al. (2019) will obtain deep, high-resolution, narrow-band Lyα *and* OVI images of two luminous, obscured AGN at z~0.26: one radio-loud, and the other, radio-quiet.

*The Outskirts of Massive Galaxies* **Oppenheimer #309**

Oppenheimer et al. (SciWP #309) argue that the UV/optical spectral region is useful for sampling cool to warm-hot gas in the CGM, but :"the majority of baryons in halos more massive than $M_{halo} \sim 10^{12} M_\odot$ along with their physics and dynamics remain invisible because that gas is heated above the UV ionization states … Information on many of the essential drivers of galaxy evolution is primarily contained in this "missing" hot gas phase."

We are persuaded that the halos of massive, red-sequence galaxies must be hot ($T>10^6$ K), but not that gas in the hot phase is out of reach of the far-UV spectral region. [Fe XXI] λ1354 emission has been detected in the COS spectrum of a filament of M87 (Anderson, Sunyaev et al. 2018). This emission line is indicative of a 1.4 keV ($T \sim 1.6 \times 10^7$ K) plasma. **Figure 2-8** shows a portion of the COS spectrum of a small patch (2.5" in diameter) of the filament showing the multi-phase spectrum ranging from cool (C II) to warm (Si IV, O IV) to hot ([Fe XXI]). The detection of [Fe XXI] suggests that not only can CETUS be used to study gas in all phases but that with its 6'-long slit, CETUS can map all phases of plasma around very massive galaxies or in the intracluster medium of galaxy clusters such as the Perseus, Virgo, or Coma cluster.

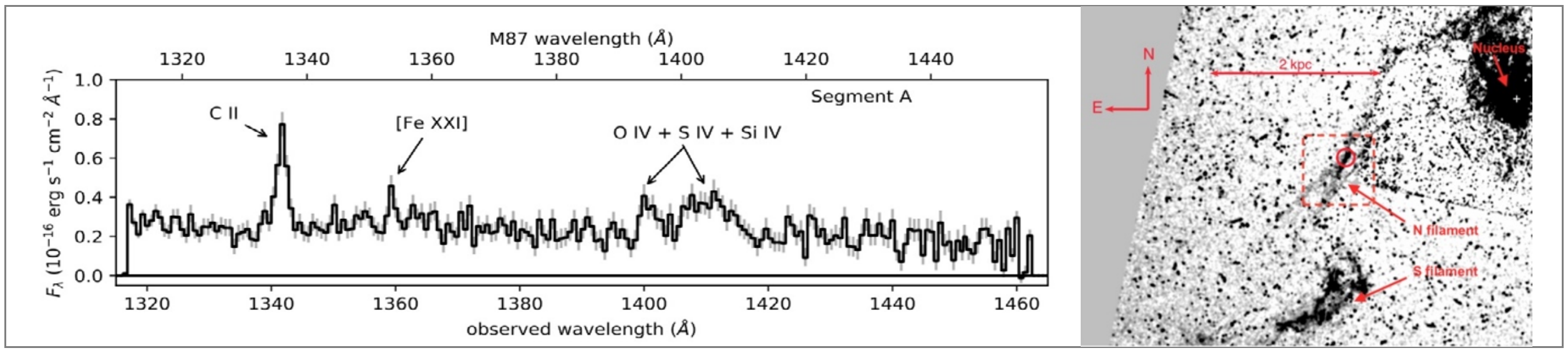

**Figure 2-8. CETUS will study the hot circumgalactic medium as revealed by [Fe XXI] λ1354 emission (left). Hubble's COS only viewed a small region (right; circled in red), while CETUS' far-UV spectrograph will show how the [Fe XXI] line and other spectral lines vary with position along the 6' slit. Figure credit: Anderson, Sunyaev et al. 2018**

**The Circumgalactic Medium** **Peeples #409, Lehner #524, Tumlinson #421, Tripp #345**

The Warm-Hot Circumgalactic Medium (CGM): The circumgalactic medium (CGM) comprises gas and dust surrounding a galaxy. It contains more baryonic (normal) matter than does the galaxy itself. Whether it contains all the matter needed to account for the "missing baryons" at low-redshift is still open to debate. Most matter in the CGM is "hidden" in a warm-hot phase (100,000 – 500,000 K) or hot phase ($T>10^6$ K). The only observable signature of the warm-hot interstellar medium (WHIM) is the O VI doublet at 1032 Å, 1038 Å, which is out of reach of Hubble's COS spectrograph for targets with redshifts, z<0.10. COS observations of O VI and other FUV resonance lines arising in the CGM of galaxies at z>0.10 have yielded important statistics on the frequency, column density, velocity, and velocity dispersion of O VI in the CGM of galaxies (e.g. Tumlinson et al. 2017, Keeney et al. 2017). However, they have not led to definitive conclusions about the missing baryons problem or any evolutionary connection of the CGM to the host galaxy.

CETUS FUV spectroscopic observations will greatly improve our knowledge about the CGM, its interplay with the host galaxy, and the missing baryon problem in several ways:

*Access to the WHIM surrounding nearby galaxies*. In this report, we refer to the Lyman-UV or LUV as the spectral region, 1000-1150 Å, and the spectral region probed by the spectrograph as the LUV/FUV (1000-1800 Å). The CETUS LUV/FUV spectrograph can reach the O VI 1032, 1038 doublet arising in the WHIM of nearby galaxies, whereas COS is restricted to galaxies at z>0.10, so it cannot study the CGM of the large population of low-luminosity galaxies, and its sample is biased toward massive galaxies. Based on Bowen's (2018) *QSO-Galaxy Pairs* catalog, we find that CETUS can obtain full LUV/FUV spectra (1,000-1,800 Å) of the CGM surrounding over 400 nearby galaxies backlit by a QSO or AGN. Of these pairs, 160 have a background AGN/QSO brighter than m(FUV)=17.4 (**Figure 2-9**), which puts them within reach of CETUS. As can be seen from **Figure 2-10**, the foreground galaxies are often dwarf galaxies ($M_B$> -18), which are not well sampled by COS.

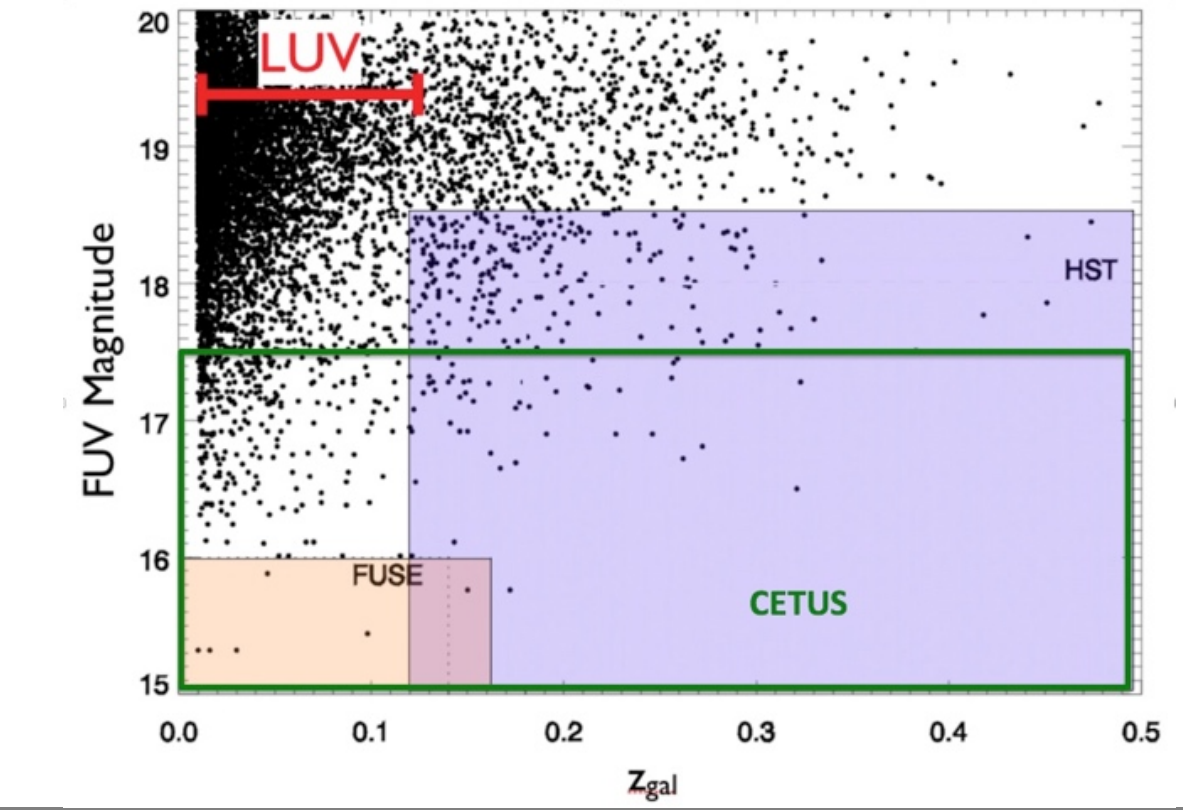

**Figure 2-9. CETUS will observe the WHIM of nearby galaxies, which are out of Hubble's reach. (Figure adapted from N. Lehner)**

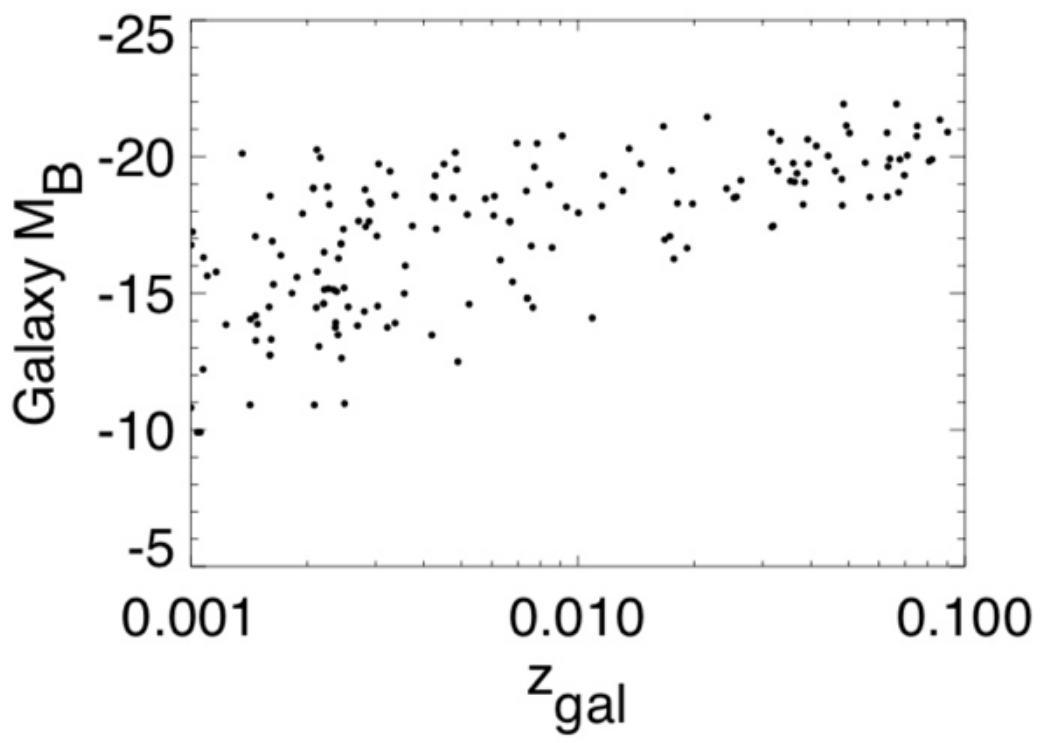

**Figure 2-10. Most known galaxies less luminous than $M_B$= -18 are nearby so access to the Lyman-UV is essential.**

*Access to UV observations of the host galaxy.* What CGM spectra don't yield is information about the galaxies themselves. In fact, we usually know more about the CGM than we do about its host galaxy. This is a great loss, because understanding galaxy evolution depends on knowing how the CGM influences the properties of the host galaxy, and vice versa. CETUS will help us understand the evolutionary connection between a galaxy and its halo (CGM). With a resolution of only tens to a few hundred parsecs in nearby galaxies but a field of view covering the whole galaxy, the CETUS FUV Camera will study the microphysics of nearby galaxies (**Figure 2-11**). Many of these galaxies are bright enough for direct spectroscopy with CETUS (after binning). The CETUS FUV spectrograph can obtain FUV long-slit spectra of the cross-section of galaxies in the CETUS sample (**Figure 2-12**).

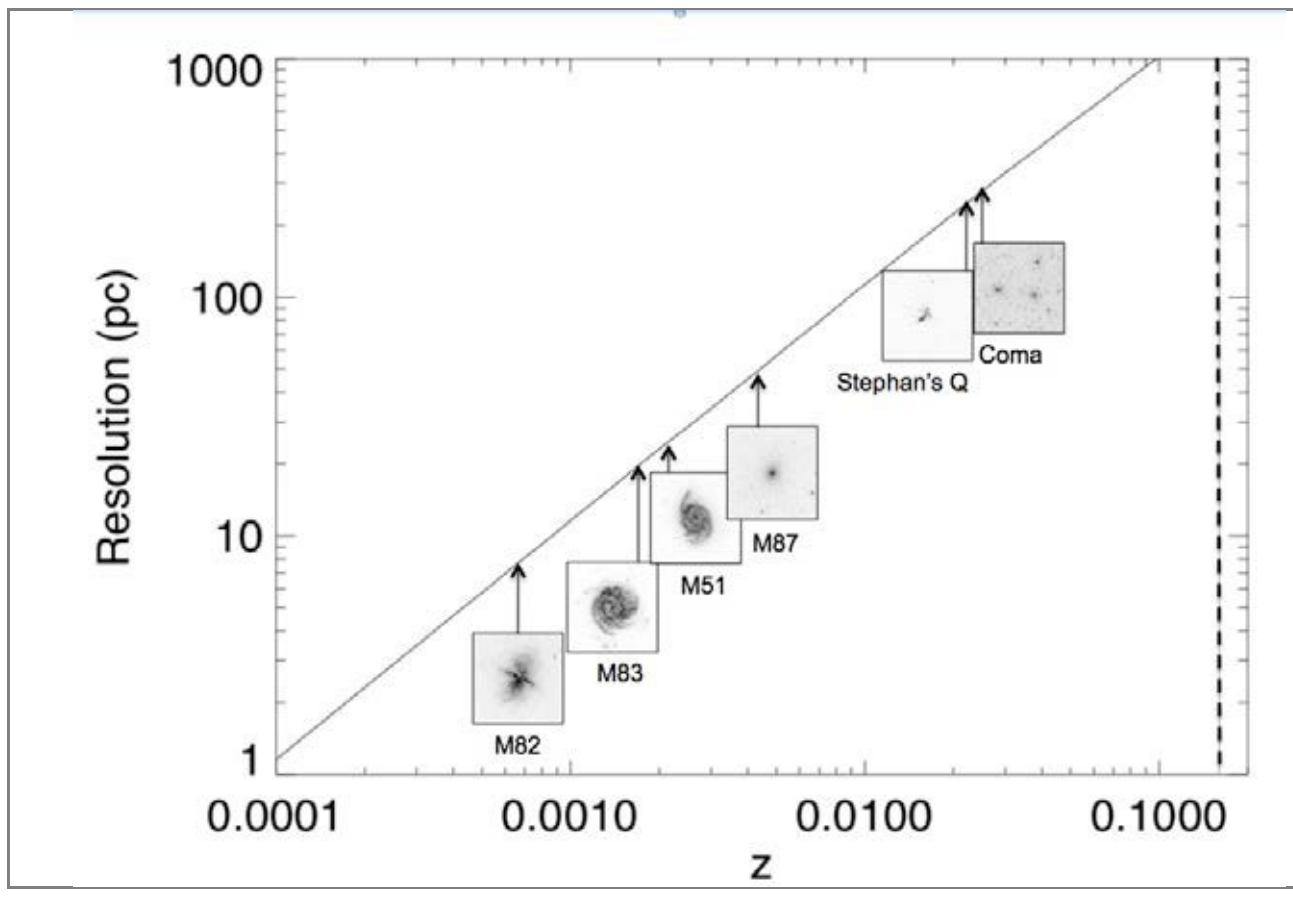

**Figure 2-11. CETUS images have a field of view (17.4'x17.4') sufficient to cover a nearby galaxy at a resolution (0.55" in FUV; 0.40" in NUV) for studying the microphysics of nearby galaxies.**

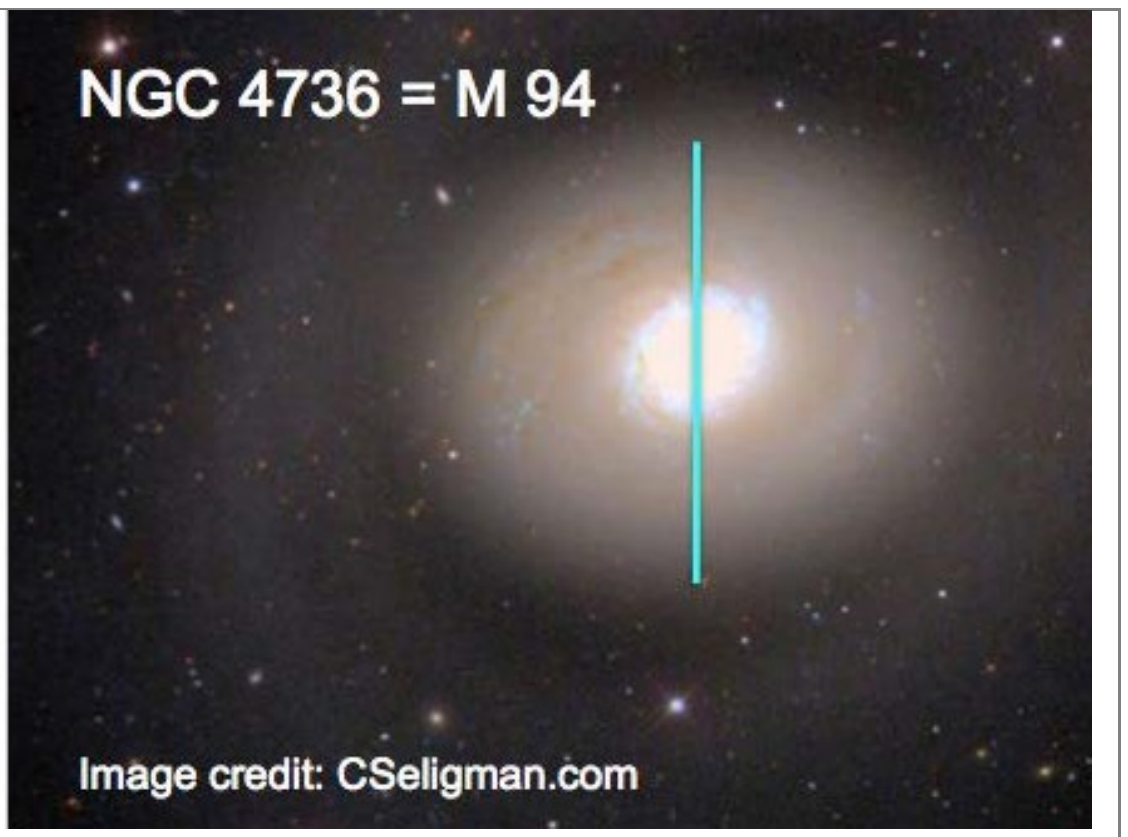

**Figure 2-12. CETUS FUV long-slit spectra (turquoise) can extend along the diameter of nearly all galaxies.**

*Reliable Measurements of FUV Line Strengths.* Measurements of the strength of spectral lines appearing in a Hubble/COS spectrum are vulnerable to systematic errors due to what is formally called "mid-spatial frequency wavefront errors" of Hubble's primary mirror (*COS Instrument Handbook*). Measurements of the COS line-spread function indicate that ~40% of the total strength of a FUV spectral line is hidden in the broad, low-level wings of the line spread function (LSF) as shown in **Figure 2-13**. This is a particularly insidious problem for measurements of absorption lines like the O VI doublet since the spurious low-level wings of the spectral lines cannot be distinguished from the continuum flux. Since the O VI doublet is the only spectral diagnostic of the warm-hot medium, and since the warm-hot CGM may be the dominant component of the nearby universe, this is a vexing problem. A systematic 40% loss in line strength is on the order of the uncertainty in the "missing-baryon" problem, so it is essential that the CETUS primary mirror have better image quality than when the Hubble telescope mirrors were made. Fortunately, as will be described in Section 3, modern mirror-polishing techniques in industry have greatly improved since what was possible at the time for Hubble, so we can expect to obtain reliable measurements of spectral lines in CETUS spectra.

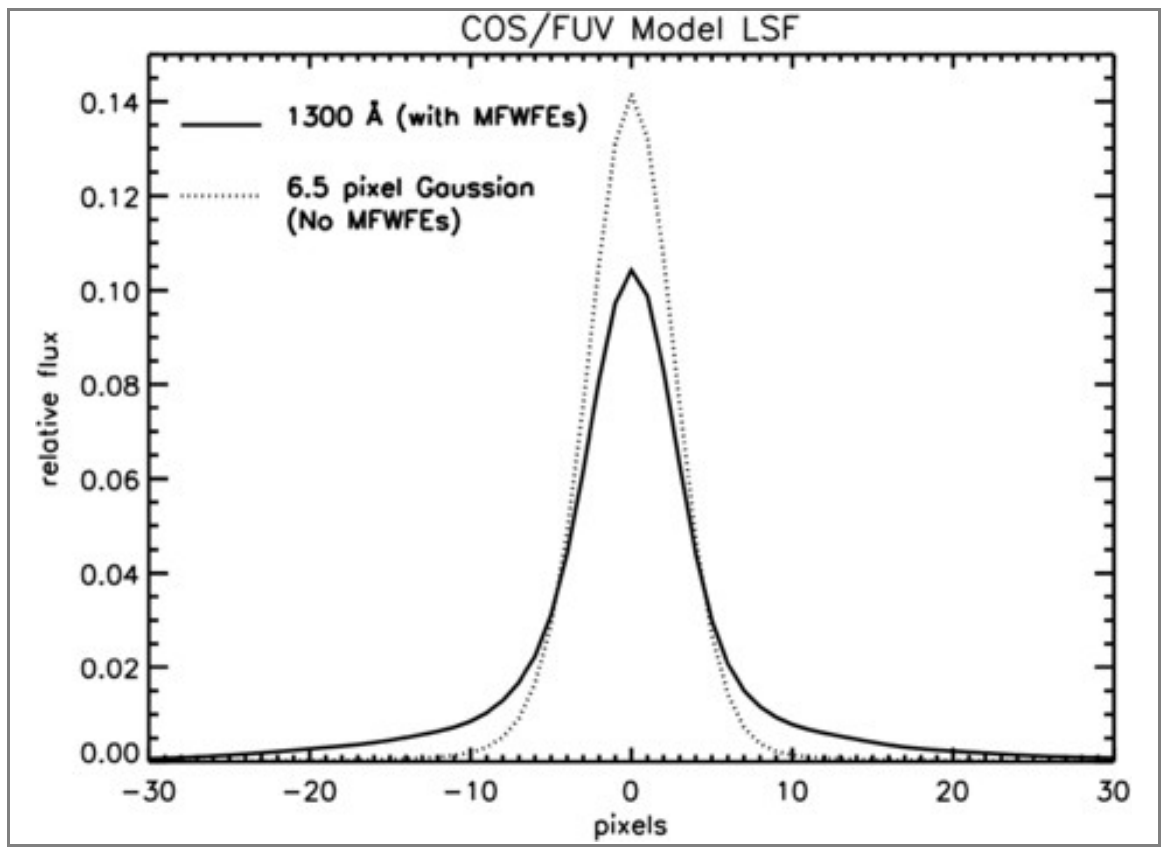

**Figure 2-13. Monochromatic spectral lines in CETUS spectra will have Gaussian profiles leading to accurate measurements of line strengths. (from HST COS ISR 2009-1)**

**Role of the CGM in galaxy evolution** **Bryan Terrazas**

Recent results from cosmological models (e.g. EAGLE, IllustrisTNG, FIRE-2) indicate that feedback from stars and black holes determines how effectively CGM gas can cool as it is falling back onto the galaxy, thereby providing fuel for star formation. The CGM needs to be continually prevented from cooling to prevent star formation from occurring in the galaxy (Terrazas priv. comm., 2017, 2019). How cooling is prevented is not observationally constrained well enough to inform models, so this physical process differs in all models. Illustris-TNG uses mechanical energy, EAGLE uses thermal energy, and other models use a combination of these. None of the models does an adequate job at matching observations, so better observational constraints are needed. CETUS will help constrain feedback from black holes and stellar winds and supernovae in two ways: (1) It will continue and extend observations by COS to understand the properties of winds from supermassive black holes (see below); and (2) CETUS will search for correlations between the properties of the CGM and that of the host galaxy that might show how feedback affects the CGM. For example, understanding the ionic phase and temperatures of the gas around star-forming versus quiescent galaxies could tell us about whether the feedback is heating gas or whether it is pushing gas away (thermal vs kinetic).

**Quasar and AGN Winds** **Sylvain Veilleux**

Galactic winds in gas-rich mergers are an essential element of galaxy and supermassive black-hole evolution (e.g., Hopkins et al. 2006; Cicone et al. 2014; Rupke et al. 2017). The most powerful of these outflows are driven by quasars and likely feed the circumgalactic medium. The outflow energetics are often dominated by the outer (> kpc) and cooler dusty molecular and neutral atomic gas phase, but the driving mechanism is best probed by the inner (sub-kpc) highly ionized gas phase. While current X-ray observatories are not sensitive enough to carry out a systematic survey of these inner winds, results from recent and on-going FUV studies with COS on HST indicate that CETUS will be ideally suited for this task. Prominent, highly blueshifted (1000 km/s) Lyα emission has been detected in most ultra-luminous infrared galaxies (ULIRGs), often accompanied by blueshifted absorption features from N V and O VI (Martin et al. 2015). The internal kinematics of ULIRGs seem to be the single most important factor determining the profile and escape fraction of the Lyα emission. However, the trends so far are entirely driven by the few AGN-ULIRGs in the current sample and are therefore highly uncertain.

CETUS will enable us to study with unprecedented statistics the gaseous environments of nearby gas-rich mergers as a function of host properties and age across the merger sequence, ULIRG → QSO (e.g., Veilleux et al. 2009). The excellent spectral resolution (~15 km/s) of the FUV spectra will enable us to distinguish between quasar-driven outflows, starburst-driven winds, and tidal debris around mergers. CETUS spectra covering 1000-1800 Å will also provide new constraints on the critically important warm-hot gas phase associated with the cooling shocked ISM predicted in some quasar feedback models. Perhaps more importantly, the excellent sensitivity of CETUS down to ~1000Å will make it possible to explore, for the first time, the molecular gas phase of quasar-driven winds. These data will provide a crucial test of some quasar wind models where the outflowing cold molecular material is proposed to condense out of the shocked ISM.

There are 56 $z<0.16$ AGN in the Milliquas Catalog (2018) whose line of sight pierces the CGM of a foreground galaxy. CETUS can easily obtain the FUV spectrum of such close AGN's (100-180 nm) and the absorption spectrum of the CGM of the foreground galaxy. And in its long-slit mode, the CETUS FUV spectrograph can obtain the FUV spectrum of the foreground galaxy's cross-section.

**The Fossil Record of the First Stars** **Ian Roederer**

The nucleosynthetic signatures of the first stars and supernovae are imprinted in the elements observed in second-generation stars, which are likely found among the most metal-poor stars. These ancient stars have less than 1/3000 of the solar iron abundance ([Fe/H] ≤ 3.5). About 80 such stars are known today, and hundreds more are expected to be found among ongoing and future surveys (e.g., LAMOST, Sky-Mapper, Pristine, 4MOST, WEAVE, LSST). Only a few tens of absorption lines are commonly found in the optical and near-IR ($\lambda > 310.0$ nm) spectra of these stars, so only ~5-10 elements are regularly detected. This limits the utility of these stars for understanding the nature of the first stars and first supernovae. Many other elements are expected to be present, but they remain undetected. The strongest transitions of these elements are in the UV, below the atmospheric cutoff, requiring space facilities for detection (**Figure 2-14**).

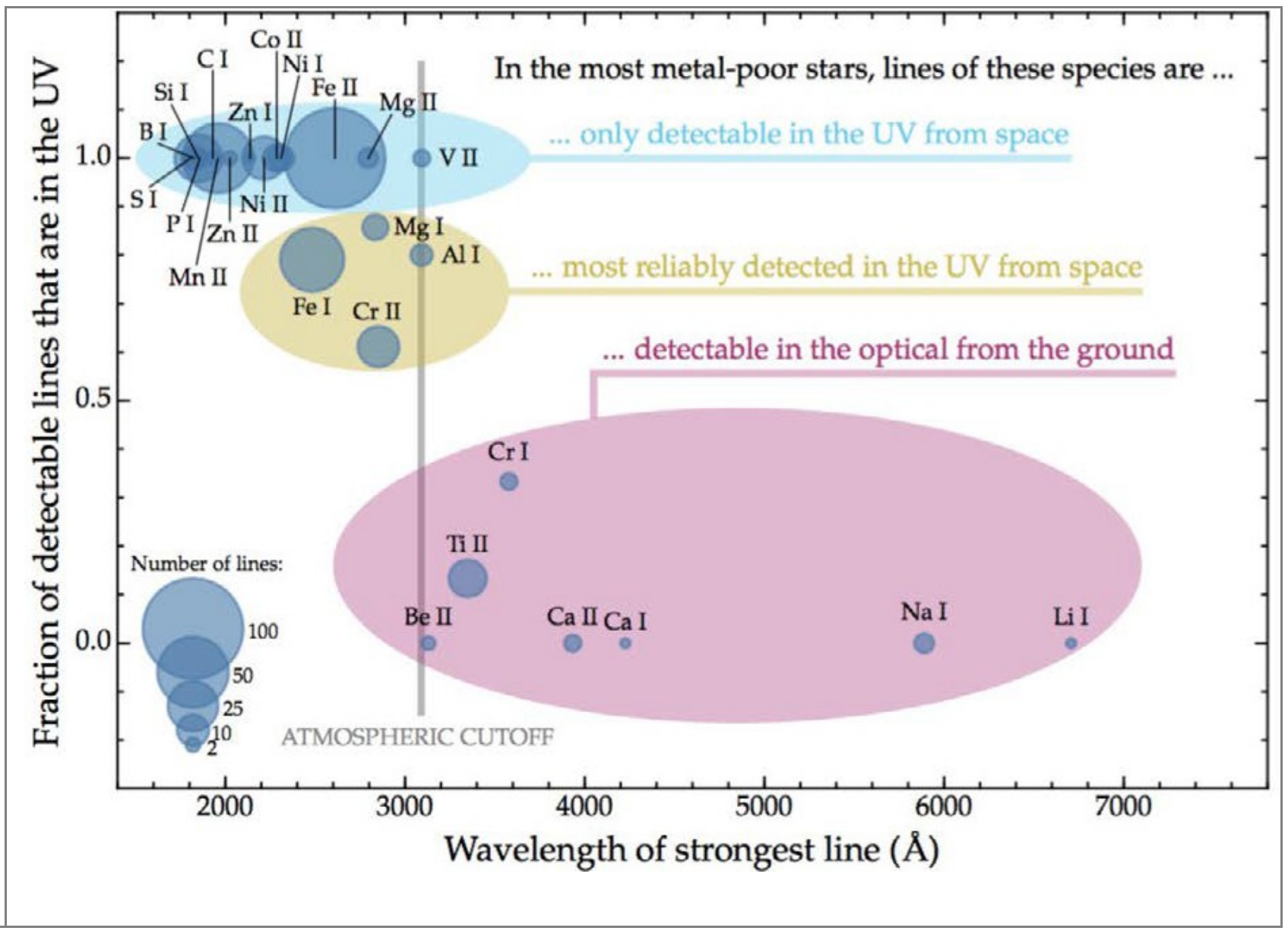

**Figure 2-14. Improvement in metal detections in the most metal-poor stars is enabled by UV spectra. Typically, ~8-10 elements can be detected in the optical alone, but UV spectra can enable the detection of ~20 elements, probing supernova physics (carbon through zinc, $6 \leq Z \leq 30$) and Big Bang nucleosynthesis, stellar evolution, and spallation reactions (lithium, beryllium, and boron, $3 \leq Z \leq 5$).**

With the Hubble Space Telescope (HST), however, we are limited to studying only the brightest stars, and only one star with V < 10 and [Fe/H] < -3.5 is known at present (BD+44 493; Roederer et al. 2016). Many of the most metal-poor (or iron poor) stars are too faint for HST, as shown in **Figure 2-15**.

The CETUS R~40,000 NUV echelle spectrograph offers a new opportunity to expand our capability to observe the sample of the most metal-poor stars known by orders of magnitude, which would begin to reveal the true diversity of the first stars and first supernovae. High spectral resolving power (R~ 40,000) and high S/N ratios (S/N ~ 50/1 or greater) are ideal to detect and accurately measure the relatively weak absorption lines produced by these elements in FGK-type stellar photospheres. CETUS would enable long, *uninterrupted* exposures of these fainter stars for the first time. This would revolutionize our understanding of the first stars, the first suernovae, and the first metals in the Universe.

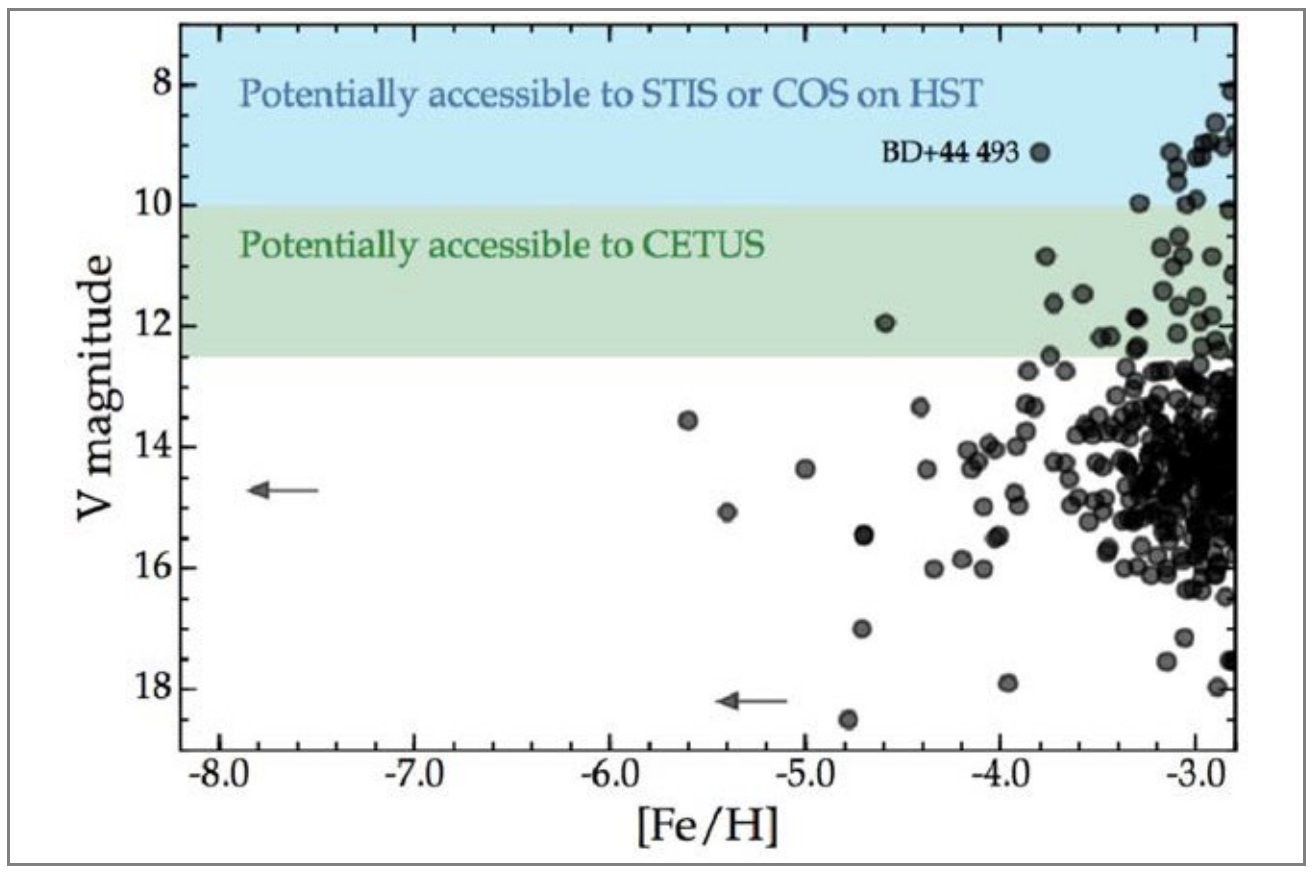

**Figure 2-15. Relatively bright metal-poor stars will be observed by CETUS. Data collected from the JINAbase abundance database (Abohalima & Frebel 2018).**

## 2.2 STAR-FORMING GALAXIES AT Z=1

We now know the star-formation history of the universe (**Figure 2-16**) thanks to surveys by many telescopes, but we still don't understand it, because: "[The evolution of the star-formation rate density] says little about the inner workings of galaxies, i.e., their "metabolism" and the basic process of ingestion (gas infall and cooling), digestion (star formation), and excretion (outflows). Ultimately, it also says little about the mapping from dark matter halos to their baryonic components. Its roots are in optical-IR astronomy, statistics, stellar populations, and phenomenology, rather than in the physics of the ISM, self-regulated accretion and star formation, stellar feedback, and SN-driven galactic winds."… (Madau & Dickinson, 2014)

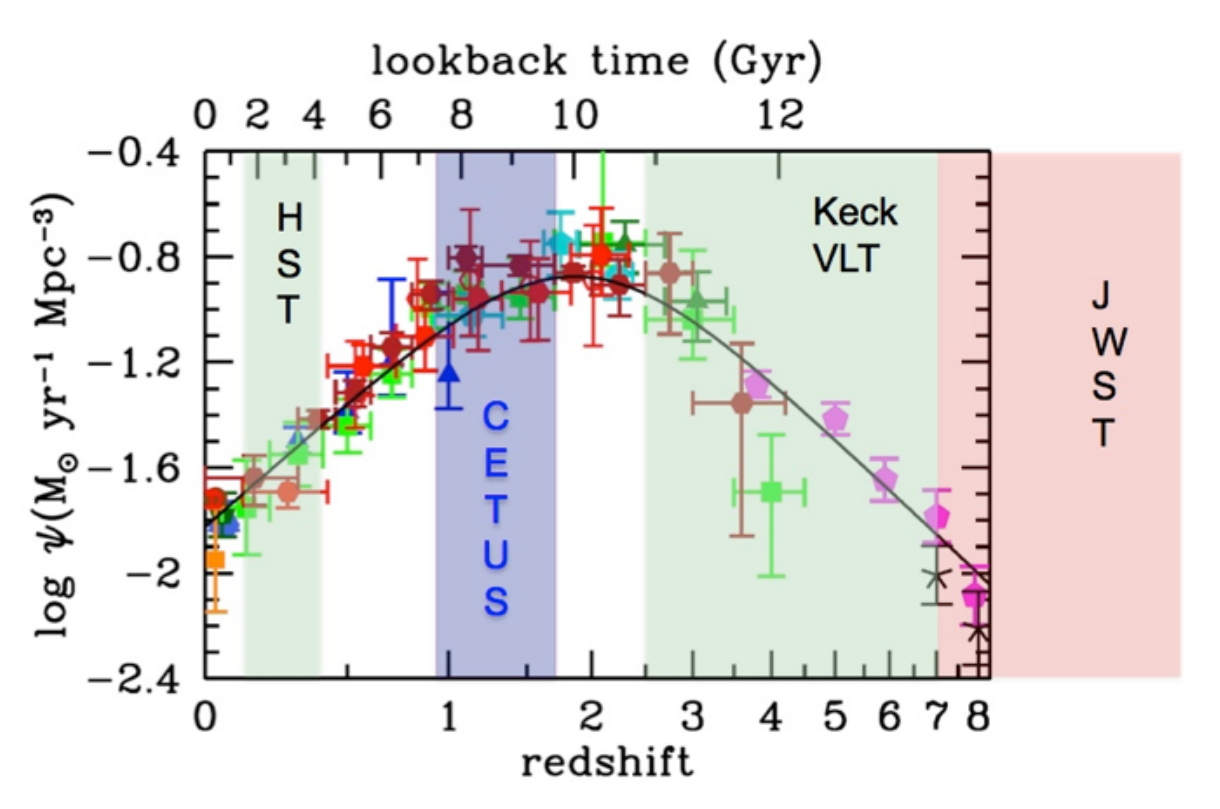

**Figure 2-16. The history of the rate of star formation raises questions, particularly the reason for the turn-over in star-formation rate at z~2. Figure adapted from Madau & Dickenson (2014).**

Rest far-UV (FUV) spectra have proven to be very useful observations for deriving information about the physics of the ISM, accretion and star formation, stellar and AGN feedback.

- *Physics of the ISM*. The rest FUV is exceptionally rich in spectral-line diagnostics, particularly resonance lines of abundant elements in a variety of ionization states. UV spectra from *Copernicus*, IUE, FUSE, and Hubble's GHRS, STIS, and COS, have refined and deepened our understanding of the ISM. Like *Copernicus* and FUSE, CETUS is sensitive to the "Lyman UV" down to 100 nm and is expected to make new discoveries about the properties of the ISM and circumgalactic medium (CGM).
- *Accretion and star formation*. Star formation rates based on UV flux require significant correction for UV dust attenuation. CETUS will map the dust attenuation parameters in nearby galaxies and derive correlations with the far-IR flux, galaxy type, stellar mass, and other galaxy properties that can be applied to z~1 galaxies.
- *Feedback & Galactic Winds*. The FUV lays bare stellar and AGN feedback processes e.g., stellar and AGN winds, supernovae, photo-ionization and heating – all thought to be important drivers of galaxy evolution.

To emphasize the power of far-UV spectral diagnostics, consider the Keck/LRIS rest far-UV spectrum of the z=2.7 galaxy, MS 1512-cB58 analyzed by Pettini et al. (2000). This single, rest far-UV spectrum of this galaxy yielded: a high rate of star formation, SFR~40 $M_\odot$/yr, but an even higher rate of mass-loss, $\dot{M}$~60 $M_\odot$/yr, via a galactic outflow at a velocity~ 200 km/s, protracted star formation (as opposed to an instantaneous burst), an initial mass function (IMF) consistent with Salpeter IMF with an upper stellar mass limit, $M_u$>50 $M_\odot$, a metallicity, Z~1/4 $Z_\odot$ (both stars & gas), a column density in H I, $N_{HI}$=7.5x10$^{20}$ cm$^{-2}$, and reddening by dust, E(B-V)~0.1-0.3.

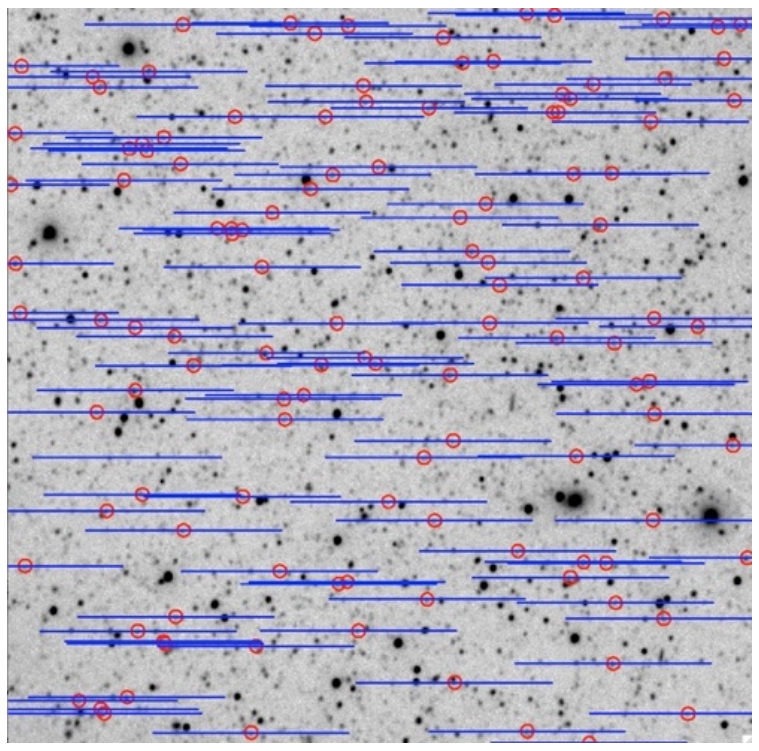

**Figure 2-17. The CETUS MOS will obtain spectra (blue lines) of up to ~100 galaxies brighter than m=24.3 AB (circled in red) in the field of view of the MSA.**

As shown in **Figure 2-16**, large ground-based telescopes like Keck and VLT have obtained rest far-UV spectra of galaxies at z>2.5, and JWST promises to obtain rest far-UV spectra of very high redshift galaxies. However, there is a gap in the redshift region, z~0.5-2. CETUS will at least partially fill this gap by making a massive, near-UV (rest FUV) spectroscopic and imaging survey of galaxies at z~1. As shown in **Figure 2-17**, the CETUS MOS will obtain NUV (rest FUV) spectra of ~100 z~1 galaxies in a single exposure.

CETUS spectra will then be combined with optical spectra of the same galaxies as obtained by Subaru's Prime Focus Spectrograph (PFS) and Very Large Telescope MOONS, and possibly other ground-based observatories to obtain a full picture of the properties of z~1 galaxies.

The figures below show details about the target z~1 galaxies as drawn from Cosmos Mock Catalog (Jouvel et al. 2011) based on multiple-telescope surveys of the COSMOS field. All the selected targets are classified as star-forming galaxies in the redshift range, z=0.8-1.3, and all are much brighter than L* (**Figure 2-18**). In order to have ~100 galaxies in the CETUS MOS field of view, CETUS must be able to obtain R~1,000 spectra of star-forming galaxies whose NUV flux is $4x10^{-18}$ $erg/s/cm^2/Å$ or brighter. This requirement sets the telescope aperture (D=1.5 m), the field of view of the MOS microshutter array (17.4'x17.4'), and f-ratio (f/5). As shown in **Figure 2-19**, the selected targets have a surprisingly wide distribution in stellar mass.

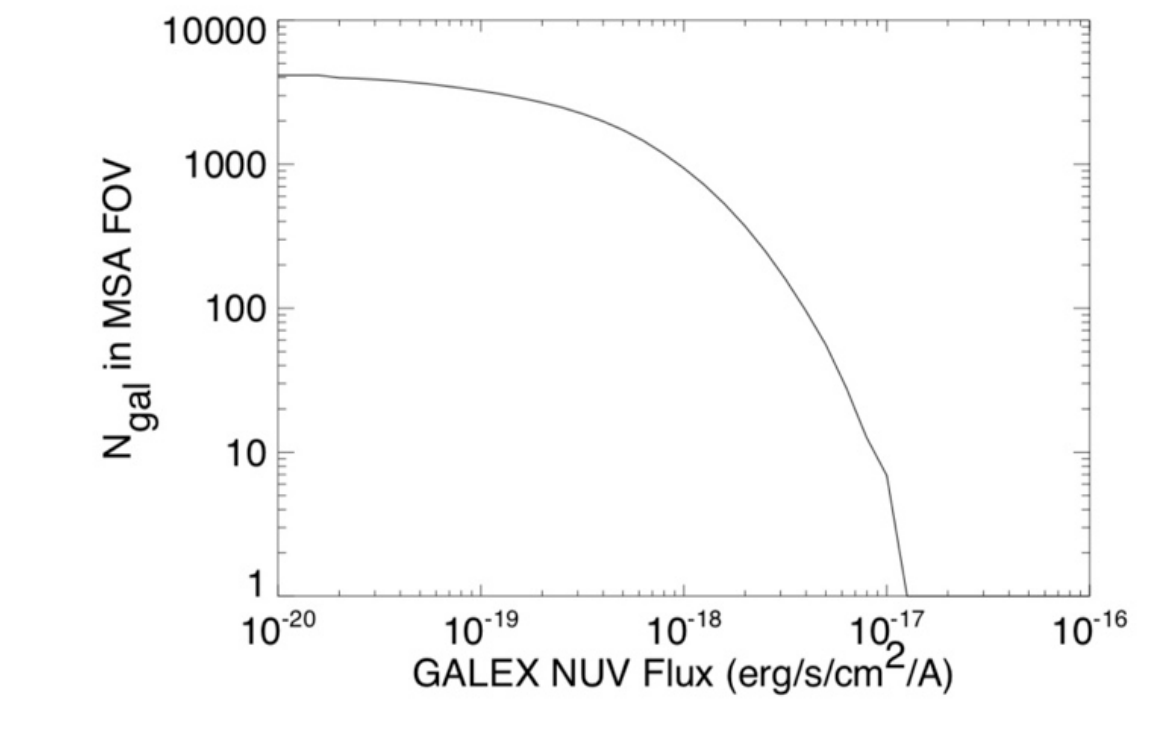

**Figure 2-18. CETUS will obtain NUV spectra of up to 100 star-forming galaxies in the field.**

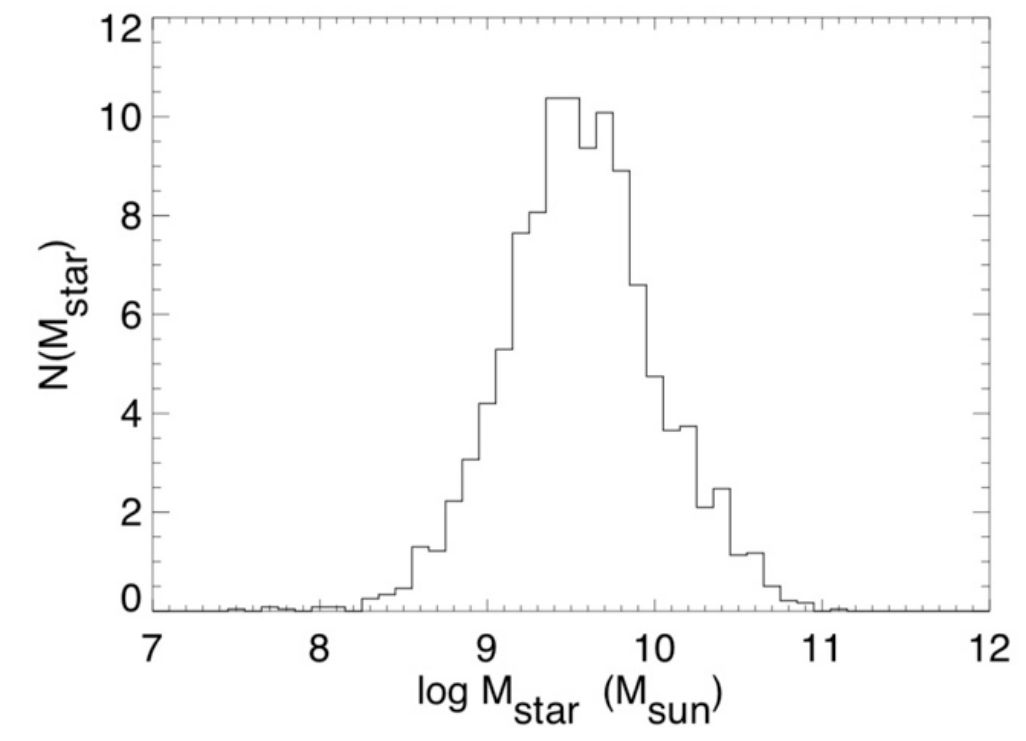

**Figure 2-19. CETUS target galaxies range in stellar mass from those of dwarf galaxies to very massive galaxies ($logM_{star}$>10.5).**

## 2.3 TRANSIENTS

**Neutron Star Binary Mergers** **Metzger SciWP #342**

**Brad Cenko**

The discovery of gravitational waves (GWs) and associated electromagnetic (EM) radiation from the binary neutron star (BNS) merger GW170817 was a watershed moment for astrophysics (Abbott et al. 2017). A short gamma-ray burst coming 2 seconds after the collision established that BNS mergers are the progenitors of these systems; light curves and spectra of the associated UV/optical/NIR counterpart revealed evidence of heavy-element material in the merger ejecta (a "kilonova"), with a total mass sufficient for BNS mergers to be the dominant production sites of heavy elements in the Universe. The UV/blue and red/IR light curves (**Figure 2-20**) suggest two distinct components of the ejecta from the merged object (Metzger 2015, SciWP #342). Free neutrons in the outer part were ejected so fast that they escaped capture, and the UV luminosity plummeted. The inner part (red) was ejected more slowly and underwent rapid (r-process) neutron-capture processing to produce heavy elements.

Not all neutron-star binary mergers are expected be like GW170817. In particular, if the total mass of the neutron-star binary were high or one component were a black hole, then the merged object would collapse immediately to a black hole, and the observed result would be a red kilonova. If the total mass of the neutron-star binary were very low, then a long-lived very rapidly rotating supramassive neutron star would be created, which would produce an abnormally powerful GRB jet, and the lag time between merger and GRB would be much longer than 2 s. In the intermediate mass case to which GW1701817 belongs, the merger creates a high-mass neutron star, and the observed result is first a blue kilonova and later, a red kilonova.

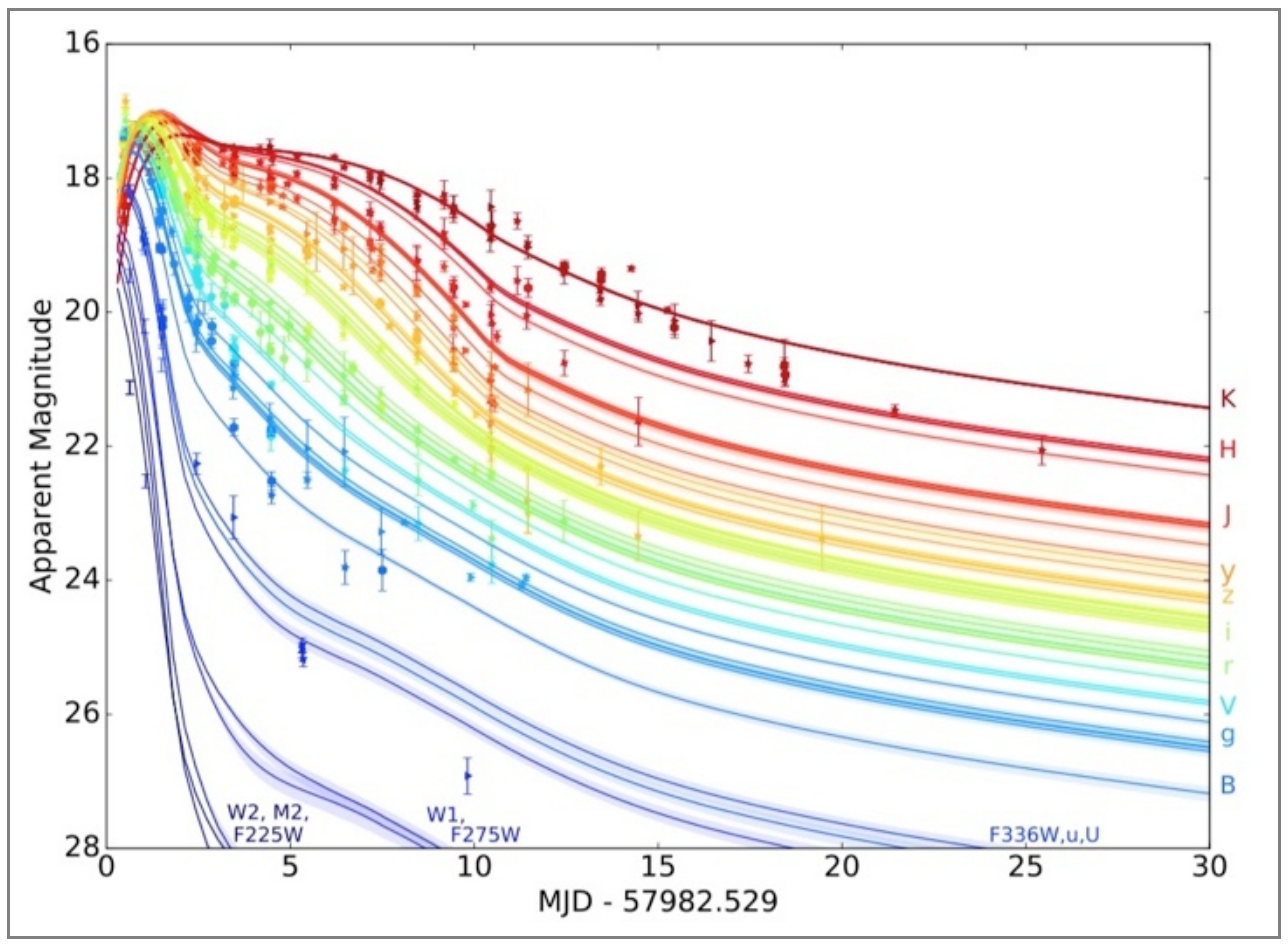

**Figure 2-20. The near-UV brightness (labelled W2 or W1) drops precipitously in a few hours, so CETUS will be on constant alert for commands from the ground and will slew to the target in ≤15 minutes. Figure credit: Villar et al. 2017**

The prompt-response capabilities and UV sensitivity of CETUS (Section 4.1) make it a unique facility to detect a blue kilonova if present, and hence to help identify the mass and type of merger. CETUS will work well with future, more sensitive LIGO detectors, as the UV camera should be able to detect a GW170817-like object at 200 Mpc and follow it for 10 hours after the merger. In the 2020's, LIGO is expected to detect on the order of a BNS merger each month. The accumulation of GRB-UV-optical-near-IR observations should enable us to explore the diversity of kilonovae, and to put more precise constraints on the total ejecta mass and composition.

## Tidal Disruption Events — Wheeler et al. #10, Pasham et al. #37

A Tidal Disruption Event (TDE) occurs when a star comes close to an inactive black hole and is tidally distorted and then shredded. Some of the debris escapes the black hole; the remainder orbits the black hole to form an accretion disk, which is then viewed as an active galactic nucleus (AGN) **(Figure 2-21).** As pointed out by Wheeler (SciWP #10), TDE's are interesting for several reasons: (1) TDE's can be used as markers for SMBH's that would otherwise go undetected; (2) TDE's are excellent probes of relativistic effects in regimes of strong gravity, and they provide a new means of measuring SMBH masses and spins; and (3) TDE's are signposts of intermediate-mass BH's, binary BH, and recoiling BH's.

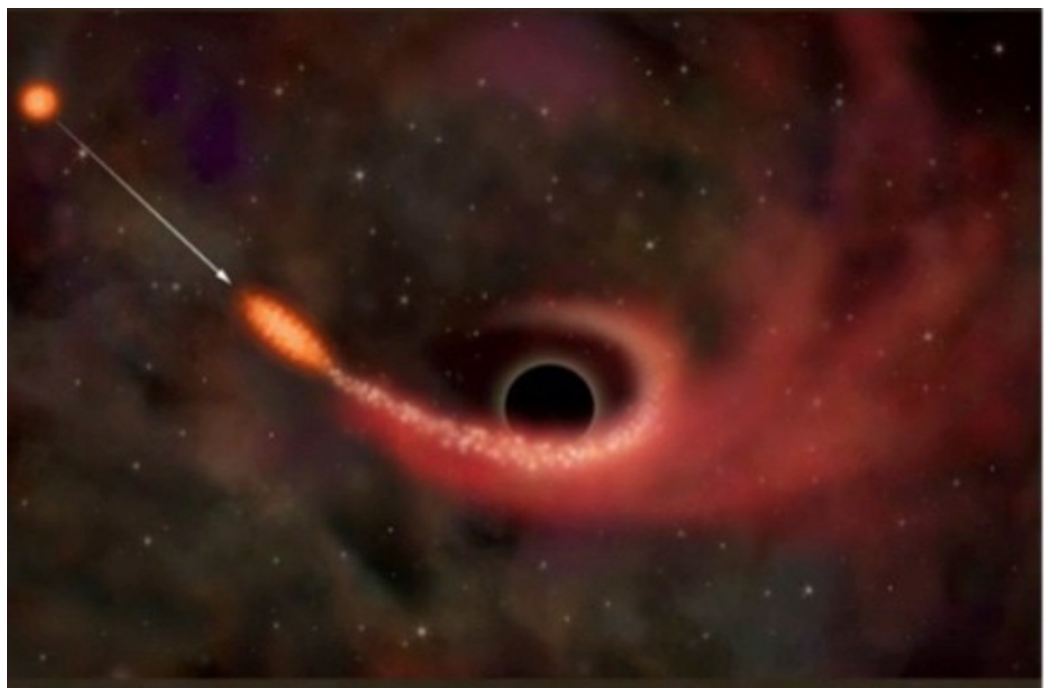

**Figure 2-21. Schematic of a tidal disruption event (TDE). CETUS will monitor the FUV and NUV spectra of TDE's to faint magnitudes. (Image credit: M. Weiss, CXO)**

With the advent of the Zwicky Transient Facility (ZTF) and LSST, the number of TDE detections will grow rapidly. We expect a yield of ~50 TDEs per year from the ZTF Northern Sky Survey and 4,000 per year from LSST. Both telescopes will obtain *g-r* colors, but UV spectroscopic follow-up and UV and X-ray follow-up imaging will be needed for classification.

CETUS will join X-ray, optical and IR observatories to observe many of the brighter TDE's. It will not only obtain NUV and FUV light curves, but also spectra with the FUV Point/Slit spectrograph similar to those from Hubble's STIS shown in **Figure 2-22** (but with higher resolution and S/N), and low-resolution NUV spectra with the MOS. These spectra will yield information on the physical properties and composition of the shredded stellar remnant and nebular material. Monitoring of the tidal disruption flare will reveal spectral evolution in the velocity, Doppler width, and strength of the emission lines.

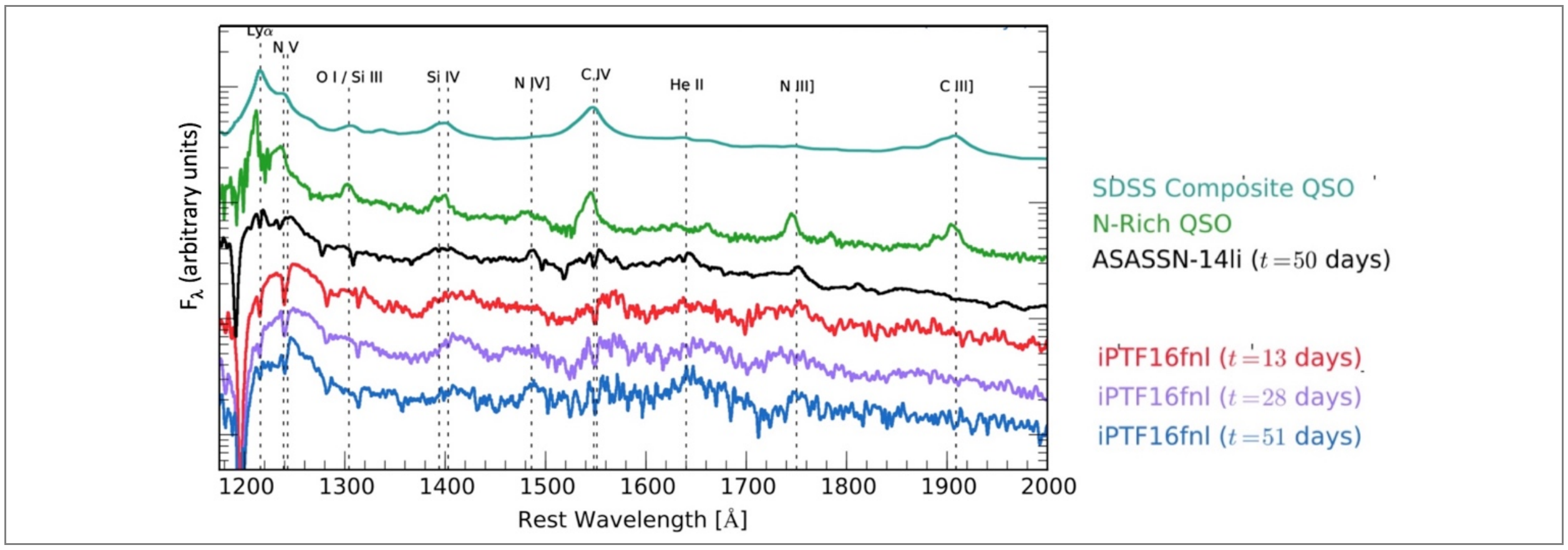

**Figure 2-22. CETUS will survey the UV spectra of TDE's (bottom 4 curves) and follow their evolution as was done for iPTF16fnl. Figure credit: Brown, Kochanek et al. (2017).**

## Core-Collapse Supernovae

In most cases, the progenitors of core-collapse supernovae are not known. A massive star near the end of its life may be a red supergiant (RSG), a blue supergiant (BSG) or a Wolf-Rayet (WR) star. As shown in **Figure 2-23**, Nakar & Sari (2010) have found that the FUV light curve of a supernova can be used to identify its progenitor. The CETUS FUV camera is well suited for this task as it is designed to respond rapidly to reports of a nascent supernovae and can follow the FUV light curve down to an apparent magnitude, $m_{AB}$~27.

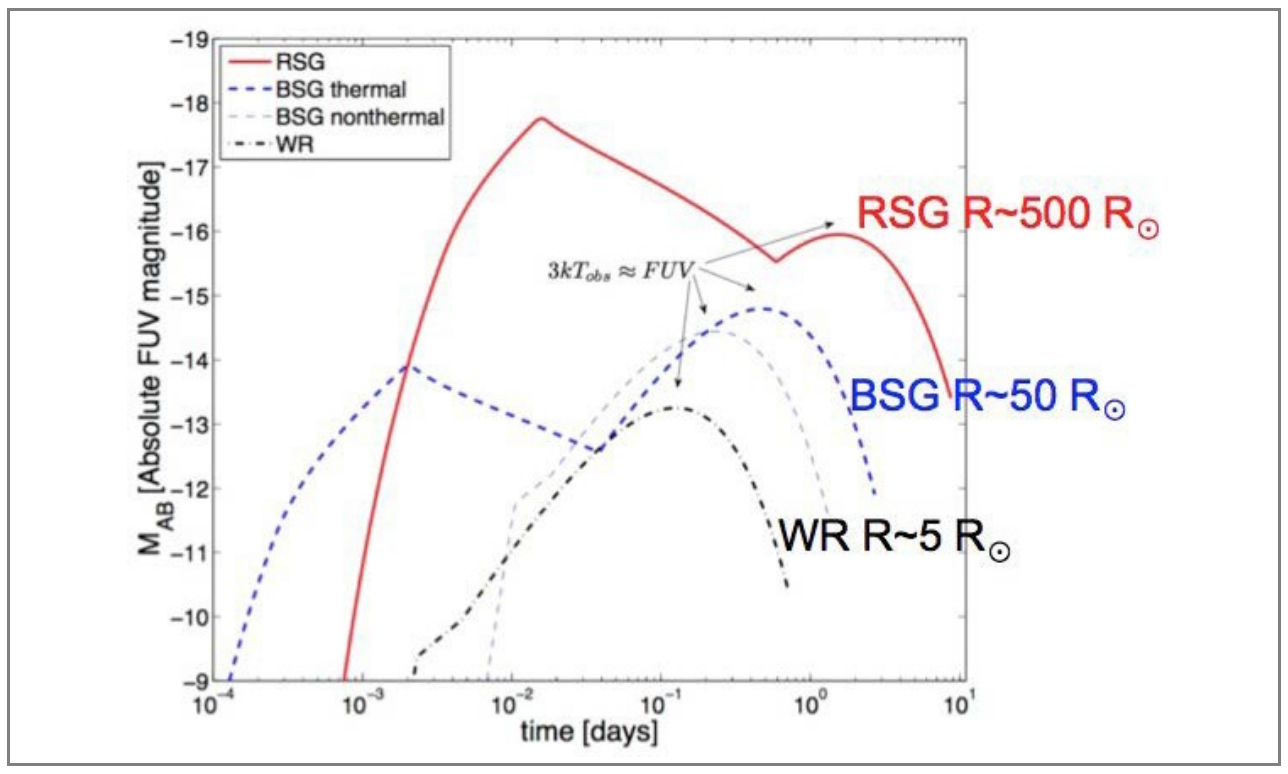

**Figure 2-23. CETUS will infer the progenitors of core-collapse supernovae from their FUV light curves. (Figure credit: Nakar & Sari, 2010)**

## Stellar Flares — Kevin France

Kepler scientists have found that, on average, every star in the Milky Way hosts a planetary system. Some of these systems contain one or more planets in the "habitable zone" (HZ). M-dwarfs, the most common stars in the galaxy and solar neighborhood, are expected to host most HZ planets that will be found by TESS (Sullivan et al. 2015). The habitable zone is where the planetary surface temperature is such that liquid water may be sustained for some portion of the "year". The effective surface temperature alone, however, is insufficient to establish habitability or to accurately interpret spectral features in terms of biosignatures (e.g. Meadows et al. 2018). Other information is needed, particularly the EUV-NUV stellar energy distribution because it drives the atmospheric heating and chemistry of terrestrial planets. The stellar flux distribution is also critical to the long-term stability of terrestrial atmospheres. Stellar flares can alter the UV luminosity by factors of 10 on timescales of seconds, and even non-flaring states are characterized by stochastic fluctuations of ~30% on minute timescales (Loyd & France 2014).

CETUS will monitor the UV spectrum of selected M-type stars in order to get better statistics on flares such as frequency, intensity, duration, ionization level, etc. Spectrally resolved monitors are critical for properly modeling M-star atmospheres including the thermal structure of the transition region (from O VI 1032Å, 1038Å), which gives rise to much of the EUV emission, and the corona (from [Fe XXI] 1354Å, [Fe XIX] 1118Å, and [Fe XII] 1242Å).

CETUS data will provide the necessary observational constraints to predict the physical processes involved in atmospheric heating and chemistry of their exoplanets. Spectral features of $O_2$, $O_3$, $CH_4$, and $CO_2$, are expected to be the most important signatures of biological activity on planets with Earth-like atmospheres (Des Marais et al. 2002). The chemistry of these molecules in the atmosphere of an Earth-like planet depends sensitively on the strength and shape of the UV spectrum of the host star. $H_2O$, $CH_4$, and $CO_2$ are sensitive to LUV+FUV radiation (1,000 – 1750 Å), while atmospheric oxygen chemistry is driven by a combination of FUV and NUV (1,750 – 3,200 Å) radiation.

## 2.4 MULTI-WAVELENGTH SURVEYS

No matter how good a CETUS image or spectrum is, its value will increase significantly when combined with observations in other spectral regions. Virtually all CETUS observations discussed so far are meant to be combined with X-ray, optical, infra-red, and/or radio observations.

- *Dust in Galaxies*: CETUS far-UV and near-UV imagery of galaxies will be used to measure the 2175-Å bump and UV slope, $\beta_{UV}$, as a function of position within the galaxy. These data will be combined with IR data from Herschel and ALMA to understand the connection between UV extinction and IR emission. CETUS will also take part in multi-wavelength surveys such as PHANGS (described earlier) to understand the physical processes involved in star formation.
- *Outskirts of Galaxies*: CETUS long-slit spectra of the outskirts of numerous star-forming galaxies will be combined with Hα and X-ray emission to help understand the outflows from galaxies and to identify what's driving them.
- *Lyman-α Emission Halos Around Galaxies*. Lyman-α halos of galaxies derived from CETUS far-UV imagery will be combined with Hα from ground-based telescopes and WFIRST, and 21-cm images of the same galaxies in order to derive the physical properties of halos around different types of galaxies and to understand the physical origins of Lyman-α.
- *The Circumgalactic Medium (CGM)*. Far-UV absorption spectra arising from the CGM will be combined with UV-optical-IR imagery and far-UV long-slit spectra of the associated galaxies to derive correlations between the properties of CGM and the properties and environment of the galaxies.
- *Star-Forming Galaxies at z~1*. Targets for CETUS near-UV MOS spectroscopy of z~1 galaxies will be selected from Subaru's Prime Focus Spectrograph (PFS) archive of optical-NIR spectra of z~1 galaxies. CETUS MOS spectra of these galaxies will be combined with the PFS spectra to produce continuous spectra from 0.18-1.3 microns for stellar, nebular, and ISM analysis.

CETUS will also contribute to multi-wavelength campaigns by numerous telescopes, such as:

- *Observations of transients*. CETUS will observe selected LIGO sources, core-collapse supernovae, tidal disruption events, important LSST sources, etc. If a transient field is defined, CETUS will contribute to it.
- *"Deep drilling fields"*. CETUS will contribute far-UV images, which are sensitive to UV sources at z=0-0.5, and near-UV images, sensitive to sources at z=0.0-2.0.
- *Hubble Deep Fields.* Hubble's WFC3 has already contributed near-UV imagery to legacy Hubble deep fields. The CETUS camera will be used to contribute far-UV observations of these same fields.

## 2.5 SCIENCE FLOWDOWN

The science objectives of CETUS have been established based on inputs from the CETUS Science Team and general science community through their Science White Papers submitted to Astro2020. **Table 2-1** summarizes the four key science programs and their principal targets, measurements and specific instruments to carry out these programs. **Table 2-2** gives an overview of a (notional) Design Reference Mission (DRM) that would carry out these programs. The DRM covers the first four years of science operations. We assume that science observations start about 6 months after launch, after travel to and insertion into orbit about Sun-Earth L2, and Orbital Verification and Science Commissioning of CETUS have been completed. Four years of science operations is equivalent to ~35,000 hr, but not all those hours can be devoted to science observations due to observation overheads and a healthy calibration observing program. We therefore allot a total of 30,000 hr to the various science programs on CETUS. The DRM table lists the major science programs and gives for each its allocation (in % and hours), the nature of the target and the CETUS scientific instrument (SI) to study it, the time on target whether the target is an individual source or a field (e.g. the 2 sq. degree COSMOS field, which requires 24 pointings to cover it), the number of pointings per observing program (ranging from ~5 to 600 for fields), and the observational yield.

**Table 2-1. CETUS Science Program by Science Instrument**

| | Objectives | The Low-z Universe | Galaxies at z~1 | Transients | Surveys |
|---|---|---|---|---|---|
| **Science** | Principal Targets | galaxies, stars, gas, dust | galaxies and Lyα at z~1 | images, spectra | images, spectra |
| | Primary Measurements | images, spectra | spectra, images | light curves, images | image properties |
| | Primary science instrument | All SI's | NUV MOS spectra | Camera | Camera |
| **OTA/Obs** | Telescope diameter 1.5m | ✷ | ✷ | ✷ | ✷ |
| | Mirror Coatings Al/LiF/ALD MgF2 | ✷ | | | |
| | FoR > anti-solar hemisphere | | | ✷ | ✷ |
| | Slew time<15 min for 180 deg | | | ✷ | |
| | Parallel operation of SI's | ✷ | ✷ | ✷ | ✷ |
| **NUV MOS** | RP~1,000; λλ180-350 nm; MSA covers 17.4'x17.4' | ✷ | ✷ | | |
| **FUV spec** | RP~20,000, λλ100-180 nm; long slit covers 2"x360" | ✷ | | ✷ | ✷ |
| **NUV spec** | RP~40,000, λλ180-350 nm | ✷ | | ✷ | |
| **FUV cam** | FOV 17.4'x17.4'; λλ115-180 nm; 5 long-pass filters; Res 0.55" | ✷ | | ✷ | ✷ |
| **NUV cam** | FOV 17.4'x17.4'; λλ 180 350 nm; 5 long-pass filters; Res 0.40" | ✷ | ✷ | ✷ | ✷ |

**Table 2-2. Design Reference Mission (4 years)prin**

| Observing Program | Allocation (%/hours) | Target / SI | $T_{obs}$ (hr/target) | Targets per Prog. | Observation Yield |
|---|---|---|---|---|---|
| **Low-z universe** | 33%/ 10,000 hr | | | | |
| Nearby galaxies | 2000 hr | Galaxy dust / FUV/NUV CAM | 10 | 200 | UV maps of 100 galaxies, 10,000 SF regions |
| Galaxy outskirts | 1000 hr | Galaxy outskirts / long-slit FUV PSS | 100 | 10 | Emission maps of 10 galaxy outskirts, ICM |
| Reflection nebulae | 300 hr | Dusty outflows & gal. / CAM, PSS | 20 | 15 | Dust properties in outflows of 15 galaxies |
| Emission-line sources | 2000 hr | Ly$\alpha$ halos / FUV CAM, PSS | 20 | 100 | Maps of Ly$\alpha$ and O VI in 100 fields |
| CGM and host galaxies | 4000 hr | CGM & Galaxy / FUV PSS | 20 | 200 | Properties of CGM around 100 galaxies |
| Stars, ISM, Solar system bodies | 2000 hr | Stars / PSS, CAM | 10 | 200 | Properties of 200 stars, ISM, SS bodies, nebulae |
| **z~1 Galaxies** | 37%/ 11,000 hr | Galaxy Field / NUV MOS | 20 | 550 | Rest far-UV spectra of 55,000 galaxies |
| **Transients** | 10 %/ 3,000 hr | Transient/FUV/NUV CAM | 100 | 30 | Light curves & spectra of 30 transients |
| **Surveys** | 17 %/ 5,000 hr | Field/ FUV/NUV CAM | 100 | 50 | Images of 50 fields |
| **Director's Discretionary Time** | 3 % / 1000 hr | | | | |

Even **Table 2-2** does not give the full picture of the enormous yield of CETUS observations. We need to include parallel observations that will be obtained while the prime instrument is observing its target(s). Below, we consider three observing modes corresponding to when the MOS, Camera, or spectrograph is prime. We assume that the observing efficiency of each instrument is 90% including overheads (small slews, read out of detectors, setup of instrument for next integration, etc.) Hence, the total observation yield contributed by the prime and parallel instruments is well over 100% and the discovery potential is very high.

*The MOS is prime:* In this case, the MOS will typically be in survey mode, obtaining the NUV (rest FUV) spectra of ~70 z~1 galaxies simultaneously. The NUV camera will observe in parallel to obtain images of galaxies at z=0-2 before or after the MOS observes that region. The FUV camera will obtain images for studies of UV morphology of galaxies at z=0-0.4. Observing efficiency: 180%.

*The Camera is prime:* This will be the case when CETUS participates in surveys of selected deep fields. See https://www.astro.princeton.edu/~dns/deep.html for more information. The NUV MOS will observe in parallel in long-slit mode to obtain imaging spectra in search of Lyα (and possibly O VI) emitters at z~1 as described earlier. The MOS configured as multiple long slits will scan the sky during long exposures by the camera. The FUV point/slit spectrograph will observe in long-slit mode in search of Lyα (and possibly O VI) emitters at low redshift. Observing efficiency: 270%.

*The FUV point/slit spectrograph (PSS) is prime:* The PSS will typically (but not always) be obtaining LUV/FUV point-source spectra of a background AGN/QSO whose line of sight pierces the CGM of a foreground galaxy or obtaining an imaging spectrum of the galaxy itself. In either case, both wide-field instruments will be observing inparallel. The NUV/FUV camera may study the UV morphology of galaxies in its field of view, and the NUV MOS will observe wide-fields (302 sq. arcmin) with multiple 17.4'-long slits to search for Lyα emitters. Observing efficiency: 270%.

### Maximizing the science yield of CETUS observations

Maximizing scientific return from NASA missions is clearly of concern to NASA as it held a workshop in October 2018 on this very topic. In the case of CETUS and other survey telescopes, the main route to maximizing scientific return is to make measurements on CETUS calibrated images and spectra and to make these measurements immediately available to the astronomical community. Quantitative measurements are essential for deriving distributions of various parameters and examining possible correlations with other parameters.

Following typical NASA practice, the CETUS science operation center will perform Level 1 and Level 2 data processing. Level 1 processing consists of receipt of telemetry with checks for and correction of telemetry errors, and formatting the packetized data into an image or spectrum. Level 2 processing includes removal of the instrument signature and conversion of the data to physical units. NASA typically does not require Level-3 processing, which in this case would consist of measurements and data archival in a publicly accessible relational database, but it is essential if CETUS is to have a strong scientific impact. The work of the CETUS science center(s) is described in **Section 5.1** (Science Operations).

### Science Traceability

Table 2-3 describes the primary science drivers that dictated the engineering and performance requirements. These science drivers and performance requirements are taken up in Chapter 3.

**Table 2-3 The science drivers and derived engineering requirements**

| # | Science Driver | Derived Engineering Requirements |
|---|---|---|
| SD 1 | Wide field imagery and multi-object spectroscopy over 1045”x1045” field | 0.4” (~1-pixel) resolution over wide field of view of camera & MOS; Opto-mechanical stability |
| SD 2 | Massive NUV spectroscopic. survey z~1 galaxies; Lyα emission survey at z~1 | NUV wide-field multi-object spectrograph with Next-Generation micro-shutter array |
| SD 3 | LUV/FUV survey of the circumgalactic medium (CGM) | Al/LiF/ALD $MgF_2$-coated mirrors of telescope & FUV spectrograph |
| SD 4 | Robust FUV spectral features | Precision polishing of OTA primary Ò << HST mid-spatial frequency errors |
| SD 5 | High sensitivity of UV optics | Large aperture<br>Control of: molecular, dust & humidity contamination<br>» AI&T; Faring & Launch; Flight |
| SC 6 | Survey of extended sources by imaging spectroscopy | Extensive FOV; long slits |
| SD 7 | NUV/FUV survey of rapid transients: neutron-star binary mergers; supernovae | UV light curves potentially starting within an hour of the event |
| SD 8 | Highly productive surveys via parallel observations | Internal mechanism to dither while holding telescope pointing steady |
| SD 9 | Science-ready CETUS data for astronomical community | Measurements of calibrated images and spectrograms |
| SD 10 | Sensitivity to faint diffuse sources | Fast f/ratio |
| SD 11 | Sensitive low-noise UV detection | NUV selection of CCD; FUV selection of MCP; Low noise, radiation resilient electronics for both NUV/FUV |

# 3. INSTRUMENTATION – ENGINEERING IMPLEMENTATION TO FACILITATE SCIENCE

## 3.1 THE CETUS CONCEPT IS MATURE AND WHY MATURITY MATTERS

Collectively, team members have made use of their extensive experience to design CETUS for performance to meet the science requirements and for minimal technical, schedule, and cost risk. One significant risk for an UV instrument is molecular, dust, and humidity contamination. The team comes with a thorough knowledge of best practices in material selection, out-gassing budgets, isolation of the payload from the spacecraft, purged volumes, and a plan for contamination facilities through observatory AI&T and a management plan in pre-launch operation.

It is generally known that >70% of the cost of a product originates from decisions made during concept development and design. Hence, *a mature CETUS concept is essential for identifying and avoiding potential technical, schedule, or cost risk.* Among the cost risks we identified is complexity. We have worked to avoid complexity by:

- Imposing on the three science instruments a common technology in optical configurations, mechanisms, and detectors;
- Designing each instrument for installation in (or removal from) the payload with minimal interference to the other instruments;
- Establishing a single Interface Control Document (ICD) governing the entire space element including science payload and spacecraft including a shared electronic architecture for the detectors and devices used in CETUS; and
- Planning for contamination control starting with assembly of each component and continuing through subsequent AI&T stages and launch.

We have worked to avoid technical and hence, cost risk by:

- Selecting high-TRL components, e.g. the space-qualified Euclid CCD. The only component below TRL 5 is the Next-Generation MicroShutter Array;
- Ensuring systems engineering is included in the conceptual design phase. The greater CETUS engineering team worked together, literally working side by side, to develop the optical-mechanical design of CETUS;
- Selecting materials known for their temporal and thermal stability. M55J structural material closely matches the coefficient of thermal expansion (CTE) of extremely stable ZERODUR® OTA mirror substrates. Thus, the design which is nearly athermal by passive means, requires only mild heater controls to ensure robust optical metering performance.

In the following sections, we describe the CETUS design at a high-level (§3.2), instrument level (§3.3), and component level (§3.4). In §3.5, we describe the mechanical – electrical -thermal design of the science payload; and in §3.6, we describe our plans for alignment, integration, and test of the telescope and science instruments.

## 3.2 TOP-LEVEL DESCRIPTION OF THE CETUS DESIGN

### 3.2.1 Meeting the Science Requirements

**Table 3-1** combined with **Table 1-1** summarize the science requirements of CETUS as they affect the Optical Telescope Assembly (OTA), science instruments, and spacecraft. In the remainder of this section, we demonstrate that the design of the CETUS mission concept meets the science requirements.

**Table 3.1 The Science Drivers Define the CETUS Implementation Requirements**

| # | Science Driver | Engineering Requirements | Status |
|---|---|---|---|
| SD 1 | Wide field imagery and multi-object spectroscopy over 1045"x1045" field with a resolution of 0.4" or better | • 0.4" (~1-pixel) resolution over wide FOV of camera & MOS; mechanism for dithering to recover resolution<br>• Opto-mechanical stability | • Wide field by design; resolution maps and distortion maps predicted<br>• Best focus via hexapod on telescope secondary |
| SD 2 | Massive NUV spectral survey z~1 galaxies; Lyα emission survey at z~1 | • NUV wide-field multi-object spectrograph with Next-Generation micro-shutter array (NG-MSA) | • SAT program on track to raise TRL of NG-MSA from TRL 3-4 to TRL5 by Dec. 2021<br>• JWST MSA can be used if TRL insufficient |
| SD 3 | LUV/FUV survey of the circumgalactic medium (CGM) around nearby galaxies | • Al/LiF/ALD MgF2-coated mirrors of telescope & FUV spectrograph | • SAT program in progress on mirror coatings<br>• Mirror coating facilities available for all but telescope PM. Planning of upgraded/new PM mirror-coating facility in progress |
| SD 4 | Robustly measured FUV spectral features | • OTA primary mirror having much reduced mid-spatial frequency wavefront errors that are present in the Hubble PM | • Contemporary deterministic polishing techniques now well developed<br>• Proposals from qualified suppliers |
| SD 5 | High sensitivity of UV optics | • Large aperture<br>• High UV reflectivity of mirrors<br>• Control of molecular & dust contamination and humidity at all phases of mission | • OTA 1.5m diameter<br>• Minimal number of reflections in design<br>• Hot deposition of high-reflectivity coatings<br>• Detailed plans for fabrication and alignment, integration<br>• Testing from component to system level |
| SC 6 | Survey of extended sources by imaging spectroscopy | • Extensive FOV<br>• Long slits | • NUV imaging spectra via 1044"-long MOS slits<br>• LUV/FUV imaging spectra via 360"-long slit in design |
| SD 7 | NUV/FUV survey of rapid transients, neutron-star binary mergers, supernovae, etc. | • UV light curves potentially starting within 15 minutes from an alert from the ground<br>• FUV spectra with the FUV PSS<br>• NUV spectra with the NUV PSS (R~40K) or NUV MOS (R~1,000) | • Upgrades to spacecraft & solar panels in design<br>• Enlarged FoR covering >2π steradians of sky<br>• Dual-axis high-gain antenna continuously pointed at Earth<br>• Upgrades to Near Earth Network (NEN) planned<br>• Reaction wheel assemblies (RWAs) sized to enable 180 deg. slew in ~15 min.<br>• Thermal control of optical metering |
| SD 8 | Highly productive surveys via parallel observations | • Capability for dithering while holding telescope pointing steady | • TRL 8-9 mechanism for M2 optic (mirror or grating) of Offner relay in wide-field instruments |
| SD 9 | Science-ready CETUS data for astronomical community | • Measurements of calibrated images and spectrograms<br>• On-line relational databases | • In initial planning |
| SD 10 | Sensitivity to faint diffuse sources | • Fast f/ratio | • OTA baseline is f/5, and relays for both MOS and CAM are 1:1, preserving f/5 |
| SD 11 | Sensitive low-noise UV detection | NUV selection of CCD<br>FUV selection of MCP<br>Low-noise, radiation-resilient electronics for both NUV/FUV | • NUV detectors: e2v "Euclid" CCD + e2v electronics + enhanced UV sensitivity (JPL)<br>• FUV detectors: Berkeley SSL CsI MCP + SSL electronics |

### 3.2.2 Top-Level Architectural Features of CETUS

Development of the architecture of the CETUS payload (**Figure 3-1**) started with accommodating the science drivers and has advanced to a level typical of mid-Phase A. The architecture is the result of optical and mechanical designers working together with high-end tools, realistic flight hardware components, and experience in achieving simplicity, low risk and cost. We have conducted our study of CETUS following NASA Risk Management Handbook NASA SP-2011-3422. All risks have been or will be retired to the domain of low Probability of Failure and low Consequence of Failure. Elements of the CETUS architecture include:

1. *Orbit.* The observatory will operate in a sun-earth L2 halo orbit with a <85° solar keep-out angle.
2. *Telescope.* The 1.5-m primary mirror assembly and all optics will use mature, high-TRL substrates and optical fabrication methods resulting in much better control of phase errors over mid-spatial frequencies than was possible with Hubble. We expect a substantially better (gaussian) line profile than is yielded by Hubble's COS and STIS instruments (see **Figure 2-13**).
3. *Science Instruments*. Each of the three instruments – camera (CAM), a point/slit spectrograph (PSS), and multi-object spectrograph (MOS) -- has its own aperture at the telescope focal plane, and each operates independently (with the exception that the prime instrument controls the telescope pointing and roll angle). Each instrument can be removed or inserted into the instrument bay without disturbing the others. Together, the instruments are managed under a single governing Interface Control Document (ICD), which makes use of commonality of detectors (CCDs and MCPs), thus having similar electronics, packaging, drivers and software. Commonalities of Offner relays and devices in the camera and MOS are recognized in the ICD. (See **Figure 3-1**.)
4. *Maximization of Optical Efficiency*. To maximize UV efficiency, the number of optics in the optical path is minimized. Al/LiF/ALD AlF3 coatings on the telescope and PSS mirrors give unparalleled sensitivity in the Lyman-UV (100-115 nm). Delta-doping of CCD windows yields highly enhanced sensitivity of the three NUV instruments down to 180 nm.
5. *Thermal Stability and Control.* Thermally critical opto-mechanical components will be maintained between 280K and 300K. The observatory will be stable enough against thermal transients that it can be slewed anywhere in the anti-sun hemisphere in 15 minutes and be ready to observe. To protect the optics from contamination, all mirror and window surfaces will be biased slightly warmer than surrounding structures and can be elevated in temperature for periodic redistribution of any contamination.
6. *Hardware Implementation.* Key contributors to the greater CETUS team bring heritage-hardware experience to CETUS and access to facilities of appropriate cleanliness and capacity. The CETUS team has carried out design-to-cost methods including addressing the choice of "better" vs. "good enough."
7. *Potential for Exo-Planet Observations.* A science white paper (Lisman #250) notes that CETUS might be the participating telescope for the Occulting Ozone Observatory (O3). In fact, the CETUS architecture can be made "starshade ready". While this possibility has not yet been specifically studied, there is no aspect of the current CETUS architecture that prevents it from participation in a starshade mission, albeit with some increase in cost.

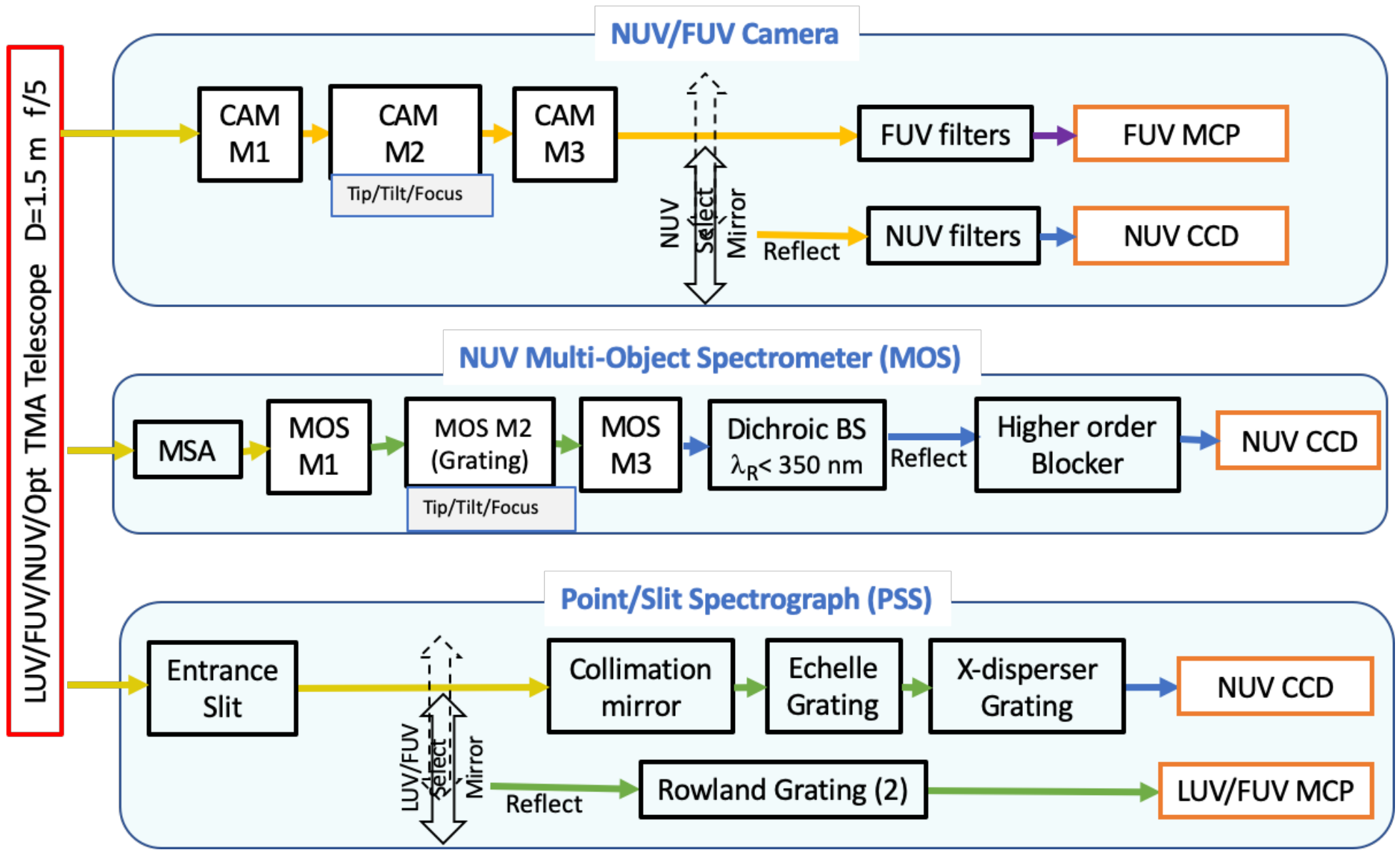

**Figure 3-1. Each of the 3 science instruments on CETUS is simple in design and has several aspects in common with other science instruments.**

## 3.3 INSTRUMENT-LEVEL DESCRIPTION OF THE CETUS SCIENCE PAYLOAD:

### 3.3.1 The Optical Telescope Assembly (OTA)

We have selected an all-reflective Three Mirror Anastigmatic (TMA) telescope producing a flat focal plane. This design gives a minimum number of reflections needed to achieve full field correction, critical in a UV payload. The 700-mm Cassegrain back focal distance supports efficient and heritage packaging of the front end of the OTA. This spacing enables a stiff Stable Member structural assembly, which supports the primary mirror (PM), secondary mirror (SM) support truss, and science instrument suite, and which connects to the spacecraft. **Figure 3-2** shows the resulting OTA design. The SM is supported by six struts **t**hat extend from the PM housing to the SM mount. The OTA is baffled to off-axis sources by a Main Baffle surrounding the 1.5-m light beam. It extends from the PM to beyond the SM with standard cassegrain central baffles: the SM cone baffle and the PM central baffle.

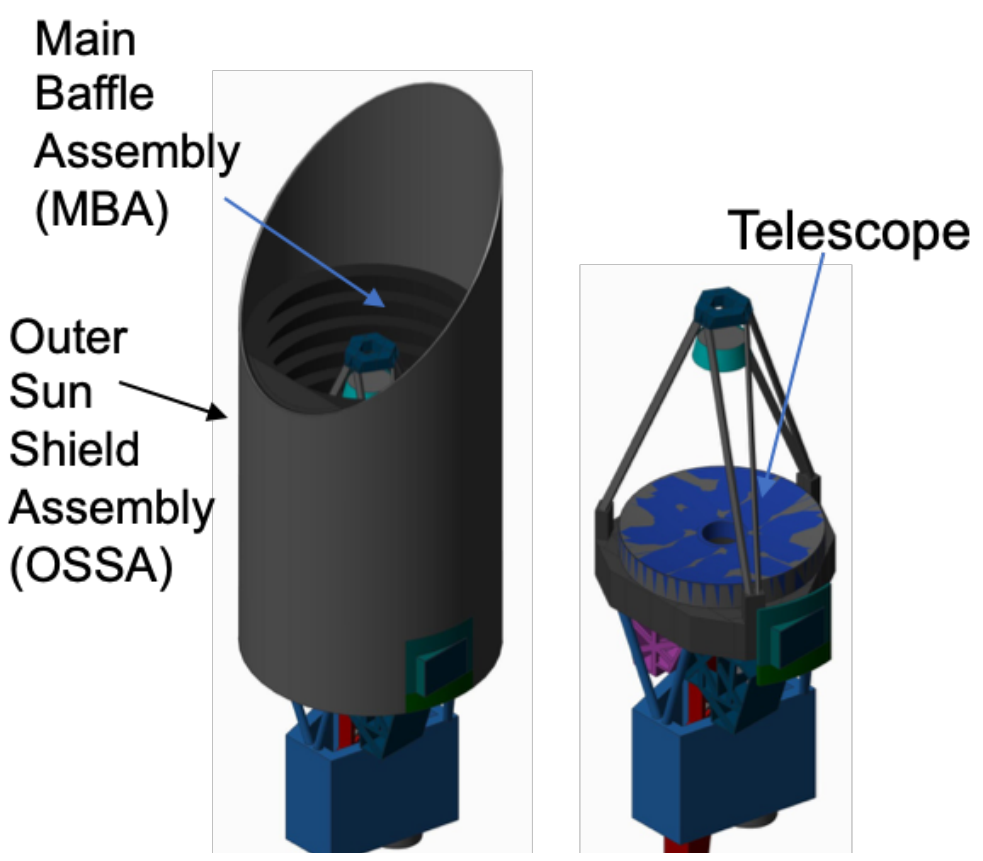

**Figure 3-2 The design of the payload is mature.**

An Outer Sunshield Assembly surrounds the Main Baffle Assembly. It extends from behind the PM to well beyond the SM. Its front end is angled at 45 degrees to exclude sunlight from entering the OTA at a solar exclusion angle of ≤ 85 degrees. A reusable door opens for science operations and closes for safety when needed. This door provides contamination control as do other design features such as pre-flight $N_2$ purge, temperature control of optics at temperatures warmer than other coldest surfaces, screening of materials for low outgassing, and other system controls.

High optical efficiency is enabled by 1) an all-reflective optical design of the OTA and instruments, 2) highly reflective mirror coatings, 3) minimal number of reflections in each optical path. The CETUS approach to mirror coatings is described in Section 6, as is discussion of detectors with high UV quantum efficiency and multi-layer UV-transmissive detector windows.

**Figures 3-3 – 3-4** illustrate the large FOV of the OTA focal plane. The OTA simultaneously feeds three separate scientific instruments (and two CMOS sensors in the fine-guidance system). Each instrument views a separate portion of the TMA image plane, so it can observe in parallel with the others. For example, the CAM and MOS can observe while the PSS is prime. The optical design of CETUS is presented by Woodruff et al. (2019, JATIS 024006, Vol 5(2)), and is briefly described below.

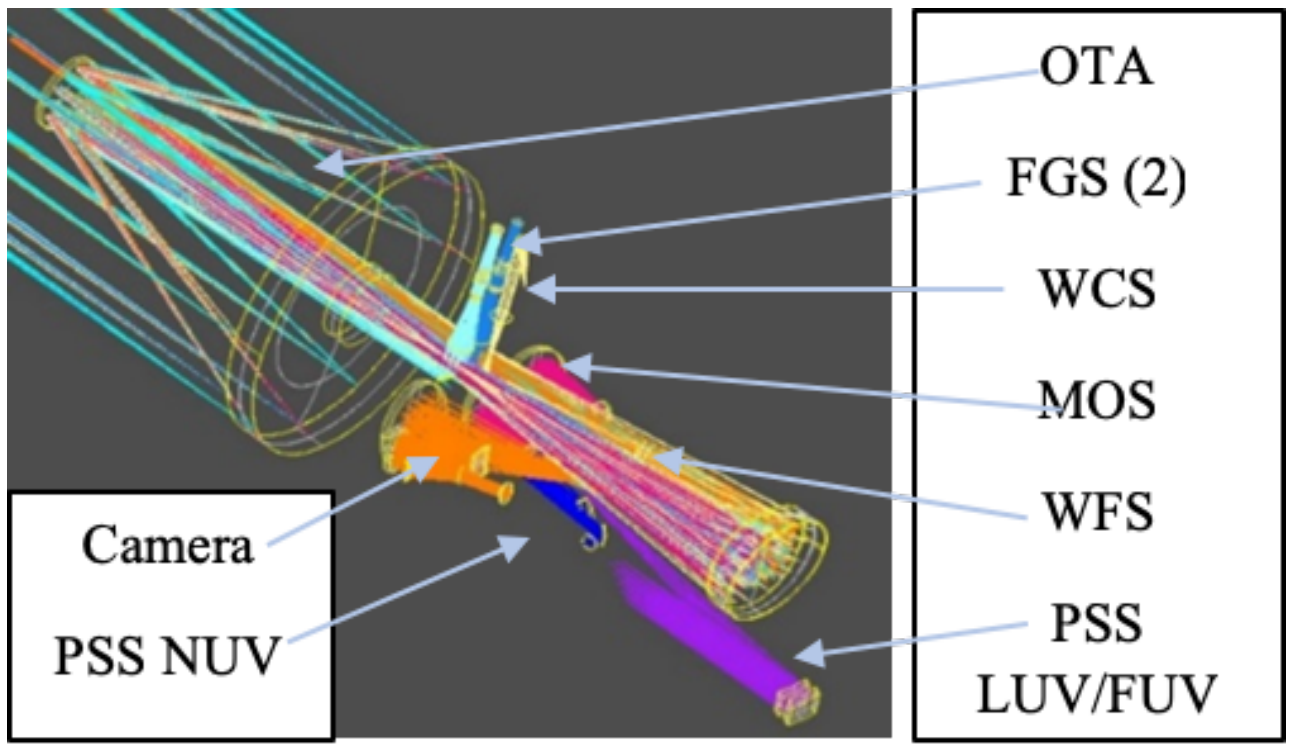

**Figure 3-3. The CETUS Telescope feeds the 3 UV instruments which have been laid out and packaged to fit within the allowable volume. Each instrument has a unique location on the TMA's wide field focal plane as shown in Fig. 3-4.**

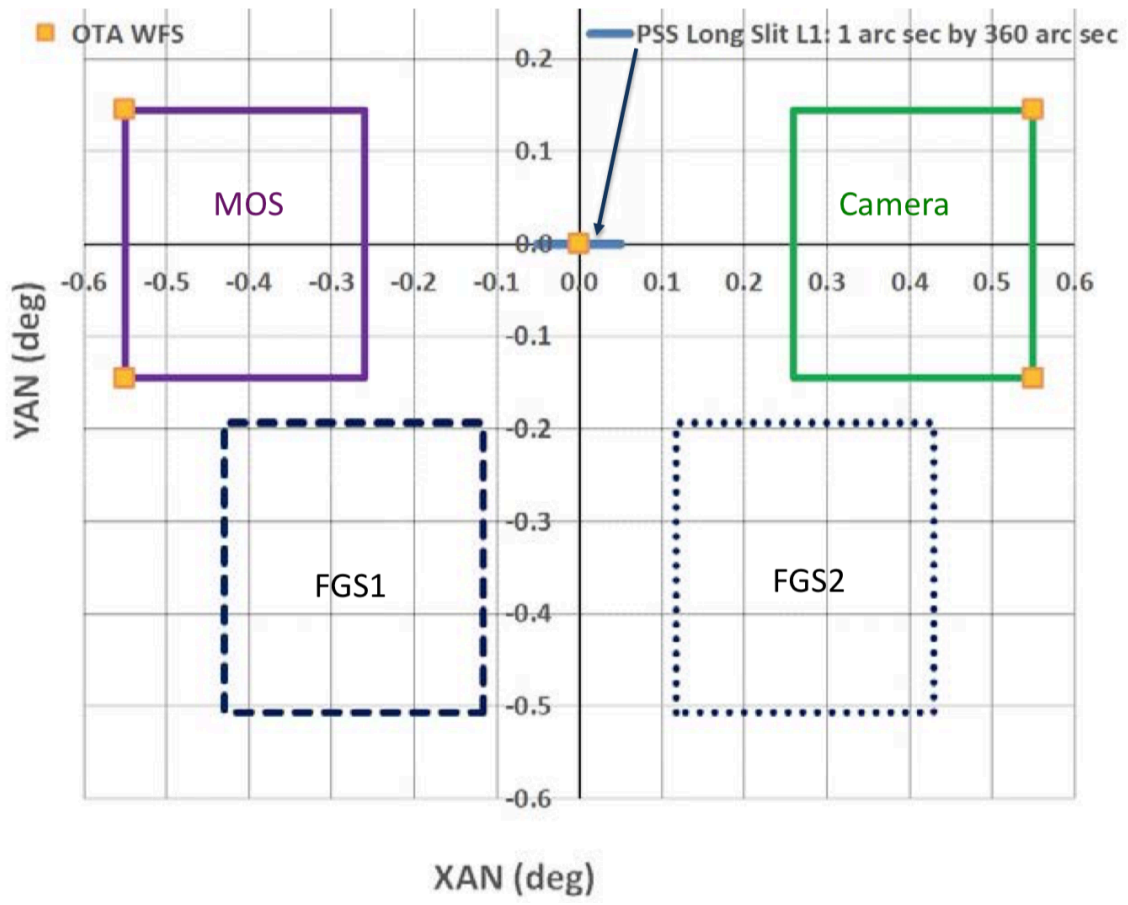

**Figure 3-4. The wide-field telescope focal plane accommodates all instruments.**

The optical designer has carried out a sensitivity analysis which produced the specifications needed to develop an error budget for the telescope and instruments as shown in **Tables 3-2** through **3-4.** The design allocates a looser tolerance for the PM since it is the most difficult mirror to fabricate due to its size and light-weighting. As found on Hubble, mid-spatial frequency (MSF) wave-front errors degrade the line-spread function (LSF) of FUV spectra. This MSF error is due not to quilting but rather, to zonal errors. Our industrial partners, Collins Aerospace and L3Harris have submitted proposals for polishing the telescope mirrors using mature processes for deterministic small-tool polishing that easily mitigate zonal errors and substructure print-through. The CETUS error budget addresses these mid-spatial frequencies.

Due to the inclusion of a hexapod behind the secondary mirror (SM), large adjustments needed due to gravity release, launch yield, and secular creep can be accommodated by deterministic methods of wavefront sensing and control (WFS&C). Small alignment corrections due to changes in thermal boundary conditions can be continuously applied via thermal sensors and proportional heating at designed nodes. We have multiple options for estimating alignment errors and will select the best approach. Costs covering these options are included in the CETUS budget. The CETUS approach has emphasized optical, mechanical, and thermal characteristics that enhance long-term stability, thereby greatly reducing the need for realignment.

**Table 3-2. OTA mirror fabrication and mounting specifications are consistent with a diffraction- limited OTA at λ=2000nm. Allocations have been made not only for optical figure but also for phase and amplitude effects at all spatial frequencies induced by the coatings.**

| | OTA Mirrors as Components (nm rms surface) | | |
|---|---|---|---|
| | PM | SM | TM |
| Low Spatial Frequency | 31.3 | 15.3 | 15.3 |
| Mid Spatial Frequency | 3.0 | 3.0 | 3.0 |
| High Spatial Frequency | 1.5 | 1.5 | 1.5 |
| Micro-roughness | 1 | 1 | 1 |
| Overall Uncoated | 31.5 | 15.7 | 15.7 |
| Reflective Coating | 3.0 | 3.0 | 3.0 |
| Overall Coated | 31.6 | 16.0 | 16.0 |

The overall error budget for the OTA and science instruments shown in **Tables 3-3** captures the allocation of wavefront error. These allocations have ramifications for the thermal and structural design both in flight and during pre-launch alignment, integration and testing (AI&T). The largest source of error is “despace” (distance between the OTA primary and secondary mirror). We have found that with easily implemented thermal control, the thermal component of the error budget is met with a 70% margin.

**Table 3-3. OTA Overall Wavefront Error Budget**

| OTA WFE 173.5 nm rms | | | | | | | |
|---|---|---|---|---|---|---|---|
| OTA design residual WFE (nm rms) | Fabricate WFE (nm rms) Overall | | | Align WFE (nm rms) | | Environment and | |
| 34.8 | PM mounted | SM mounted | TM mounted | Due to SM Alignment Errors** | 121.2 | Environment | 50.0 |
| | 63.3 | 32.0 | 32.0 | LOS jitter with roll jitter | 22.5 | Other | 50.0 |
| | | | | SM tilt jitter | 11.3 | Reserved | 50.0 |
| | | | | **Due to SM despace | 106.6 | | |

### 3.3.2 NUV Multi-Object Spectrometer

**Figure 3-5** shows the design of the MOS. The MOS imaging spectrometer observes numerous stellar objects simultaneously recording the UV spectrum of each object at a one-pixel resolving power, R~1000 over the spectral range from 180 to 350 nm. The *f*/5.24 MOS optics reimage a 1045 x 1045 arc sec portion of the OTA FOV onto the detector format via a three-mirror, all-reflective, Offner-like relay with a nearly one-to-one magnification. The three-mirrors are nominally concentric with M1 and M3 concave radius of curvature 800 mm and M2 convex with radius of curvature 400 mm. The M2 mirror is a convex spherical diffraction grating with 140 grooves per mm blazed at 250 nm. Mirrors M1 and M3 are high-order aspheric mirrors. To reduce fabrication complexity, we maintain a spherical convex figure for the MOS M2 diffraction grating. Simultaneous UV spectroscopy of up to ~100 sources is enabled by the configurable MicroShutter Array (MSAO located at the image plane provided by the OTA. The MSA has 380 by 190 individually selectable rectangular apertures of dimensions 100 µm X 200 µm (or 2.75”x5.50” on the sky). See Section 6.1 for further description of the MSA.

A single 4Kx4K, CCD (e2v CCD273-84 with 12-µm pixels registers the spectrum of the source in every open shutter, providing simultaneous observations of up to 100 sources in the field. The effective spatial resolution of the MOS is 0.39”. Hence, there is imaging within a MSA shutter. Each 2.75”x5.5” shutter contains ~7x14 resolution elements.

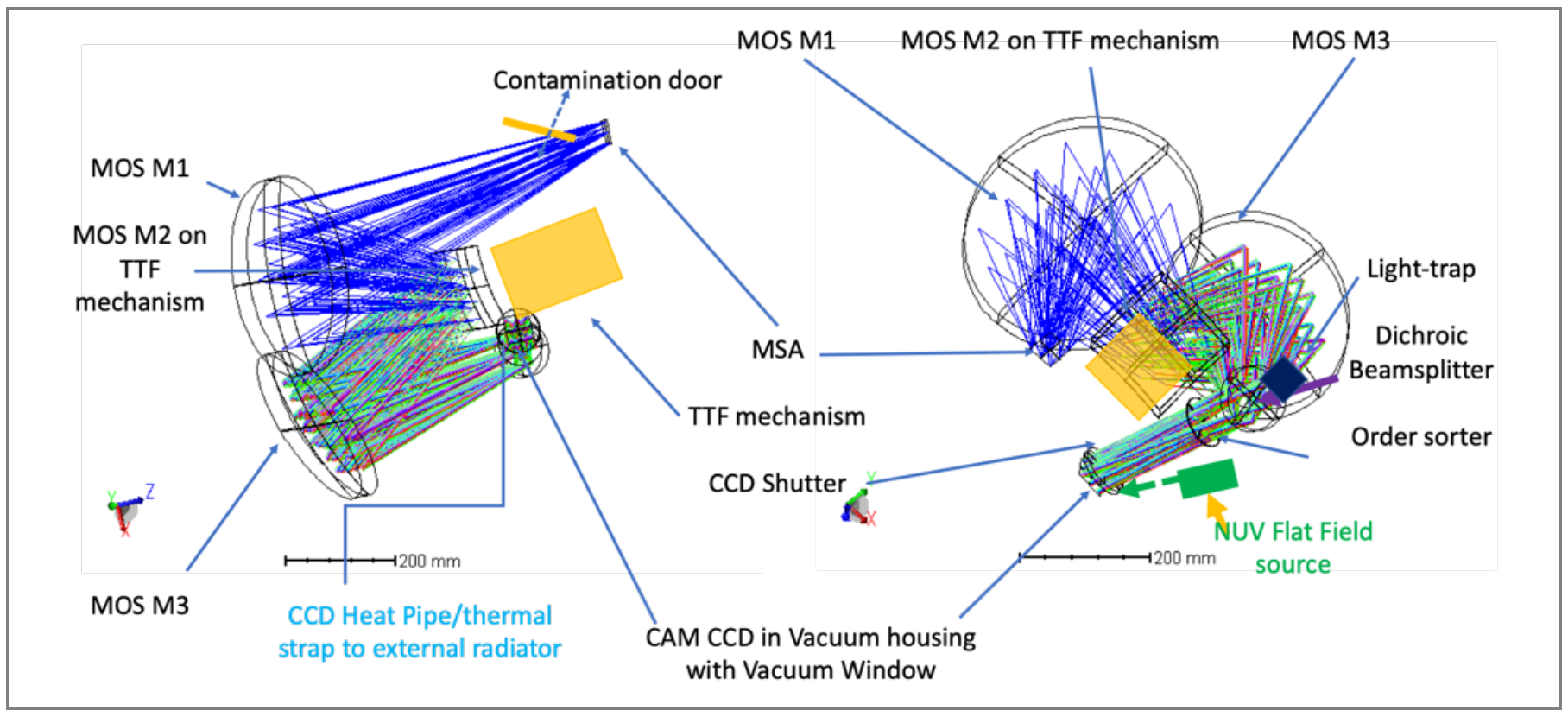

**Figure 3-5. MOS: Mirrors, mechanisms, flat-field calibration sources and FPA interface enable simultaneous spectroscopy of as many as 100 objects.**

The CCD housing is a vacuum enclosure with a $MgF_2$ vacuum window. Placed between the M3 mirror and the CCD is a 2 mm thick UV-grade fused silica order sorter that absorbs light of $\lambda < 160$ nm effectively blocking second and higher order light.

To reduce programmatic cost and risk, the same basic CCD assembly design is also used in the NUV camera and the NUV PSS. The only design differences relate to the window material type and tilt.

The M2 convex reflective diffraction grating is supported on a tip/tilt/focus mechanism. The MOS can be focused independently of the OTA and the camera. The OTA focuses at the field location of the PSS by adjusting the OTA SM based on error signals generated by WFS. The center WFS sensor is optically conjugate to the PSS slit array. Tip/tilt adjustment enables dithering the image at the FPA to sense detector pixel sensitivity variations and enhance observing efficiency.

### 3.3.3 NUV/FUV Camera

**Figure 3-6** shows the Camera, complementary to the MOS, in more detail. The FUV or NUV Camera (CAM) image numerous sources simultaneously in their respective spectral region. The F/5 CAM reimages a 1045 x 1045 arc sec portion of the OTA FOV via a three-mirror, all-reflective, Offner-like, unit-magnification imager. The three mirrors are nominally concentric with M1 concave radius of curvature 776.985 mm, M3 concave radius of curvature 657.935 mm and M2 convex radius of curvature 409.949 mm. All three mirrors are high-order aspheric mirrors.

The FUV and NUV bands share the three powered mirrors in an Offner-like relay. A Mode Select Mechanism lying between mirror M3 and the detectors is used to select the mode. In the FUV mode this fold mirror is removed from the optical path. When inserted, the fold mirror reflects the field to the NUV detector. Each mode incorporates an 8-position filter wheel slightly in front of the respective detector providing 6 spectral filters, 1 opaque (dark), and 1 for flat-field imaging.

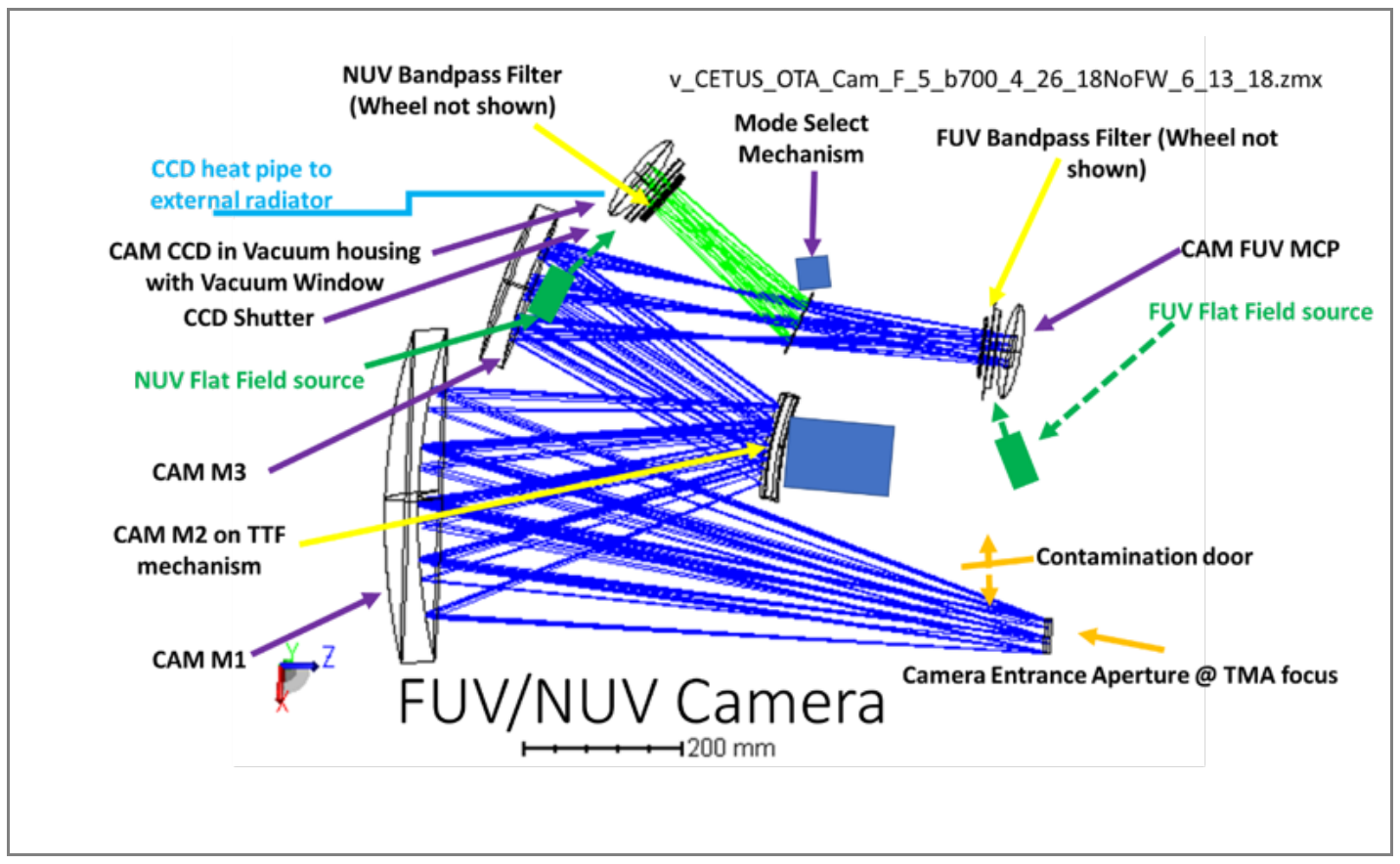

**Figure 3-6. Camera: Mirrors, mechanisms, flat-field calibration sources and FPA interface.**

Following experience with Hubble's ACS instrument, each NUV filter is an air-spaced (i.e., vented) dual element filter with coatings on each of the four plano surfaces to define the in-band spectral throughput and block out-of-band signal from the bandpass upper wavelength to the red-cutoff of the CCD and from the bandpass lower wavelength to the blue cutoff of the CCD.[12]

The image in the FUV mode is sensed by a 50mm X 50mm sealed CsI photocathode, solar-blind Micro-Channel Plate (MCP) detector with 20-µm effective resolution. The MCP housing is a vacuum enclosure with a MgF2 window. The detector operates at room temperature. The effective resolution of the FUV detector is 550 mas on the sky. The image in the NUV mode is sensed by a 4 K x 4 K, e2V Euclid CCD273-84 with 12-micron pixels, similar to the MOS CCD. The CCD housing is a vacuum enclosure with MgF2 window. As with the other two CCDs, heat pipes are used to cool the camera CCD.

As in the MOS, the M2 mirror is supported on a tip/tilt/focus mechanism, so the CAM can be focused independently of the OTA and the MOS. Tip/tilt adjustment enables dithering the image at the FPA to sample the PSF and to correct for detector sensitivity variations.

### 3.3.4. LUV/FUV/NUV Point-Slit Spectrograph (PSS)

The CETUS Point/Slit Spectrograph (PSS) comprises a LUV/FUV Rowland-like spectrograph with a 2”x6’-long slit capable of observation of a point source or greatly extended source, while the NUV echelle spectrograph assumes a point source (star or QSO). These two wavelength bands share a common entrance slit located in the telescope focal plane. The slit has dimensions: 2” x 360” but is pinched down to 0.2” near the middle to enable R~20,000 LUV/FUV spectroscopy of point sources.

**Figures 3-7** and **3-8** show the design of the PSS in more detail.

The OTA light passes through the entrance slit and then diverges at *f*/5 to one of two mirrors: a fixed NUV parabolic collimation mirror feeding the NUV echelle spectrograph or a relay mirror to the LUV/FUV spectrograph. The NUV spectrograph design is similar to designs of Hubble's GHRS and STIS. The LUV/FUV spectrograph design is similar to Hubble's COS spectrograph with a single optic, the disperser, for photon-efficient spectroscopy.

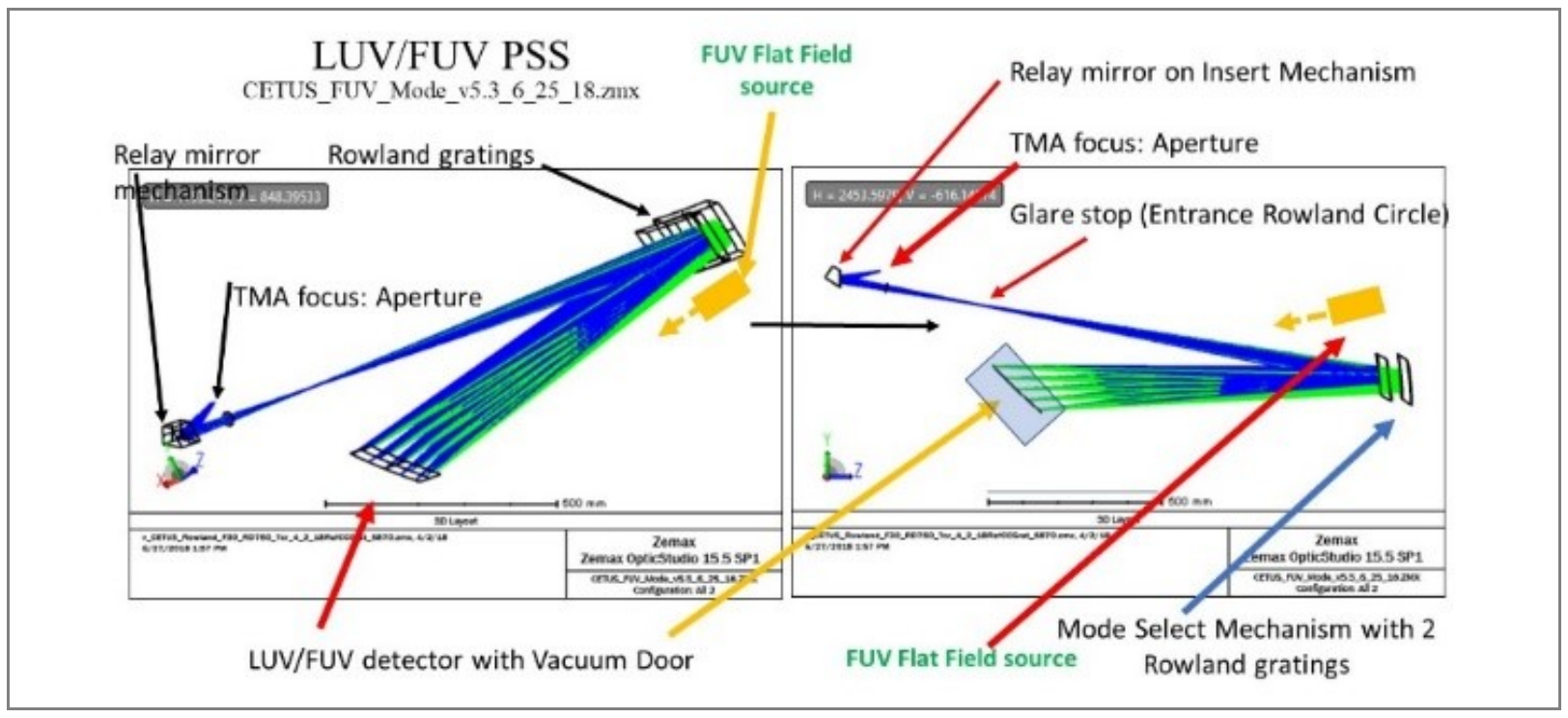

**Figure 3-7. The LUV/FUV PSS is a COS-like, Rowland-like spectrograph capable of reaching the LUV (100-115 nm).**

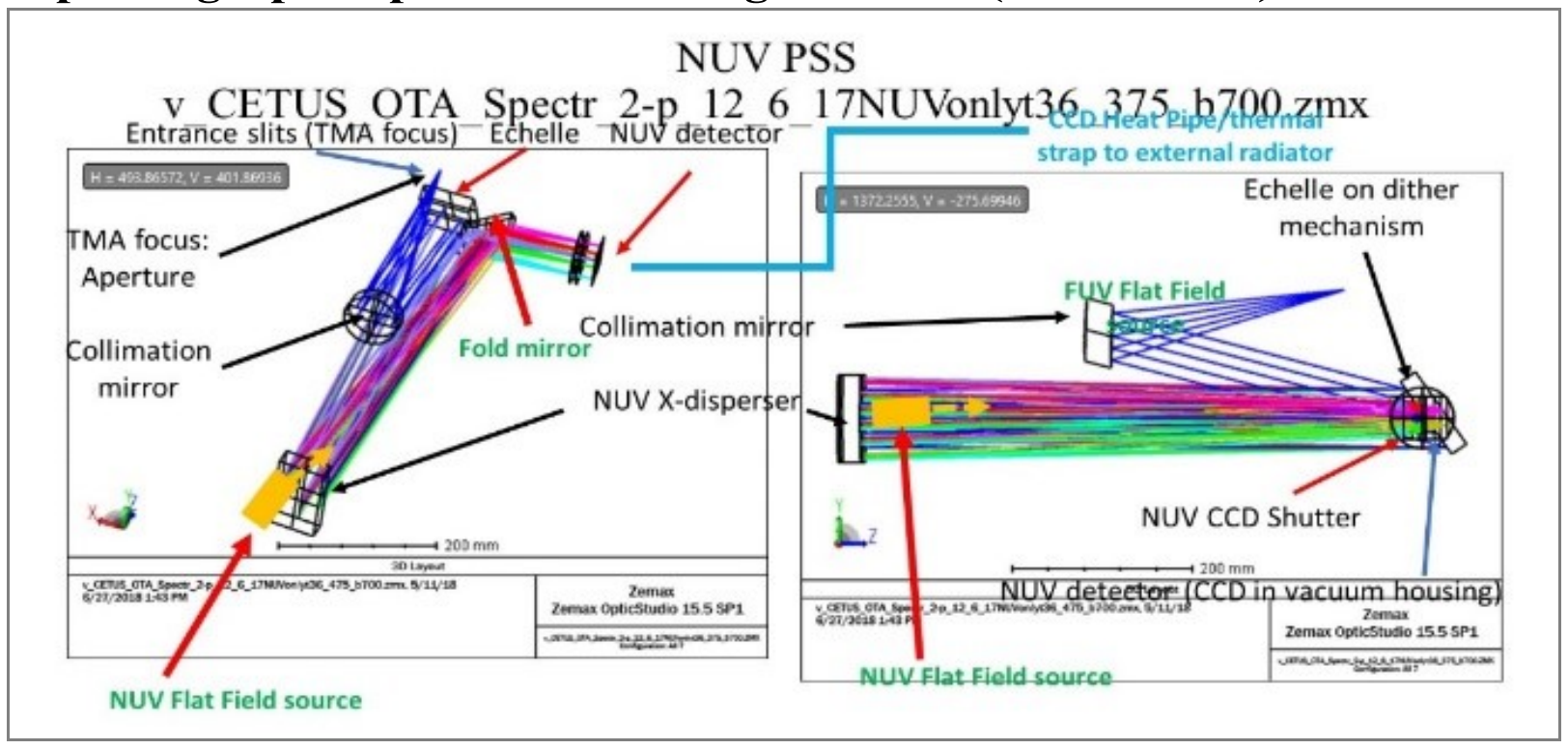

**Figure 3-8. The NUV spectrograph is efficiently packaged and has direct heritage from HST's GHRS and STIS**

To select the NUV mode, the LUV/FUV relay mirror is withdrawn from the beam diverging from the slit. The NUV parabolic mirror collimates the diverging beam, reflecting it to the plano NUV echelle grating. The spectrum is then imaged as an echellegram by the off-axis parabolic cross-disperser, as was done in GHRS and STIS. The NUV echellegram images on the 4 K x 4 K, e2v Euclid CCD273-84 with 12-µm pixels. (This is the same detector type that the MOS and NUV Camera use.) This mode achieves a spectral resolving power, R~40,000 over the spectral region 178 nm to 354 nm, with 2.5-pixel effective resolution. The CCD housing is a vacuum enclosure with a UV-grade fused silica window. The window provides order sorting, absorbing $\lambda < 160$ nm effectively blocking second and higher order light. As for the MOS and camera CCDs, the light-sensitive surface of the NUV spectrograph CCD is cooled by heat pipes.

To observe in either of the two LUV/FUV modes (G117 or G155), a mechanism inserts a relay mirror into the diverging beam from the OTA thereby blocking light into the NUV PSS. The LUV/FUV gratings are selected by a Mode Select Mechanism which inserts and registers in alignment the selected grating. The full LUV/FUV spectral region of 100 nm to 172 nm is split into two overlapping spectral intervals, one per grating: 100-134 nm (G117 mode) and 128-172 nm (G150). The design uses environmentally stable ALD-protected LUV Al+LiF for the mirrors and LUV grating. Both grating modes have a resolving power, R~12,000 in slitless mode (i.e. with the point source viewed in the 2"-wide slit), or R~20,000 when the point source is viewed through the narrow, pinched region of the slit. CETUS has viable coating options using Al+LiF reflective coatings on the TMA mirrors and LUV/FUV relay mirror and gratings with additional purge protocols. The spectral image in the LUV/FUV mode is sensed by an open CsI (solar-blind) photocathode, curved Micro-Channel Plate (MCP) detector with 22-µm effective resolution. The active area of the MCP is 200 mm long by 70 mm wide. The MCP housing is a vacuum enclosure with a vacuum door that is opened on-orbit. The detector operates at room temperature.

The CETUS design process has benefitted from lessons learned: both mitigation of risk and control of cost have weighed equally with science flow downs, environments and scientific productivity and versatility. The opto-mechanical payload design was initiated in co-located concurrent sessions, involving highly experienced optical engineers, opto-mechanical engineers, system engineers and managers. Through these sessions, a viable initial design addressed flow-down characteristics of photon collection, scale and constraints on the instruments. Gravity release, launch loads and faring dynamic envelope were also considered.

The design been accomplished with over one FTE year of expert optical design. Multiple design configurations have been developed as part of the CETUS optimization process. An independent review of the optical design, and the subsequent assemblies to be built and tested, has been reviewed with concurrence by Dr. James Burge. The sequence of designs all reference the requirements and figures of merit mentioned above, and have led to a practical optimal design.

## 3.4 MAJOR CETUS COMPONENTS

### 3.4.1 Wavefront Sensors

Wavefront sensing will be performed periodically to assess the optical alignment quality and image fidelity of the OTA. If alignment tune-up is indicated, the SM alignment will be adjusted deterministically to correct the OTA optical alignment and therefore the image quality of the OTA. Adjustment of the OTA SM allows correction of a limited number of image errors: third order coma, defocus, and image plane centering and tilt. With a wide FOV OTA, care must be exercised when adjusting the SM. An approach using use five Shack-Hartmann wavefront sensors distributed over the full reach of the OTA FOV has been designed and costed. We will further examine phase diversity methods of sensing wavefront error correctable by the 5 degrees-of-freedom adjustability available OTA secondary mirror.

### 3.4.2 Wavelength Calibration and Flat Fielding

The in-flight Wavelength Calibration System (WCS) images a large number of known wavelengths into the PSS for wavelength calibration. When wavelength calibration of the PSS is performed, light to the science instruments is blocked by a Calibration Insert Mirror (CIM) that rotates into the region near Cassegrain focus. This directs light from the wavelength calibration source to and through the PSS entrance slit. The spectrograph optics then disperse and image the calibration spectrum. This design provides wavelength calibration input using a STIS/GHRS-type Pt/Cr-Ne hollow core lamp. Again, FGS fine pointing control is maintained during this operation.

In addition, flat field source light floods each FPA to sense pixel/resels sensitivity variations and stability. Flat field lamps are mounted within each science instrument, respectively, to directly illuminate the subject detector. For FUV channels we plan to use Xe line lamp like we used on GHRS. For NUV channels we plan to use Kr or deuterium line lamps like we used on STIS.

### 3.4.3 Fine Guidance Sensor (FGS)

Two Fine Guidance Sensors, each with a 1127 x 1127 arc sec FOV, provide fine pointing error signals to the attitude control system which in-turn body points the OTA line-of-sight (LOS). The two fine guidance sensors view OTA fields-of-view (see Figure 3-1a) that are separate from those viewed by the science instruments (and WFS and WCS operations), so sensing does not interfere with science observing. This enables LOS control during science observations.

Each FGS uses an H4RG CMOS 4096 x 4096 pixel FPA with 10 μm x 10 μm pixels at F/5. The angular pixel width is 275 mas/pixel. The FGS FOVs are selected by fixed plano pick-off mirror near Cassegrain focus. A series of lenses image the FOV on its FPA. The FGS outputs provides error signal to body point the OTA LOS jitter to less than 40 mas (1 sigma).

### 3.4.4 Detectors

The different scientific instruments use detectors with pixel width of 12 μm (CCD), and resel's (resolution element) width of 20 μm (FUV MCP), and 22 μm (LUV PSS). A feature of this design is the use of a single CCD camera architecture, reused from instrument to instrument. This follows from a single observatory ICD, and allows a single CCD camera design, with common parts and common qualification. The effective pixel widths, as re-imaged by the SI optics, at the TMA focal surface are presented in Section 3.2, Image quality error budget. For reference, if widths corresponding to these pixel/resels sizes (12, 20, and 22 μm) were placed at TMA focus (i.e., at focal length of 7,500.0 mm) the angular widths are 330, 550, and 605 milliarc sec (mas), respectively.

For the near-UV, we have adopted improved CCDs developed by e2v that are better suited for CETUS. Developed and space qualified for Euclid is the CCD 273-84, a 4Kx4K CCD with 12-micron pixel, An in-family variant of the Euclid CCD, e2v's CCD272 customized for UV sensitivity has been adopted for CETUS. **Figure 3-9** shows the predicted transmission of the CCD window with a custom AR coating. The readout noise is only 2 e-. CCD272 was developed for the Solar UV Imaging Telescope (SUIT) (and other customers). It is expected to be delivered to the Indian Space Research Organization (ISRO) in 2019.

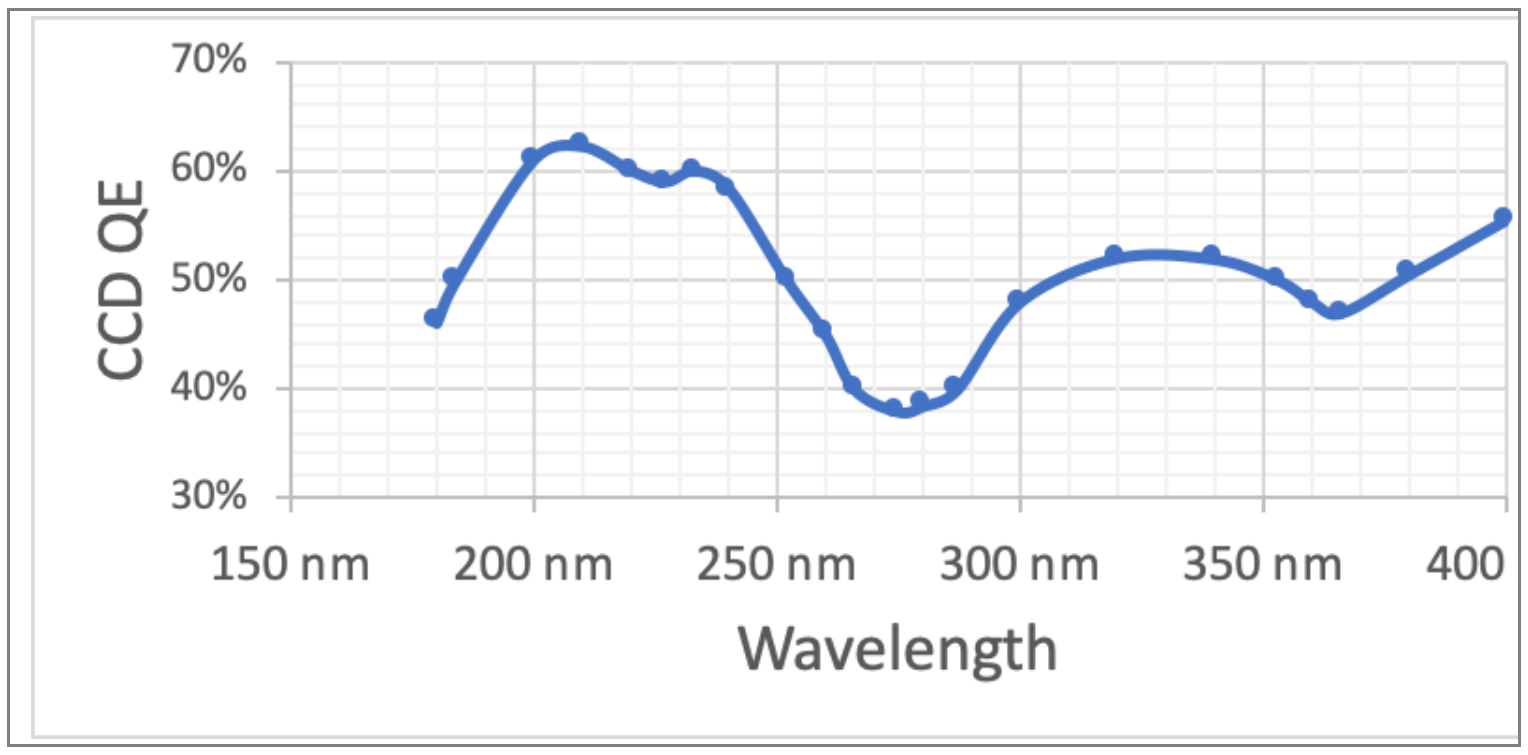

**Figure 3-9. The CETUS CCD is expected to have a high quantum efficiency at wavelengths as short as 180 nm, based on custom window coating for e2v's CCD272. e2v and JPL have started to work together on delta doping this CCD for even better QE in the UV. We expect this will be developed and qualified by the start of Phase A.**

For the far-UV, we have chosen micro-channel plate detectors (MCP's) developed by U.C. Berkeley Space Science Labs (SSL), the makers of Hubble's Cosmic Origins Spectrograph (COS) detector. Funded by NASA's Strategic Astrophysics Technologies (SAT) program, SSL has made major, numerous improvements in its far-UV detectors (**Table 3-4**) that enable the CETUS camera and Point/Slit Spectrograph to meet their science performance requirements. CETUS is also tracking progress in QE of delta-doped e2v EMCCDs. However, the solar blind characteristic of the MCP may be the deciding characteristic.

**Table 3-4. Improvements in the CETUS Far-UV Detector Over HST/COS**

| | FUV CETUS MCP | FUV HST/COS MCP |
|---|---|---|
| Detector type | XS (cross-strip), CsI photocathode | XDL, CsI photocathode |
| Spatial resolution | 20 micron FWHM | 35 micron FWHM (dispersion); 65-550 micron FWHM (X-disp) |
| Low gain operation | $10^6$ | $10^7$ |
| Higher dynamic range | Multi-MHz rates | 60,000 global count-rate limit |
| Ultra low MCP background | <0.05 events/sec/cm$^2$ | 1-2 events/s/cm$^2$ |
| High UV QE | 50% @115nm, 30%@200 nm | 26% @ 133 nm, 12% @ 156 nm |
| Solar-blind cutoff | ~350 nm | > 239 nm |
| Long stable lifetimes | >4x$10^{13}$ events/cm$^2$ | 4 lifetime positions due to gain sag |
| Format size | <200 mm x 200 mm; no gaps | 2x85mm x12 mm; 9-mm gap |

**Dithering incorporated to recover resolution -** The resolution of both the near-UV MOS and near-UV camera is set by the subtense of a CCD pixel (12 microns, or 0.33 arcsec). To properly sample a pixel, it is necessary to dither, i.e. to make small movements of the field incident on the detector format. Real detectors have hot pixels, blemishes, non-uniform response, etc., so it is also necessary to move the scene by several pixels on the detector format. We can combine these offsets with the sub-sampling offsets to construct a dithering pattern.

Our implementation of dithering uses the M2 optical component of the Offner relay (i.e. the convex grating in the MOS, or the convex mirror in the camera) to achieve dithering by non-integer pixels. The target in the MOS or CAM stays at exactly the same position at the telescope focal plane throughout the exposure. The fine guidance sensor would not see any movement in the guide stars. There is no need for a fine steering mirror. There is a need, however, for a precise mechanism on M2.

## 3.5 PAYLOAD MECHANICAL/ELECTRICAL/THERMAL DESIGN

### 3.5.1 Payload Mechanical/Structural Design

We designed the CETUS science payload structure to provide rigid support of telescope and instrument optical elements and of the instruments. To minimize any mechanical and thermal distortions between the OTA and the instruments, we baselined an all-composite structure based on a flight-proven M55J fiber/Cyanate-Ester resin construction. **Figure 3-10** shows the CETUS payload structure components. The 33-cm thick Stable Member serves as the primary mirror bench and the main structure component of the forward metering structure. The Stable Member also serves as an interface structure for the aft optics tower and science instruments.

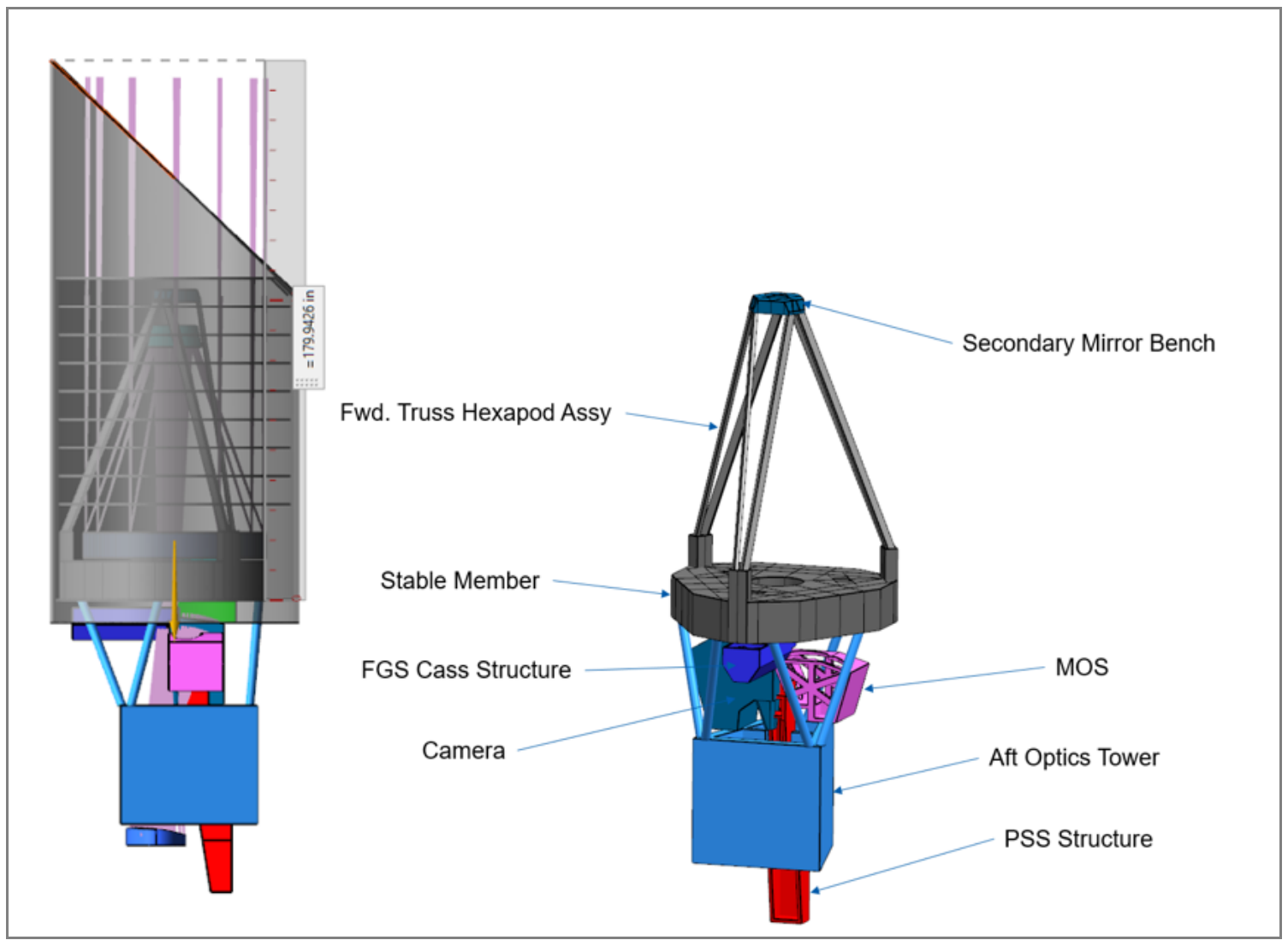

**Figure 3-10. The full Payload (left, and cut-away view showing instruments right) has been modeled at NGIS. All dynamic requirements are satisfied.**

The forward metering structure holds the primary mirror and the secondary mirror using a proven M55J composite hexapod truss assembly. Invar 36 is baselined for optical interface fittings to minimize thermal CTE mismatch between the Zerodur optical elements and the M55J structure. Also, part of the forward metering structure is the Main Baffle assembly, which uses M55J/Cyanate-Ester barrel and stray light vanes, and Invar 36 fittings. The interior of the baffle assembly is painted with Z306 black paint to provide maximum stray light protection for CETUS instruments. Surrounding the Main Baffle assembly is the Outer Sun Shield, which acts as a main thermal barrier for the CETUS OTA.

The aft optics tower holds the tertiary mirror, PSS, and Wavefront Sensor assembly, and connects to the Stable Member through M55J composite hexapod truss. The camera, MOS, and FGS subassemblies attach to the backside of the Stable Member using kinematic mounts. As CETUS instruments may be developed and built by partner institutions, the CETUS payload structure is designed to easily integrate fully qualified instrument assemblies to the payload structure with ample available space for alignment of each instrument assemblies to the optical assembly.

Each instrument housing is designed using a flight proven M55J composite construction to provide rigid support of optical elements and to maintain critical alignment tolerances using kinematic mounts. Each instrument housing also accommodates all detector assemblies, including detector Front End Electrics (FEE), and various mechanisms such as Tip/Tilt/Focus mechanisms and other required mechanisms.

A preliminary analysis of the CETUS payload structure verified the structural soundness of design. The results show that the lowest frequency first mode is 29.4 Hz, which indicates the structure is stiff enough to be viable.

### 3.5.2 Payload Electrical Design:

Each CETUS instrument is connected to a single Main Electronics Box (MEB), which provides all power and control interfaces to operate the instrument. To meet the requirements of NASA Class B risk classification, we designed each MEB to be single fault tolerant dual string redundant architecture. In addition to MEBs for Camera, MOS, and PSS, two additional MEBs are allocated for OTA control and auxiliary instrumentation including FGS and Wavefront Sensors. Each MEB is about 10 kg in mass and requires 48 Watts of power. **Figure 3-11** shows the CETUS payload electrical block diagram.

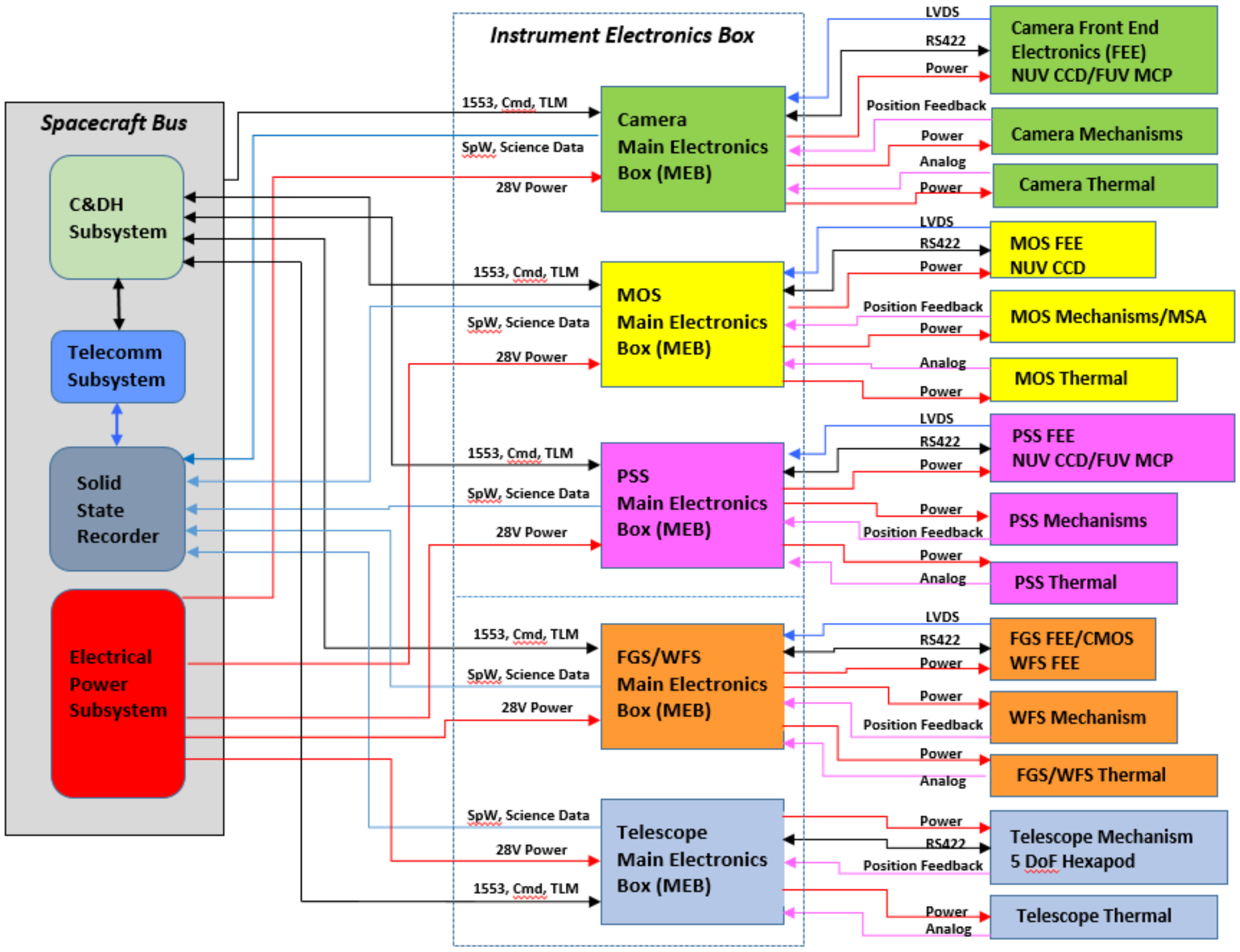

**Figure 3-11. The electronics are redundant and satisfy NASA Class B risk classification. Figure credit: GSFC/M. Rhee**

### 3.5.3 Payload Thermal Design

The CETUS payload thermal design satisfies three requirements:

1. Opto-mechanical stability of the end-to-end optical system, including mirror figure and solid body alignment: This is accomplished by minimizing the sensitivity via selection of "best materials" for thermal stability, and supplementing this with low-authority easily achieved proportional heater networks placed at sensitive nodes.
2. Driven by in-service molecular and particulate contamination issues, the optics operate "warm biased" with respect to surrounding structural components, near room temperature and under near steady state conditions. Thermal management addresses the optical wavefront error budget. Furthermore, the temperature of the optical surfaces can be increased by another approximately 10C to periodically redistribute any residual contamination. In addition, mechanisms (e.g., the Secondary Mirror Hexapod), are maintained at operational temperatures.
3. The CCDs require cold operation at approximately -100C or lower. The design cools the CCD cavities, and the associated front-end electronics (FEEs), via heat pipes and conductive paths to radiators looking at cold space.

Equilibrium temperature fields have been analyzed CETUS' L2 orbit, and looking at cold space. **Figure 3-12** illustrates the passive thermal analysis of CETUS. The front end of the Sun Shield Assembly is angled, with the high side of the scoop always favoring the sun direction. While CETUS can observe transients in the anti-sun direction, in survey mode it will point approximately 90 degrees to the sun using simultaneously the MOS and camera. There is no requirement to make the CETUS payload isothermal, but temperature stability is required. Thermal distribution in survey mode at SEL2 matches our detailed thermal node analysis. Together they indicate that a highly stable optical path can be maintained by only mild nodal heating. The optical system will rapidly adjust to transients resulting from large slews.

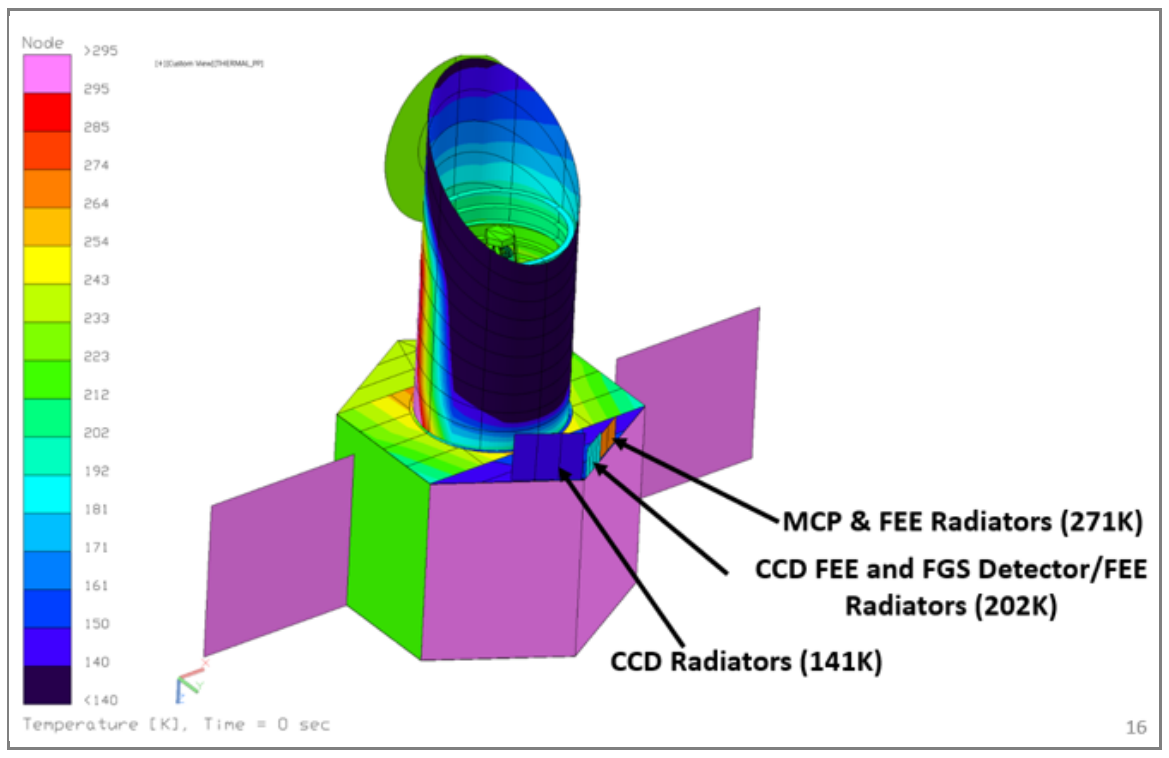

**Figure 3-12. The thermal-optical system is stable both for surveys and for large slews to reach transients. Figure credit: M. Choi**

*Requirement #1, Stability*, is achieved by first using high heritage materials that first have extremely low coefficients of thermal expansion (CTEs). The baseline material selection is ZERODUR® mirror substrates, exhibiting state-of-the art homogeneity of CTE, and M55J Carbon Fiber Reinforced Polymer (CFRP) metering structures. Not only are these material CTEs lowest available for space, but also ZERODUR® and M55J can be tailored to CETUS operational temperature band, and also matched closely to each other. This passive approach will be complemented by active heater circuits placed near metallic components in the metering path. Only easily achieved low-precision thermal control is needed, and space payloads routinely require thermal control over two orders of magnitude better than that needed by CETUS. With these "best" materials, coupled with easily achieved thermal control, the reserve in the error budget for thermal control is nearly a factor of two, which we regard to be highly robust as well as cost effective and low risk.

*Requirement #2, warm biased mirrors*, is easily achieved via radiative transfer employing thermofoil heater arrays on the back of a thin aluminum emissive plate located a short distance behind the mirror. Nominal "bias" temperature and "contamination redistribution" temperatures are achieved with modest power requirements.

### 3.5.4 CCD Coolers

Teledyne-e2v's CCD 273 has optimal performance (lower dark, higher CTE) if operated at -120 C. This operating temperature is achieved via the following design approach. Each CCD is put in a temperature-controlled dewar to minimize parasitic heat paths via radiative transfer from the cavity and window, or conduction from the walls of the instrument. While there inevitably is some parasitic heat transfer from electronics via fine-gauge wires electrically connecting the CCD to the front-end electronics, this path too will be minimized. The CCD dewar assembly is mounted on a cold plate, with the central portion relating to the device maintained at -120C. The cold plate transfers any heat gained by the CCD to thermal heat pipes, which move the heat to radiators located outside the observatory. Due to constraints of roll angle, the radiator surfaces are always shielded from the sun as well as earth and moon. Thus, the radiators will always look to stable cold space. The full performance of the CCDs will be tested in a thermal-vacuum chamber. While the payload will include tests of alignment and gravity release through 6 azimuthal rotations, the SNR characteristics of the CCDs will be measured only at the azimuth where the heat pipes are aligned perpendicular to gravity. In this orientation, the radiators will face proximate LHe-cooled cold plates closely simulating the temperatures during operation. All methods for building, cooling and testing the CCD camera assemblies are tracible to other missions; thus, they assume a high

TRL. All requirements can and will be tested robustly during the sequence of thermal vacuum (T-V) tests.

## 3.6 ALIGNMENT, INTEGRATION, AND TEST (AIT)

The CETUS study generated a comprehensive AI&T test flow. Components will be integrated into the CETUS system as shown in the overall flow diagram, **Figure 3-13**. OTA PM and SM will be integrated and aligned, as depicted in **Figure 3-14**, using a double pass wavefront test with an autocollimating flat. The OTA TM and the instruments will then be integrated and aligned.

Alignment for the CETUS OTA and instruments will utilize (Computer-generated hologram (CGH) alignment techniques. Non-symmetric three mirror anastigmatic telescopes such as CETUS present alignment challenges due to degeneracies which occur because there are different degrees of alignment that have nearly the same signature in the wavefront. For example, the aberration of coma in the wavefront could come from either a decenter or a tilt of any of the aspheric mirrors. The correct alignment state must be determined for each mirror, as correcting, again for example, coma, with the wrong mirror creates higher order aberrations, with higher order field dependence. We avoid this complexity by first aligning the primary-secondary pair using the FEFDAT test, the Front- End Field Diverse Autocollimation Test. We then align the full OTA with a similar test.

The field diversity provides unambiguous information on the state of the alignment for this pair of mirrors. Taking the primary mirror as the datum, we have only the secondary mirror tilt, decenter, and focus to determine. While a single field point will show degeneracies, measurements across several field points sort this out. We use a CGH to simultaneously provide five field points for testing across the FOV in autocollimation. Importantly, the CGH creates all five field points from a common CGH, with multiple patterns, avoiding alignment degeneracies.

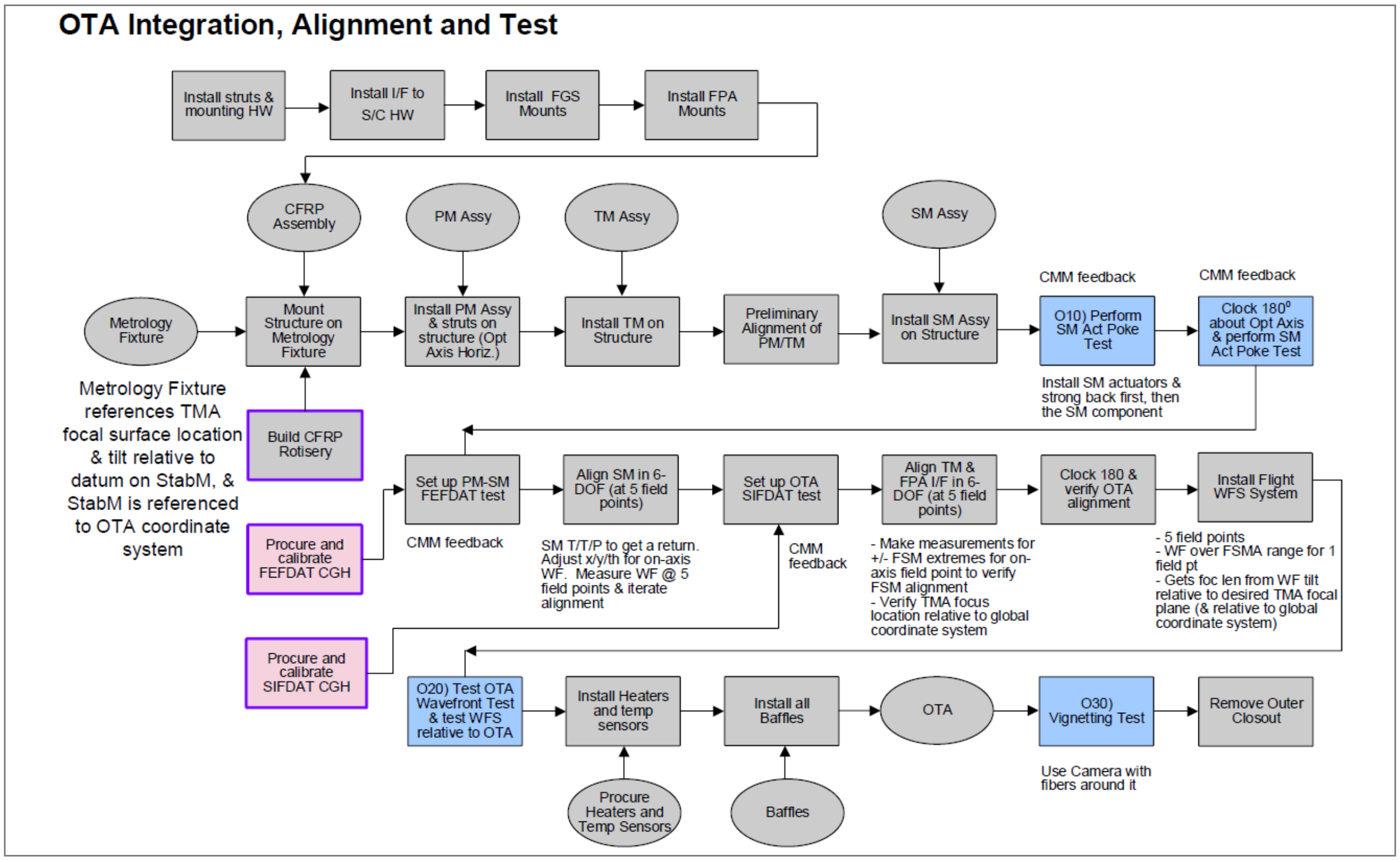

**Figure 3-13. CETUS AI&T Flow**

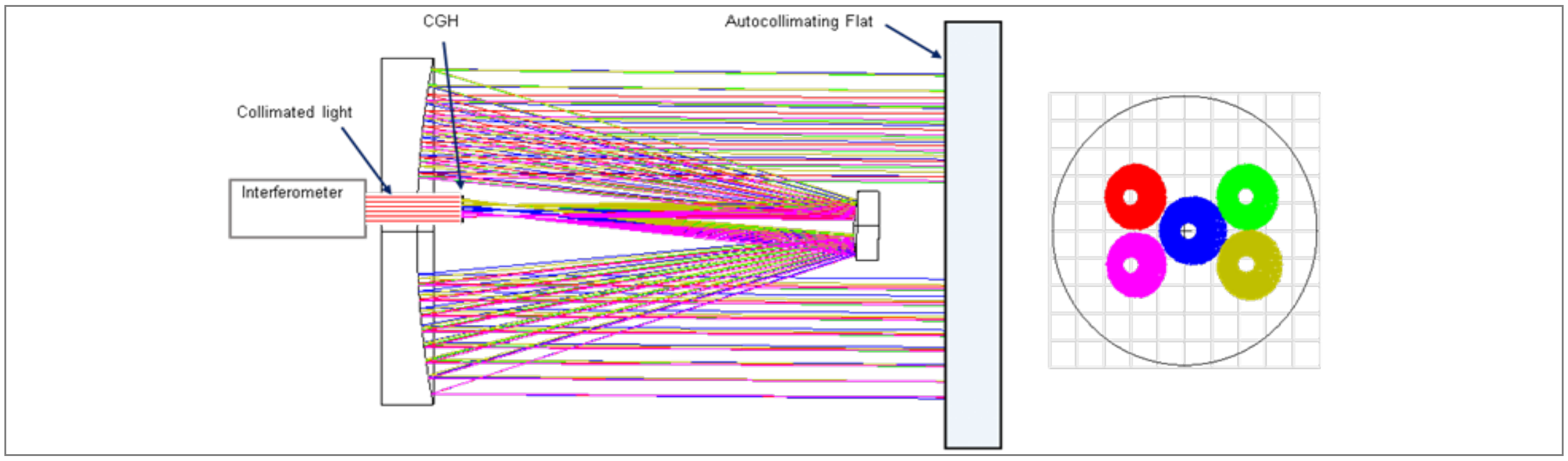

**Figure 3-14. The Korsch TMA OTA alignment is straight forward using established autocollimation, inteferometry and a CGH with patterns covering the OTA's wide field of view. The CGH pattern layout is shown at the right. Rotational shear will be used to remove gravity effects.)**

A similar alignment will be used for the instruments. An example is shown in **Figure 3-15** for the camera, utilizing a CGH providing nine field points. This again allows the various alignment degrees of freedom to be measured and avoids any field-dependent degeneracies.

System wavefront testing will take place with fiber optics aligned to the TMA focus. Single-mode fibers provide excellent point-sources, which will overfill the system aperture. These sources enable full-aperture system wavefront testing of the system wavefront, including the OTA, as well as the Camera and MOS.

The more expansive AI&T flow with the integration of the Science Instruments, FGSs, and WFSs with the telescope and subsequently the entire Science Payload with the Spacecraft bus is shown in **Figure 3-16**. The Observatory (Science Payload and Spacecraft bus) integration with the Launch Vehicle is the last step.

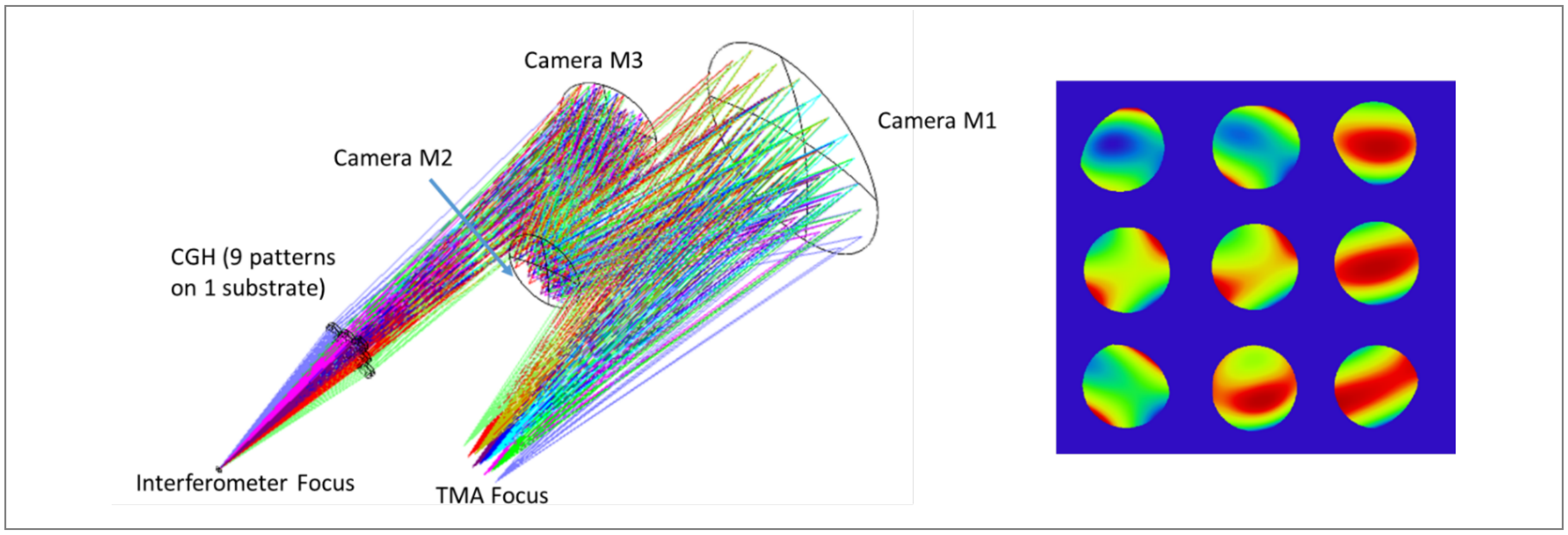

**Figure 3-15. The CAMERA's aspheric-based Offner Relay is straight-forward to align with methods established at AOS. The test CGH has multiple field points. On the right is the resulting conspicuous fringe appearance of just 10µm decenter of M2**

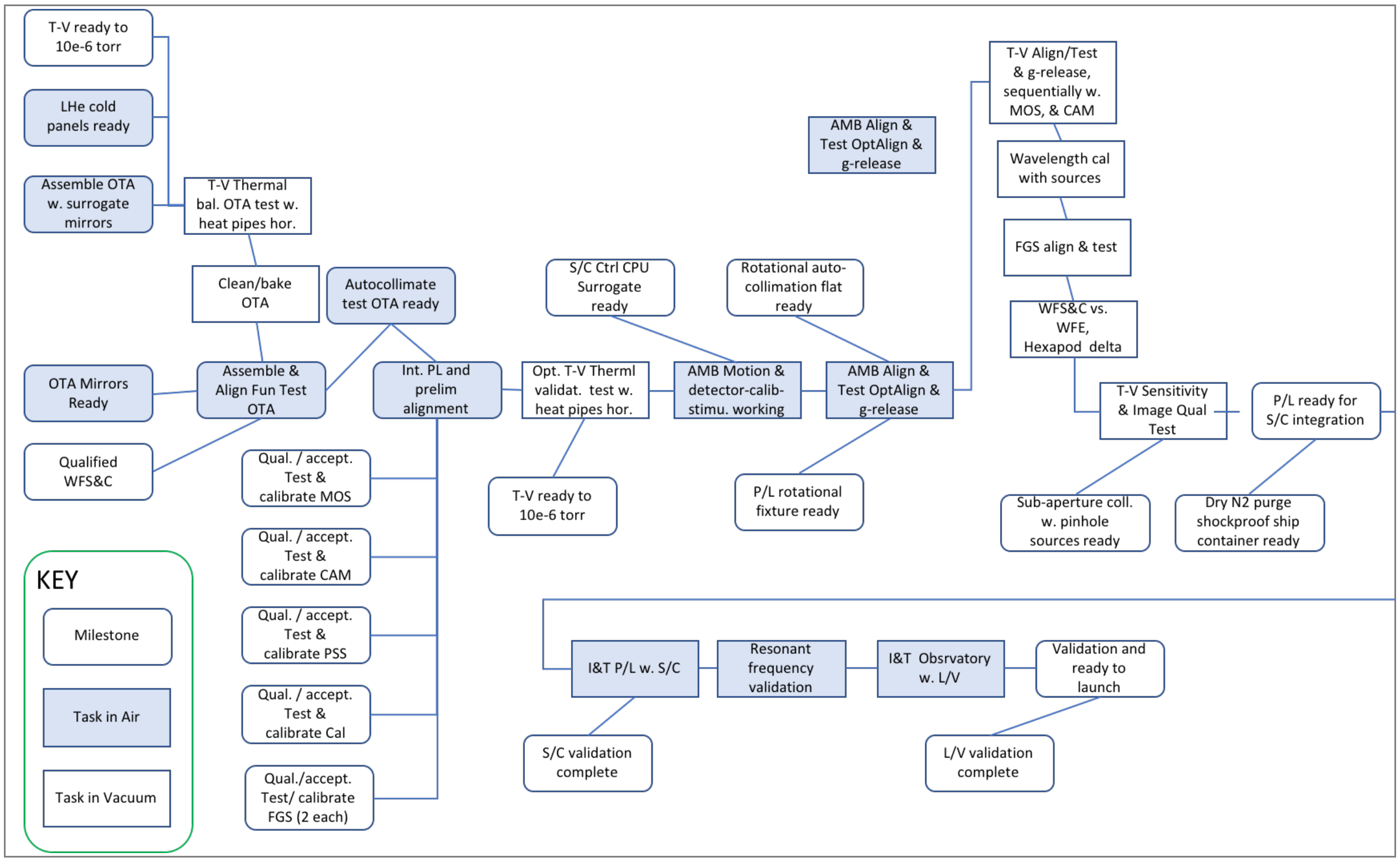

**Figure 3-21. CETUS AI&T has been planned through integration with the Launch Vehicle, including working with facilities at CU LASP for payload integration, and NGC Redondo Beach for observatory integration.**

# 4. MISSION DESIGN

The mission design for a space telescope such as CETUS basically consists of developing the lowest-cost approach for a set of instruments to successfully carry out a pre-defined set of science observations. Accomplishing this objective requires careful design of four separate but interrelated mission elements: orbit, operations, SC Bus, and launch vehicle (LV).

## 4.1 OVERVIEW

The starting point for developing this set of designs is to define the instrument "accommodation requirements", which for CETUS can be summarized as follows:

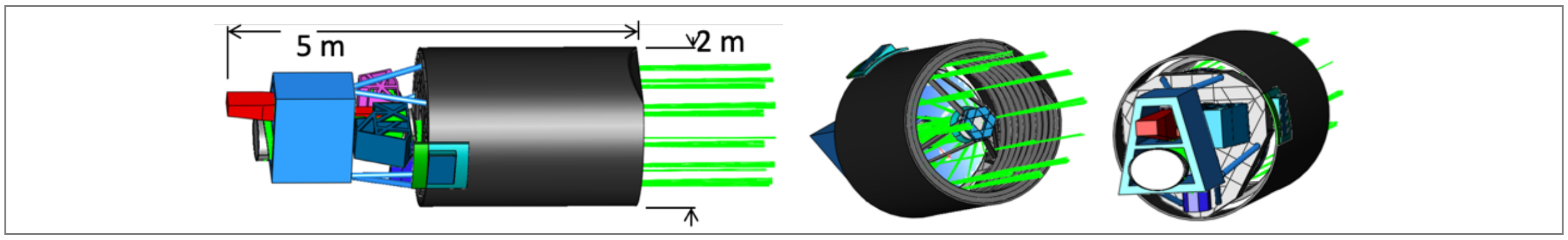

**Figure 4-1 The CETUS Science Payload's three UV instruments are fabricated and integrated independently and can operate simultaneously on-orbit.**

- Instrument envelope (Shown in **Figure 4-1**) of about 5 m long and 2 m in diameter
- Instrument mass of ~1,084 kg and power requirement of ~1,200 W
- Instrument Field of Regard (FoR) of about 2 π steradians of sky around the anti-Sun line at any time and 4 π steradians of sky in a year or less.
- Instrument pointing accuracy of ~0.1 arc-sec

- "Room" temperature thermal environment of instruments (operating at slight bias for stability)
- Science Data Downlink of about 27 Gbit/day
- Anytime command uplink to quickly slew to observe a transient event
- Instrument slew of 180 degrees in less than 15 minutes after commanded
- Duration of observations: 5-year design life with 10-year goal

Key features of the CETUS mission and the corresponding rationale are summarized in **Table 4-1**.

**Table 4-1 Key features of the CETUS mission and rationale**

| CETUS Mission Feature | Rationale |
|---|---|
| Orbit about the Sun-Earth L2 | • ~2 $\pi$steradians of sky visible at any one time<br>• 4 $\pi$ steradians visible over any 12-month period<br>• Relatively easy to reach and maintain orbit<br>• Reasonable cost of communications<br>• Thermally benign with Sun/Earth light always from roughly the same relative direction |
| 5-year mission design life with consumables for 10 years | • Longest mission duration that avoids significantly more expensive design |
| Dual wing solar arrays (SAs) with single axis articulation | • Supports Field of Regard (FoR) over 2 □steradians of sky<br>• Keeps Cp near Cg, minimizing rate of momentum buildup |
| Oversized Reaction Wheel Assemblies (RWA's) | • Slew to view any transient in FoR within 15 minutes after commanded |
| Two-axis articulated High Gain Antenna | • Can communicate with ground at high data rates while CETUS is observing anywhere in FoR |
| On-board Celestial Navigation for onboard Orbit Determination | • Eliminates costs of tracking by ground system |
| Near-Earth Network Ground System | • Adequate performance and less expensive than Deep Space Network |
| Falcon 9 Heavy Launch Vehicle | • Adequate and proven performance, adequate fairing size, low price |

## 4.2 ORBIT

The orbit selected for CETUS (**Figure 4-2)** is approximately centered on the Sun-Earth Second Lagrange Point (SEL2 Orbit). This orbit is a beneficial one for a variety of space telescopes (e.g., WMAP, Planck, Herschel, Gaia) and is planned for further observatories in development (JWST, WFIRST, PLATO).

The SEL2 orbit provides a Field of Regard (FoR) of over $2\pi$ steradians of the sky at any point in time, and 4 $\pi$ steradians over 6 months bringing the following benefits to CETUS:

- The large FoR will facilitate the CETUS GO program because there will be no shortage of opportunities for CETUS to observe any specified GO target.
- The large FoR also means that CETUS will always have a better than 50% probability of being able to observe any short-lived transient event.
- The FoR also enables Celestial Navigation (in which small star-tracker-like cameras facing in the anti-sun direction measure the positions of known Near-Earth Objects (NEOs)) which CETUS can use to determine it position and basically eliminate the expense of DSN ground station tacking.

It will be practical at any time to send CETUS a command to slew to and observe a transient event. Also, since a SEL2 orbit is about 1.01 AU from the Sun, the spacecraft (SC) will be able to use high-heritage solar arrays and thermal control technologies that have been refined by heritage SC that have flown under similar conditions.

## 4.3 GROUND STATIONS AND OPERATIONS

The mission design goal is for CETUS to be able to use the most cost-effective ground stations that can handle its regular downlink science data, plus provide on-demand command uplink capability to enable CETUS to almost immediately slew to and observe transient events. The combination of the relative proximity of a SEL2 orbit and non-extreme downlink requirements (~27 Gb per day) allows CETUS to use the Near-Earth Network (NEN) as opposed to the more expensive Deep Space Network (DSN). Also, the larger number of NEN ground stations relative to the just-three DSN stations will facilitate having a ground station available to transmit an immediate command sequence for CETUS to monitor a transient event.

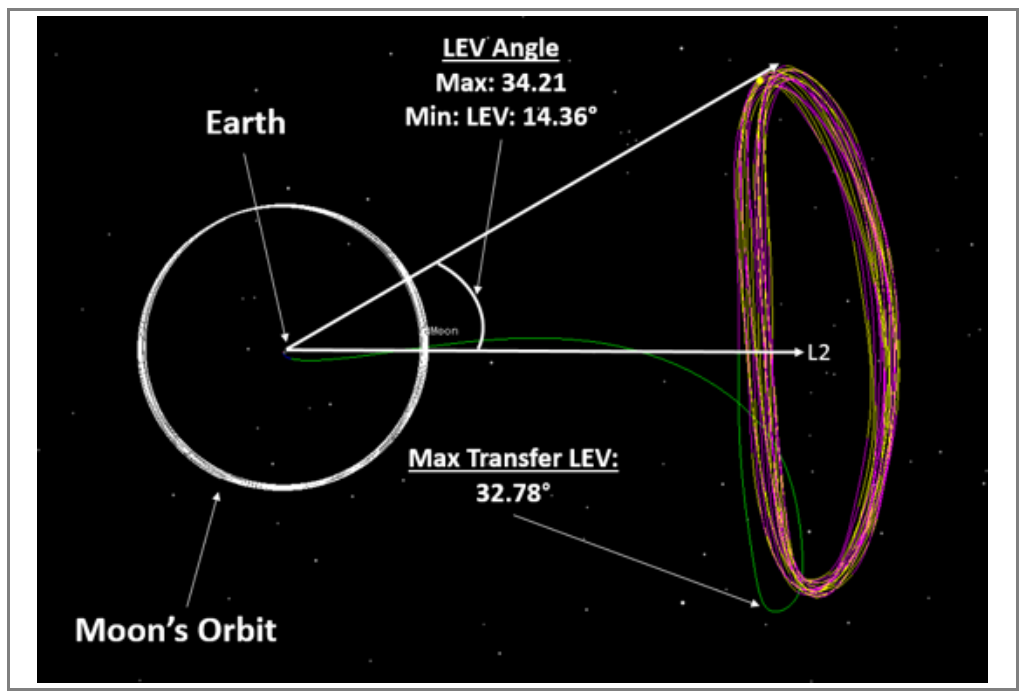

**Figure 4-2. 4x34-deg L2 Earth Vehicle (LEV) Angle of notional CETUS Orbit**

## 4.4 SPACECRAFT BUS

The CETUS Spacecraft (SC) is based on an existing Northrop Grumman Innovation Systems (NGIS) product line to reduce cost and risk. An independent bus design was also done by the GSFC MDL with most features being essentially identical. The NGIS LEOStar-3 product line is being used on Joint Polar Space System-2 (JPSS-2), Space Test Program Satellite-6 (STPSat-6) and Landsat-9 in both LEO and GEO. It is used for high value, long life missions that require a precision Attitude Control Subsystem (ACS), wideband communication, redundancy and robust fault management. The product line uses a common architecture that can accommodate mission specific requirements. Fourteen have been launched and five are in production. The CETUS SC configuration is shown in **Figure 4-3**. The design is completely redundant and uses Technology Readiness Level (TRL) 7 to 9 hardware with no new technology development required.

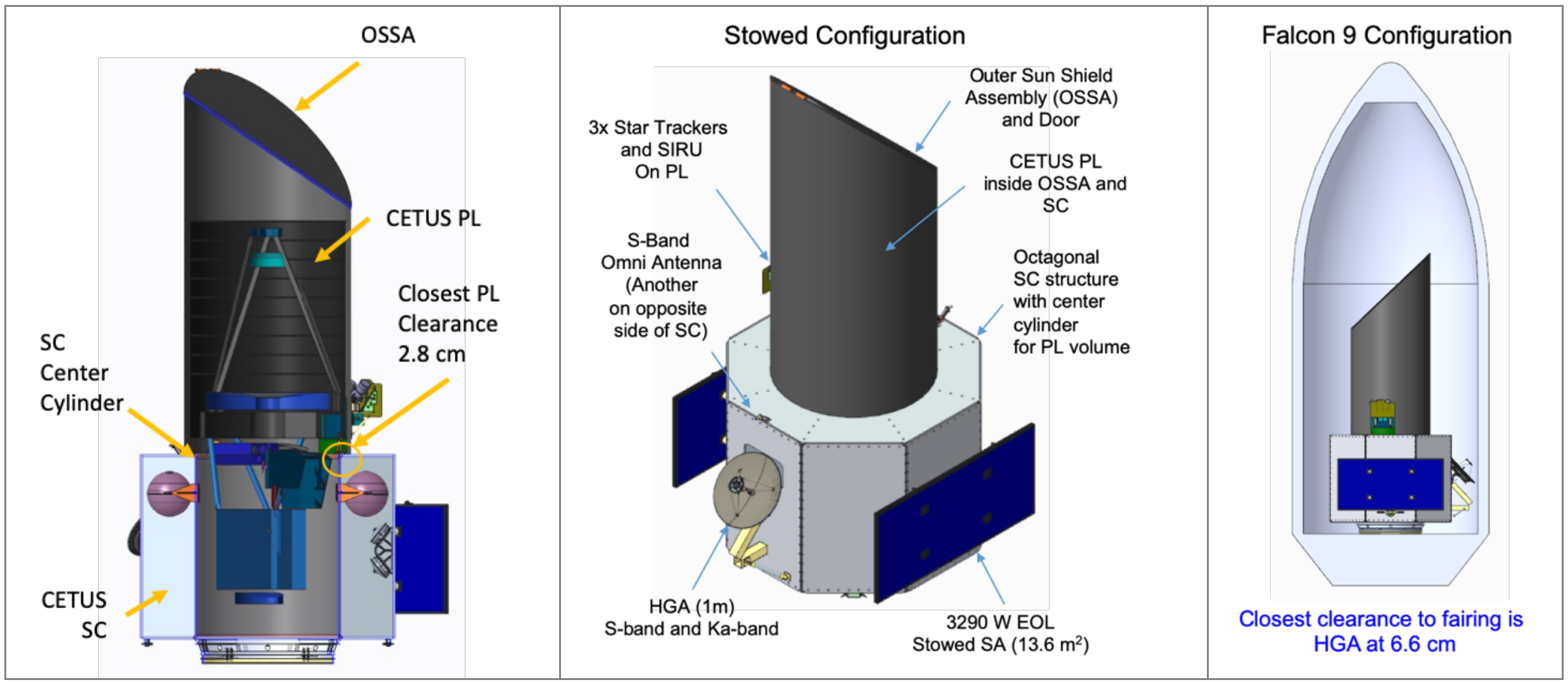

**Figure 4-3. The CETUS Observatory has a mature payload and a high heritage spacecraft based on the NGIS LEOStar-3 spacecraft. The dynamic envelope fits with margin within the Falcon 9 fairing.**

The CETUS Spacecraft Bus reserves the volume inside its center cylinder volume for the CETUS payload and uses its center cylinder to support the payload. The interfaces between the payload and spacecraft are kinematic and adiabatic. The spacecraft Outer Sun Shield Assembly (OSSA) surrounds and protects the payload. This arrangement provides a consistent thermal environment for the payload. The five payload Main Electronics Boxes (MEB) mount on a spacecraft equipment panel. The star trackers and gyro mounted on the payload Stable Member to provide a tight linkage between spacecraft pointing and payload boresight.

The CETUS spacecraft (SC) accommodates both the Payload (PL) and Launch Vehicle (LV) Dynamic Envelope (DE) as shown in Figure 4-3. NGIS high heritage SCs are the basis for the CETUS). Features of this design are:

- SC TRL levels range from 7 to 9 with no new technology development for low risk
- Redundant SC elements with 5 years of SC design life and 10 year goal).
- SC provides center volume for CETUS PL and uses composite center cylinder to support the PL like the NGIS GEOStar-2 and -3 product lines.
- SC does celestial navigation using 3x Malin ECAM50 cameras and Goddard Image Analysis Navigation Tool (GIANT) SW
- 30 days between RWA desaturation events
- Preliminary analysis indicates 40 milli-arcsec jitter and drift requirement can be met with close-loop control system using PL FGS, and SC gyro and RWAs
- The SC employs S-Band command and telemetry links using LGA and HGA, and 15 Mbps Ka-band wideband data downlink using HGA
- The SC employs redundant Integrated Electronics Modules (IEM) with a RAD750 CPU and cPCI Backplanes for SC C&DH, and a redundant Payload Interface Electronic (PIE) with 512 Gb Flash Memory Cards for the CETUS PL interface
- FSW uses VX-Works RTOS, C++ and autocoded ACS SW
- Mono-propellant blowdown propulsion system, pointing away from the PL for minimum contamination
- 3290 W (EOL) SA with two wings and 78 Ahr Li Ion Battery
- SC includes the Observatory Outer Sun Shield Assembly (OSSA) and door

The Spacecraft Bus design meets the performance requirement of the CETUS mission with significant margin as shown in **Table 4-2**. This margin reduces risk and provides flexibility to accommodate changes.

**Table 4-2. CETUS Spacecraft has appropriate margin for early stages of conceptual study. The performance characteristics can be estimated at this time for all but three of the parameters, and the CETUS team is confident that these will close easily at the next study phase.**

| Requirement Description | Rqmt Value | Perf. | Margin | Comment |
|---|---|---|---|---|
| CETUS Observatory Wet Mass | 3375 kg | 2640 kg | 22% | Falcon 9 capability to L2 |
| CETUS SC Dry Mass | 1252 kg | 1074 kg | 16% | Uncertainty based on status |
| S-Band Command UL using HGA w/ranging | 2 kbps | 2 kbps | 18 dB | 18m NEN GT and at L2 |
| S-Band Telemetry DL using HGA w/ranging | 2 kbps | 2 kbps | 14 dB | 18m NEN GT and at L2 |
| Ka-Band Wideband Downlink Using HGA to NEN | 15 Mbps | 15 Mbps | 6.8 dB | 18m NEN GT and at L2 |
| Mission Delta V | 102 m/s | 102 m/s | NA | Fuel based on 3375 kg mass |
| Fuel | 224 kg | 272 kg | 18% | Tank size margin |
| Slew and settle within 2π FoR | 15 min | 5 min | 67% | Six RWAs |
| Jitter and Drift over 30min observation (1 sigma) | 40mas | 29mas | 27% | Preliminary 1 axis analysis |
| SC Pointing Knowledge (1 sigma) | 2 arcsec | 1.5 arcsec | 25% | 2 for 3 star trackers |
| RWA Desaturation | 7 days | TBD | TBD | Six RWAs |
| 7 Days of Science Data Storage | 183 Gb | 512 Gb | 180% | Flash Memory in PIE |
| Solar Array Size | 2562 W | 3290 W | 22% | w/15% SC and 30% payload uncertainty |

CETUS uses techniques developed on other LEOStar-3 programs to reduce LOS jitter and uses existing LEOStar-3 product line avionics. Six Honeywell HR14-50 Reaction Wheel Assemblies (RWA) are baselined, so zero crossings can be controlled. They are only operated in the lower half of the speed range to minimize jitter. There are two solar array wings with one-axis drives designed for 15° cosine loss. The L2 orbital drift is only 0.08°/hr, so the solar array wings are fixed during observations. A 1-m Ka-Band High Gain Antenna (HGA) with a Ka-Band transmitter and 10W TWTA provides mission data downlink. An S-Band system provides 24/7 communications capacity for command, telemetry and ranging using an S-band Transceiver with omni antennas. The Command and Data Handling (C&DH) Subsystem uses a proven Integrated Electronics Module with a RAD750 CPU and cPCI backplane for spacecraft functions, and a Payload Interface Electronics (PIE) for payload interfaces and data storage. These avionics meet the L2 radiation requirements. Flight Software (FSW) uses the Vx-Works Real Time Operating System (RTOS), C++ code and ACS autocode. The Propulsion Subsystem uses a mono-propellant, blowdown approach with 8 thrusters for orbit changes and 8 for RWA desaturations. The Electrical Power Subsystem (EPS) uses a proven Power Distribution Unit (PDU) for power control, battery charging, and power distribution and switching, a 3290 W (EOL) solar array and 78 Ahr Li Ion batteries.

Based on our approach and the reasonable changes to the NGIS product line design, the CETUS SC is low risk.

### 4.5 LAUNCH VEHICLE (LV)

The Falcon 9 or Falcon 9 Heavy is the LV choice for CETUS due to its well-established reputation for low cost and reliability. The Falcon 9 has also successfully launched one spacecraft (DSCOVR) into a SEL1 orbit, which from a LV perspective requires the same kind of trajectory capability as a SEL2 orbit. The mass and stowed volume of the CETUS observatory (Science PL and SC Bus) fit readily within the Falcon 9 fairing size (Figure 4-3 right) and mass capability to a SEL2 orbit. The mass margin with a Falcon 9 is shown in Table 4-2 and illustrates that CETUS is compatible with the Falcon 9 capabilities. However, at this stage we prefer a higher than 22% mass margin and have therefore changed our baseline to a Falcon 9 Heavy to provide significantly higher mass margin. The cost of a Falcon 9 Heavy is now in our cost baseline as shown in **Section 8**. The length of the fairing allows a photon-efficient design that doesn't require any fold mirrors for packaging.

## 5. CONCEPT OF OPERATIONS

### 5.1 SCIENCE OPERATIONS

CETUS science operations include long-range planning of CETUS observing programs, scheduling CETUS observations, post-launch receipt, data processing, and archival of CETUS data.

**Long-Range Planning**: CETUS is essentially a survey telescope serving the astronomical community. Long-range planning is relatively simple, since surveys usually involve repeated observations over a selected field. We do not foresee a major General Observer program in the first ~4 years, although individual observing programs may be carried out via Director's Discretionary Time. We envision a small number of steering groups for major programs, e.g. survey of the CGM. The steering groups would be formed during the development of CETUS and continue into the post-launch period. In the development phase, each steering group would be responsible for reviewing hardware development, continually assessing the scientific capabilities of CETUS, and planning for science verification and calibration. In the operations phase, each steering group would formulate an overall plan for executing science and calibration to be carried out in the next six months. A major exception is the steering group for transients. That group would formulate criteria for selecting alerts to be acted upon, which involves interrupting one of the on-going surveys, and for developing canned observing procedures to be stored on-board CETUS for different types of transients.

**Scheduling**: Scheduling of CETUS observations for hand-off to Mission Ops would be similar to that of other astronomical space telescopes.

**Post-launch Data Processing:** NASA typically requires a science operations center to carry out what is called Level-1 and Level-2 data processing. Level 1 processing consists of receipt of telemetry data and correction of telemetry errors, and formatting the packetized data into an image or spectrogram. Level 2 processing includes removal of the instrument signature and conversion of the data to physical units. NASA typically does not require Level-3 processing, which for CETUS would consist of measurements and archival of these measurements in a relational database, but it is essential if CETUS is to have impact. Quantitative measurements are essential for deriving distributions of various parameters and for examining possible correlations with other parameters.

Consider the high-impact paper (2003 citations and counting) by Kauffmann et al. (2003) on active galactic nuclei (AGN) observed in the Sloan Digital Sky Survey: "We find that AGN of all luminosities reside almost exclusively in massive galaxies and have distributions of sizes, stellar surface mass densities and concentrations that are similar to those of ordinary early-type galaxies in our sample." To come to these conclusions, the authors must have made use of measurements of at least 20 parameters on 122,000 galaxies in SDSS images and spectra.

Detailed, robust measurements will be made on CETUS images and spectra. The measurements will be stored in a relational database in the CETUS science archive along with raw and processed observational data. Astronomers will access this database of measurements in order to answer scientific questions and make discoveries.

CETUS will draw from publicly available algorithms and software being developed for other surveys to derive software pipelines for CETUS data such as "The Hyper Suprime-Cam Software Pipeline" (Bosch et al. 2017), ProFound (Robotham et al. 2018) and ProFit (Robotham et al. 2017), Software Pipeline for CGM and IGM spectra (U. MD/Gatane 2018).

## 5.2 MISSION OPERATIONS

The basic concept underlying CETUS operations is that the three science instruments can operate together as described in **Section 2.5**. Joint operation of science instruments is made possible by the comprehensive optical-mechanical design of CETUS in which the three science have separate apertures at the telescope image plane (**Figure 3-4**). The middle optic of the Offner relay – a convex grating in the MOS, a convex mirror in the camera – has an attached tip/tilt/focus mechanism that enables dithering of the spectrogram (MOS) or image (camera) on the detector format while the telescope pointing remains steady. This innovation enables the three instruments to operate to their fullest extent: independently and simultaneously. Given a time on target of ~90% afforded by the orbit about L2, the total observing efficiency of CETUS 3x0.90= 2.7.

What distinguishes the prime instrument in any given observation from the resultant parallel instruments is that the prime instrument sets the telescope pointing on the sky, roll angle, and the time on target. As shown below, the prime instrument is related to the scientific campaign:

- in the massive spectrographic/imaging campaign at z~1-2, the near-UV MOS is prime;
- in the campaign on z~0 galaxies, their outskirts, and the CGM, the far-UV PSS spectrograph is usually prime;
- in UV observations for a multi-wavelength observing campaigns, the near-UV/far-UV camera is usually prime.

One possible exception to joint operation concerns transients requiring rapid response (on target within an hour of an event; ~45 min mission operation time to command; <15 min slew ane settle) such as catching the early phases of mergers of neutron-star binaries or catching supernovae on the rise. In this case, there is no time to check for bright sources in the field of view, so the near-UV camera will be the sole instrument, at least at the start. Another exception is the far-UV PSS spectrograph operating in long-slit mode. If the long slit is used to scan across a region (called push-broom mode), then it will operate alone. However, the information content of the resulting datacube ($x$, $y$, $\lambda$) more than compensates for any loss in "observing efficiency".

The objective of the CETUS Concept of Operations (CONOPS) is to enable CETUS to accomplish its observational goals from its design orbit and within its design mission lifetime. The allocation of observing time in the first four years of operation is given in **Table 2- 2**.

Table 5-1 shows a typical sequence of observations in a survey of galaxies at z~1involving the NUV MOS and FUV/NUV camera.

**Table 5-1. A typical cadence with the Camera and MOS will enable several hours of FUV/NUV multi-bandpass images and spectra of about 100 galaxies**

| | |
|---|---|
| 1. | Slew (<1 deg.) to next field |
| 2. | FGS Acquires Guide Stars |
| 3. | CETUS uses FGS to adjust pointing to place pre-selected MOS targets on their assigned MSA shutters |
| 4. | Assigned MSA shutters (~100) are opened and assigned camera filter is rotated into place |
| 5. | Camera and MOS in parallel take an exposure that lasts up to 30 min. The camera exposure time may differ from the MOS exposure time. |
| 6. | If desired, camera and MOS images may be dithered and a new camera filter may be rotated into place while detectors are read out |
| 7. | Steps 5 and 6 are repeated 20 times to get a total MOS integration time of ~10 hours which is needed for an adequate SNR. The camera, operating independently, may cycle through the 10 FUV/NUV filters/ |

**Table 5-2** provides the basis for an approximate upper bound on the amount of data that will be generated over a 24-hour period. It assumes that the maximum data volume is generated when using the MOS in parallel with the camera. Later consideration of operational inefficiencies and possible data compression algorithms is expected to reduce the data volume, but this upper limit will be used for the initial definition of CETUS downlink communi-cations requirements. Given a down-link data rate of 8.5 Mpbs, about 30 Gb of science data can be down-linked in ~1-hour to a ground station, which is an amount that provides a margin on the maximum amount of data that the detectors are expected generate daily.

**Table 5-2. A conservative estimate of the amount of science data generated in a day has been used to size storage and downlink rates**

| Description | Value |
|---|---|
| Number of MOS CCD Pixels in each Row and Column | 4096 |
| Bits per CCD pixel after A2D Conversion | 16 |
| Number of Bits produced per CCD readout | $2.68\times10^{8}$ |
| Time between CCD readouts/integration time (sec) | 1800 |
| Number of CCD readouts per 24 hours | 48 |
| Number of CCD bits per 24 hours | $1.29\times10^{10}$ |
| Number of CCD bits per 24 hours (Gb) | 12.88 |
| Increase due to operating camera in parallel | 2 |
| Total number of bits in 24 hours (Gb) | 26 |

One of the benefits of a SEL2 orbit is that it is close enough to Earth to potentially be able to use a NEN ground station, which is less costly than a DSN ground station. In the case of CETUS, the communications subsystem on the CETUS SC-Bus is sized to support this down-link data rate to a NEN ground station with an 18 m dish. Currently 18-m NEN ground stations are in White Sands and are also expected to be installed at Hartebeesthoek, South Africa.

In parallel with the daily science data downlink, the NEN ground station will also be able to up-link commands and collect CETUS ranging data. CETUS Flight Software (FSW) will combine this ranging data with information from the onboard Celestial-Navigation (CelNav) cameras, thus permitting CETUS to perform its own Orbit Determination (OD) and avoid the cost of additional GS tracking. The OD data will in turn be used to plan the orbit maintenance burns. These maneuvers are combined with momentum unloading burns and are performed once per week.

A final aspect of the CETUS CONOPS is the provision for receiving “any-time” alerts of transient ToOs. Because the CETUS HGA antenna is mounted on a two-axis gimbal, it can be pointed toward Earth at all times. The beam-width of the CETUS HGA is sufficient for it to receive at any time a fast-enough (>=100 bps) command uplink from any NEN ground station. There are enough NEN ground stations around to world for there to always be one in sight of CETUS. The CETUS attitude control system has a large enough reaction wheel assembly (RWA) to slew to any point in its FoR in less than 15 minutes. Thus, CETUS should begin observing a ToO within 1-2 hours of the approval of the TOO.

# 6. TECHNOLOGY DRIVERS AND ROADMAPS

The science drivers affecting technology requirements for CETUS are: 1) high system throughput through use of a microshutter array in the MOS, 2) reflective coatings for the OTA and FUV PSS extending down to 100 nm (Lyman UV) that are also uniform in phase and amplitude, and 3) detectors with high quantum efficiency in the UV and solar blind/ minimizing red leak. We have technology roadmaps for maturation of items 1 and 2; item 3 requirements are already met. We continue to actively monitor and drive relevant technology advancements for future incorporation if warranted by improved performance (while maintaining low risk).

The critical technologies for the CETUS OTA and its UV instruments are listed in **Table 6-1**. The NG-MSA is the key technology for further development, and that is proceeding at GSFC under a multi-year Strategic Astrophysics Technology (SAT) program. Should this not be matured in time, the TRL 8 JWST NIRSpec MSA will be utilized.

**Table 6-1. The pertinent UV optical technologies are currently mature or will be advanced to meet the CETUS schedule requirements**

| Technology | Heritage/Comments |
|---|---|
| Next-Generation Micro-Shutter Array (NG-MSA) | • The JWST NIRSpec MSA is space-qualified with a 365x172 array with 100 μm x 200 μm shutters. This MSA could be an off-ramp if NGMSA not TRL 5 at the start of Phase A.<br>• As part of an APRA program, a NG-MSA pilot 128x64 array has been constructed, so the current TRL is 3-4. A NASA/JHU sounding rocket experiment with this NG-MSA is planned for October 27, 2019.<br>• GSFC currently has a 3-year SAT grant for maturing the NGMSA to TRL 5. The SAT PI says that the program to develop a 840x420 array at TRL 5 is on track to be finished by 31 Dec 2021. CETUS can accommodate this larger array. |
| Mirror coatings of Al/LiF/ALD MgF2 providing high reflectivity from 100 nm to 1100 nm | • Laboratory research (Fleming et al. 2018) shows that Al/LiF overcoated with a very thin (1-2 nm) ALD coating of $MgF_2$ or $AlF_3$ is effective in protecting Al/LiF from humidity. Rocket payloads will be testing these mirror coatings in 2019-2020.<br>• Facilities exist at JPL for ALD coating of mirrors smaller than 0.5m. However, no such facility is known for the 1.5-m telescope primary. Collins Aerospace is planning a large coating facility which might be modified to apply ALD coating. We are working with JPL to identify other facilities. |
| Micro channel Plate (MCP) detector with high quantum efficiency in the FUV | • The CETUS far-UV MCP detectors made by U.C. Berkeley Space Science Lab are the same technology as flown on the Hubble COS spectrograph.<br>• CETUS FUV Camera MCP uses CsI photocathode (~50x50 mm) with $MgF_2$ window – TRL 6+<br>• CETUS PSS FUV MCP uses CsI photocathode (200x70 mm) windowless. This is just a scale-up requirement.A 200x200 mm MCP has recently flown on a University of Colorado rocket experiment. Sounding rocket programs have and continue to provide verification of comparable MCPs – CU's DEUCE (2017, 2018) and NASA/JHU's rocket payload planned for launch in October 2019. |

## 6.1 NEXT-GENERATION MICRO-SHUTTER ARRAY (NG-MSA)

The Micro-shutter Array (MSA) planned for MOS is based on the prior work done by Goddard Space Flight Center for the JWST NIRSpec instrument (Li, et al., 2017). However, the implementation has evolved as the Goddard team of scientists and engineers have continued to develop the micro-shutter array technology (Kutyrev, et al., 2019). The size of each individual shutter is 100 µm x 200 µm. Compared to the MSA in JWST's NIRSpec spectrograph, the NG-MSA will have 100 times faster actuation of the shutters, greater robustness with fewer mechanical failures, and a lower operational duty cycle on individual shutters.

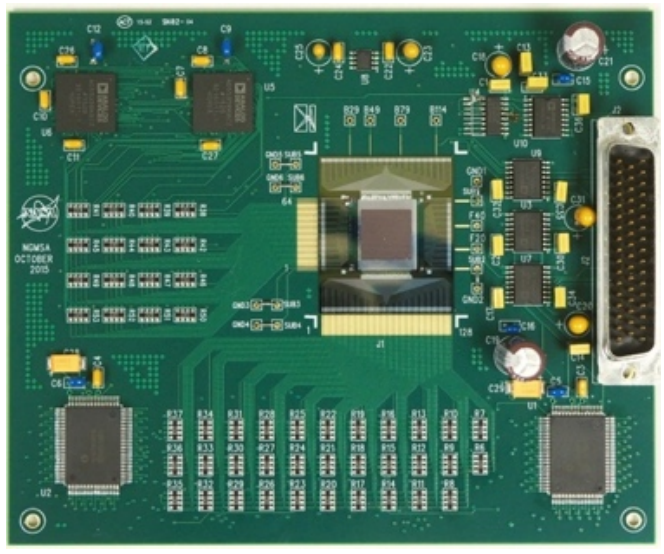

**Figure 6-1. An integrated NG-MSA will be procured from GSFC. A pilot design array with 128x64 shutters on a PC board with integrated drivers has been fabricated.**

NG-MSAs are electrostatically actuated MEMS devices that can be used as multi-object selectors and filters in spaceborne applications. After the group at GSFC successfully built micro-shutter assemblies for JWST, they began working on the NGMSA development (starting in 2009) to incorporate operational simplifications and higher reliability. The NG-MSA has a simplified, scalable design and is fully electrically actuated (unlike the JWST micro-shutters which require magnetic actuation). In addition, only the shutters to be opened have to be actuated so their duty cycle is lower than the earlier generation MSAs which significantly reduces the number of actuations per shutter, thus extending the lifetime and reducing the risk of failure.

GSFC has demonstrated 128x64 full array actuation and 2-D addressing on a fully integrated NG-MSA assembly, which brings the technology to TRL 4. **Figure 6-1** depicts an MSA packaged assembly. The CETUS UV MOS will make use of the 420x840 NG-MSA with a shutter size of 100µm x 200 µm now being developed although only a part of it (380x190) will be used. The SAT program on the NG-MSA is on track to produce a 420x840 NGSA at TRL by December 2021.

## 6.2 OPTICAL COATINGS, STRAYLIGHT, AND CONTAMINATION CONTROL

The dielectric optical coatings for the MOS optics will be designed to reflect the NUV and transmit the longer wavelengths. The design is based on coatings previously demonstrated by Materion (MacKenty, 2016) on small samples. The cut-off wavelength will be optimized for 350 nm rather than the 320 nm shown in **Figure 6-2**. The long wave cutoff can be shifted as desired. Similarly, the dielectric optical coatings for the Camera Offner optics will be designed to reflect both the FUV and NUV bands and transmit the longer wavelengths.

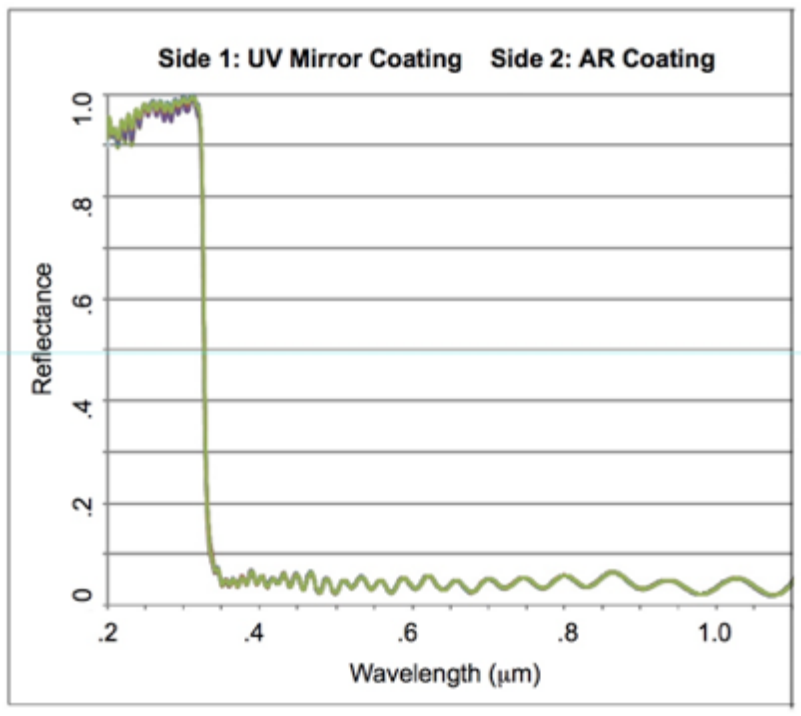

**Figure 6-2. The coatings are designed to sharply reduce reflectance of light longer than 350nm. [Reflectance data from MacKenty, J., 2016]**

Exceptional stray light control of out-of-band radiation will be accomplished by directing the light to a highly absorptive light trap comprised of carbon nanotubes (Hagopian, et al., 2010; Butler, et al., 2010). Carbon nanotubes are extremely dark due to their electronic structure and vanishingly low density. Carbon in its bulk form has an index as high as 4 resulting in a strong reflectance. By tailoring the nanotube growth process, the fill factor of carbon can be reduced to less than one percent creating an effective index approaching unity. Therefore, the light sees no impedance mismatch. A carbon nanotube forest only 25 microns tall will strongly absorb light from the UV (only 0.87% to 0.52% reflectance) to the far infrared (<0.2% reflectance).

**Figure 6-3** shows the hemispherical reflectance of carbon nanotubes grown on a metal substrate; compared to the ~ 4% reflectance of Z306 paint often used by NASA for stray light control. Hemispherical reflectance quantifies the total amount of light that is scattered by light striking a sample that is collected over a hemisphere. Measurements of the materials bi-directional reflectance function (BRDF) also indicate very low reflectance at glancing angles making this an ideal material for a light trap. The implementation of the light trap behind the dichroic beamsplitter in the MOS path will greatly reduce out-of-band stray light in the system. Depending on the final design of the system, additional measures may be required to reduce stray light from the structure that holds the dichroic or within the Offner relay assembly.

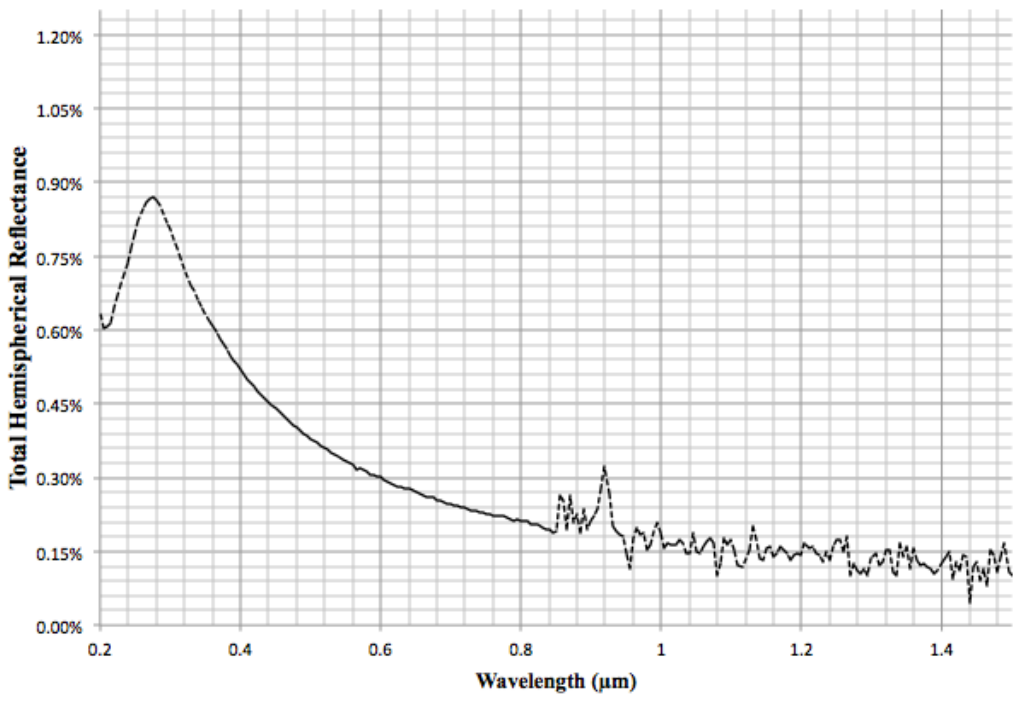

**Figure 6-3. A nanotube light trap behind the beamsplitter will greatly reduce out-of-band scatter**

The Point/Source Spectrograph (PSS) is designed for observations at wavelengths as short as 100 nm. The Hubble Al/$MgF_2$ coating would satisfy a 115-1100 nm requirement, and GSFC has pioneered hot deposition processes to increase UV reflectivity. However, to extend the range down to 100 nm, Al/LiF-coated mirrors are needed for the FUV PSS and telescope. To preclude reflectance degradation from water absorption, the LiF needs to be kept in a vacuum or inert atmosphere or preferably protected by another coating. Recent results from laboratory experiments (Fleming et al. 2018) show that a thin (1-2 nm) overcoat of $MgF_2$ or $AlF_3$ can protect LiF from humidity, but such a thin coating calls for atomic layer deposition (ALD). JPL can and has applied ALD coatings on mirrors up to 0.5m in diameter for rocket payloads. We are in discussions with JPL about ALD coatings for the CETUS primary mirror. Also, Collins has plans for a new, large coating chamber which might be modified to apply the ALD coating to the CETUS primary mirror.

Contamination control is a concern in any program, but of critical importance for UV instruments. Molecular and particulate contamination on the optics would impact UV throughput and cause additional scatter. Therefore, contamination control during the entire fabrication, assembly, and integration process will be essential to minimize molecular contamination of the optics. Careful selection of materials and outgassing thermal cycles are planned. Heaters will be incorporated near optical surfaces to allow on-orbit reduction of volatile condensable materials. Also, the telescope has a re-usable protective cover that will be closed during launch and any sensitive maneuvers where contamination might be generated.

## 6.3 CCD DETECTOR

We make use of identical 4K x 4K CCDs with 12-μm pixels for all three NUV instruments on CETUS: the NUV MOS and the NUV channels of the camera and point/slit spectrograph. The e2v CCD 272-84, which is the Euclid sensor whose window is coated for UV sensitivity, is the baseline. This sensor has been qualified for launch. A shutter in front of the CCD controls the exposure time. **Figure 3-9** shows the expected quantum efficiency over the 180-400 nm spectral band.

## 6.4 MCP DETECTOR

Micro-Channel Plate (MCP) detectors are baselined as the sensor for the FUV Camera and FUV PSS. The detectors are made by U.C. Berkeley Space Science Lab (SSL) and use the same basic technology as flown on the Hubble Cosmic Origins Spectrograph. Funded by NASA's Strategic Astrophysics Technologies (SAT) program, SSL has made several, major improvements in its far-UV detectors that enable the CETUS FUV Camera to meet its science performance requirements. (Note that the same MCP detector is also used for the Point/Slit Spectrograph, however, without a window.)

The image in the FUV mode is sensed by a 50mm x 50mm sealed CsI photocathode solar-blind MCP detector with 20-μm effective resolution element (resel). The photocathode is housed in a vacuum enclosure with a MgF2 window. The detector operates at room temperature. The effective resolution element width of the 20-micron resels as images at the TMA focal surface is 550 mas in field.

Sounding rocket programs have and continue to provide verification of comparable MCPs. The University of Colorado has flown DEUCE missions in 2017 and 2018 and incorporated a 200 x 200 mm MCP. NASA/JHU has a planned July 2019 sounding rocket flight on 36.352 UG.

For the FUV detectors, in addition to the MCP technology, we are following the progress of the joint JPL/e2v development of 4Kx4K electron-multiplying CCDs (EMCCDs) for WFIRST. The devices will eventually be delta doped and have a sensitivity down to 100 nm or lower which would make them candidates for the CETUS FUV detectors with potentially higher UV throughput than existing MCPs.

# 7. ORGANIZATION, PARTNERSHIPS, AND CURRENT STATUS

**Figure 7-1** (left) shows the current CETUS organization with identification of partners who have helped to formulate essential mission parameters and derive engineering requirements, system architecture, and detailed designs as well as provide cost inputs. The right portion of **Figure 7-1** shows the planned post-launch organization of CETUS. The Core Team that developed the CETUS science goals, detailed design, and implementation approach will continue to support the realization of the CETUS mission.

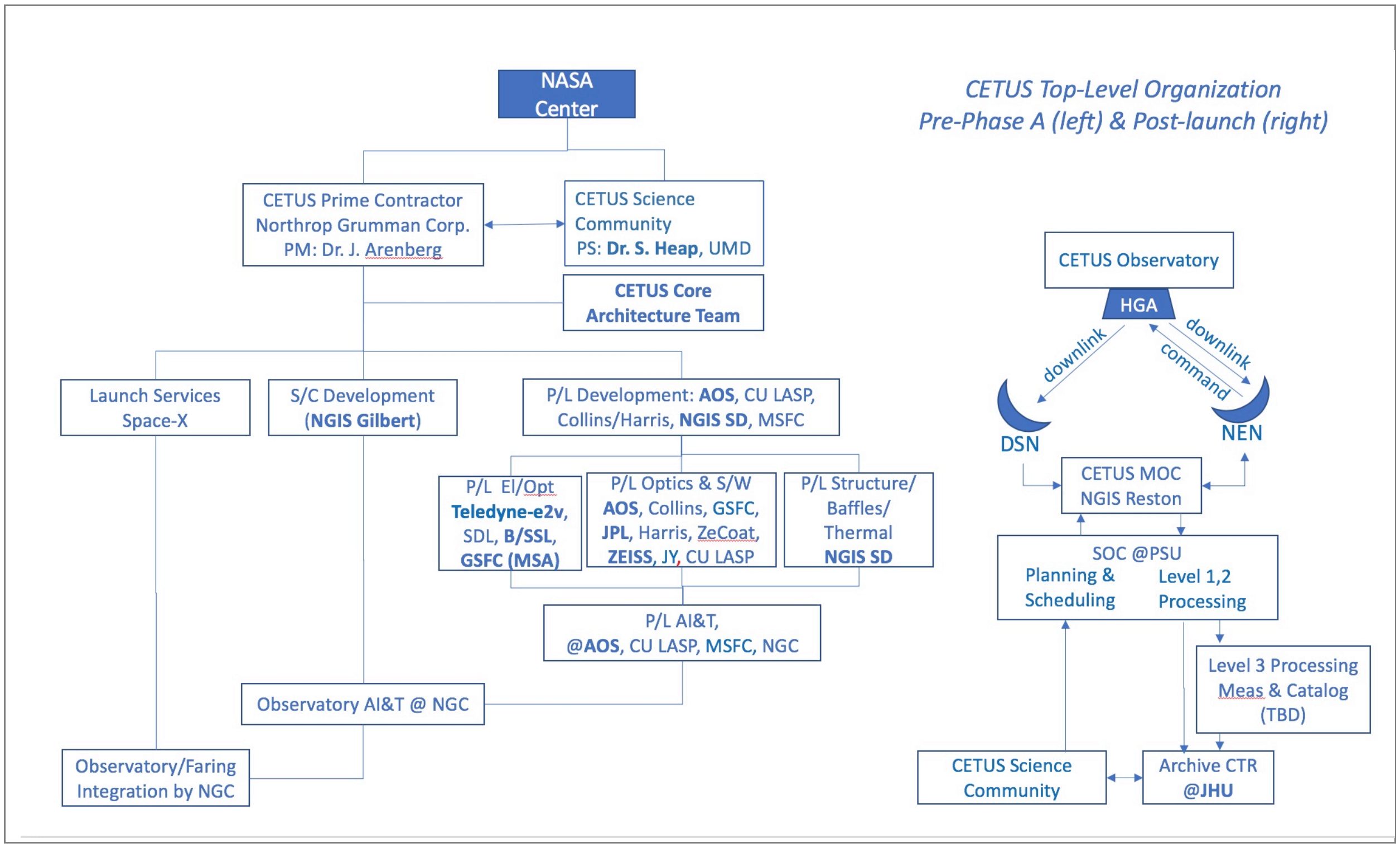

**7-1. The CETUS organization provides science, engineering, and management expertise to achieve successful CETUS development and mission operations. Those persons or institution in bold font were part of the original proposal to study the CETUS mission concept.**

The original (core) CETUS team that proposed the study and developed the design presented here are Heap, Hull, Kendrick, and Woodruff. They will remain engaged as well as our industrial/academic partners. Although not specifically shown in the figure, NASA centers will also contribute to CETUS: Marshall with its expertise in telescopes, Goddard with its expertise in highly UV-reflective mirror coatings and micro-shutter array, and JPL with its expertise in delta-doping CCD detectors for UV sensitivity and in ALD coatings of mirrors.

Over the past 40 years, UK and European institutions have worked closely with NASA on developing and observing on UV astrophysics missions such as the International Ultraviolet Explorer (IUE) and the Hubble Space Telescope (HST) in partnership with NASA. There is still keen European interest in UV astrophysics today. Of the UV-related science white papers submitted to Astro2020, about a third were from Europeans. In addition, Neiner (WP #244) has submitted a proposal to CNES to study adding a polarimeter to the CETUS NUV spectrograph. In the coming months, we plan to explore partnerships on CETUS with European institutions.

**Figure 7-2** shows the projected schedule for development of the CETUS mission. It assumes a Phase A start in October 2023 and a launch in August 2029. Operations are planned for a 5-year mission, but consumables are planned to allow a 10-year mission. Continued technology development of areas discussed in Section 2.2 prior to 2023 are fundamental.

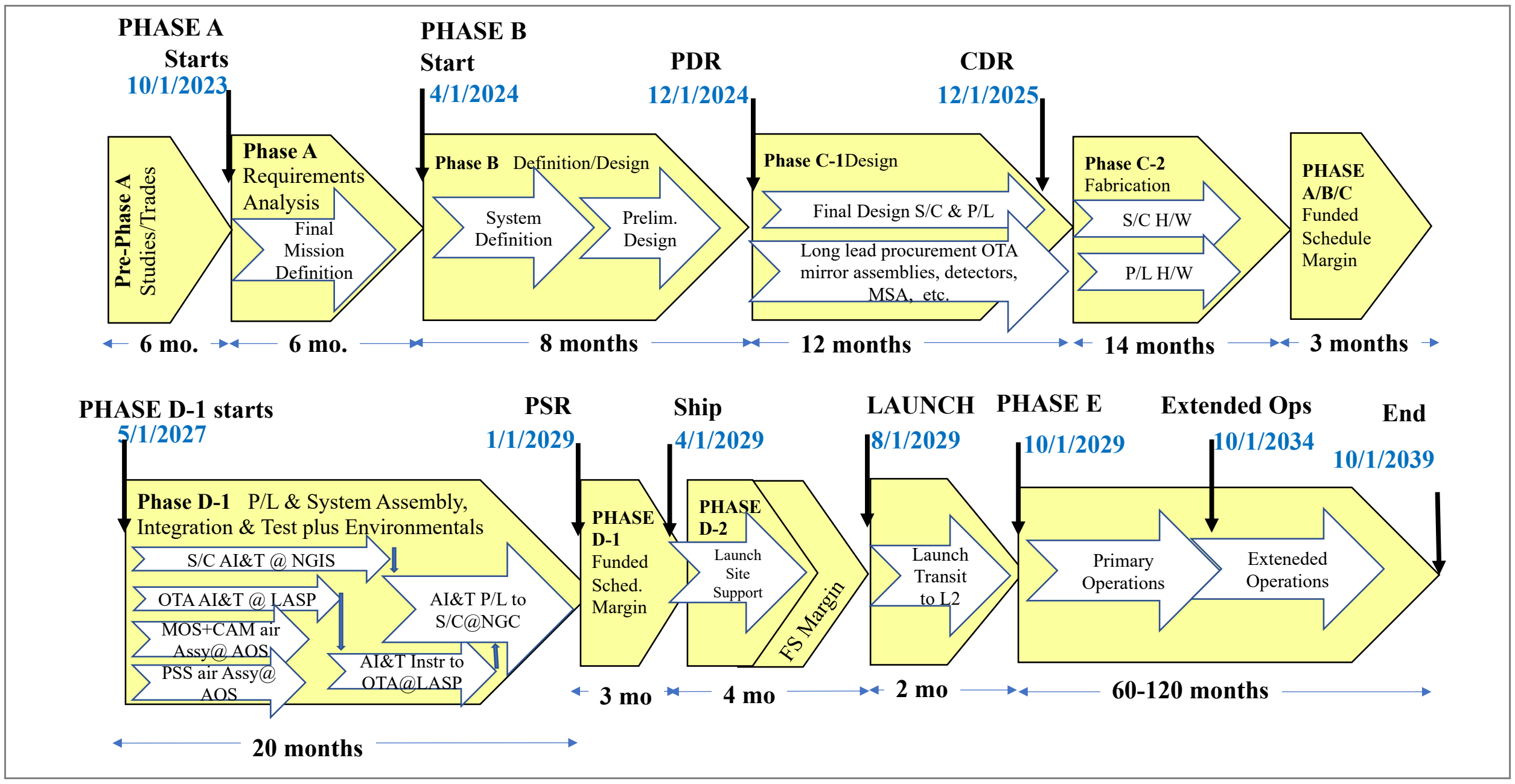

**Figure 7-2. The CETUS schedule defines a realistic mission timeline for the 1.5-m telescope, scientific instruments, spacecraft, and AI&T at high-cleanliness UV-compatible facilities. It recognizes the efficiencies of expert and experienced industrial and academic facilities and allows funded schedule margin.**

After our industrial and university partners vetted this schedule, we used this schedule, industry-norm labor rates and FTE loading to generate a cost estimate for CETUS as described in Section 7.

# 8. COST ESTIMATION

CETUS costs have been derived from industry/university input following significant design effort. High-heritage, high TRL components have been used throughout. Industrial partners were selected with the approach "go to the experts," and each has well-established expertise in their area of engagement. **Table 8-1** includes all costs that will be funded by NASA. Direct funding to the science community is described in WBS 4 for both pre-launch (Phase A-D) and post-launch (Phase E) periods.

**Table 8-1. The cost of a CETUS 5-year mission is within the range of a Probe-class**

| WBS # | Phase | Cost Estimate | Notes |
|---|---|---|---|
| | **Phase A-D** | | |
| 1.0-3.0 | Management, SE, MA | $60 M | |
| 4.0 | Science Preparation | $8 M | Includes: monitoring CETUS hardware development; building/modifying s/w for levels 1,2,3 science data processing & s/w for measurement & on-line catalogs; participation in pre-launch test and calibration |
| 5.0 | Payload (Instruments, Telescope) | $395 M | Based on industrial & institutional input from NGIS (Gilbert), NGIS (San Diego), Collins Aerospace, Harris Aerospace, SCHOTT, LASP (CU), Teledyne-e2v, JPL, GSFC, AOS, NGC. Multiple telescope cost models were used to derive the telescope cost estimate. |
| 6.0 | Spacecraft | $164 M | NGIS (Gilbert) based on significant TESS similarities and TRL 7-9 hardware |
| 10.0 | Observatory I&T(ATLO) | $20 M | NGIS (Gilbert) |
| | 30% Reserve Phase A-D | $194 M | |
| **WBS #** | **Phase E** | | |
| 1.0 - 3.0 | Management, SE, MA | $2.4 M | |
| 4.0 | Science | $30.0 M | Includes: planning & scheduling, post-observation data processing at Penn State Univ; archival & analysis center at JHU; measurements & catalogs by TBD |
| 7.0 | Mission Operations | $15.0 M | NGIS (Gilbert); 5 yrs mission baseline; consumables for 10 yrs |
| 9.0 | Ground Data Systems | $2.5 M | |
| | 15% Reserve Phase E | $7.5 M | |
| | | | |
| | Subtotal before Launch Vehicle | *$898 M* | |
| | Launch Vehicle/ Launch Services | $110 M | Space X Falcon 9 baselined in initial study, but to gain mass margin, we adopt the Falcon 9 Heavy ($90M plus $20M for launch services) https://www.spacex.com/about/capabilities |
| | 15% Reserve on Launch Vehicle | $16.5 M | |
| | **CETUS Total Cost** | ***$1,025M*** | |

# 9. CONCLUSIONS

Loss of the highly successful Hubble Space Telescope mission will eliminate scientific access to a U.S. large-aperture, UV-sensitive space telescope. The Cosmic Evolution Through UV Surveys (CETUS) mission concept, if approved and implemented, will significantly fill this void. The CETUS concept features a 1.5-m aperture diameter, large field-of-view (FoV) telescope optimized for UV imaging and spectroscopy with three science instruments that are sensitive to radiation in the 100 to 400 nm spectral region. The CETUS design enables simultaneous wide-field FUV/NUV imaging, wide-field NUV multi-object spectroscopy, and LUV/NUV/FUV point source spectroscopy. The CETUS design is responsive to science requirements from the astronomical community, and the implementation makes use of high-heritage hardware experience from industry and academia. Detailed error budgets have been established and the performance assessed relative to those budgets. Ease of manufacturing, alignment, and testing has been considered in the design and packaging and interfaces with the spacecraft bus and launch vehicle.

# 10. REFERENCES TO RELEVANT ASTRO2020 SCIENCE WHITE PAPERS

| # | Principal Author | Title |
|---|---|---|
| 605 | *Bonifacio | Extremely metal-poor stars: the need for UV spectra |
| 591 | Burchett | Ultraviolet Perspectives on Diffuse Gas in the Largest Cosmic Structures |
| 63 | *Chaufrau | UV Exploration of the solar system |
| 366 | Chen | Tracking the Baryon Cycle in Emission and in Absorption |
| 90 | Clarke | Solar System Science with a Space-based UV Telescope |
| 329 | Foley | Gravity and Light: Combining Gravitational Wave and Electromagnetic |
| 604 | *Garcia | Walking along cosmic history: Metal-poor massive stars |
| 607 | Grindlay | Big Science with a NUV-MidIR Rapid-Response 1.3m Telescope at L2 |
| 281 | *Gry | Far- to near-UV spectroscopy of the interstellar medium at very high |
| 593 | Hagen | Spatially Resolved Observations of the Ultraviolet Attenuation Curve |
| 276 | Hodges-Kluck | How Does Dust Escape From Galaxies? |
| 177 | *Lebouteiller | ISM and CGM in external galaxies |
| 524 | Lehner | Following the Metals in the Intergalactic and Circumgalactic Medium |
| 180 | *Marin | The role of AGN in galaxy evolution: Insights from space ultraviolet spect |
| 565 | Martin | IGM and CGM Emission Mapping: A New Window on Galaxy and |
| 592 | McCandliss | Lyman continuum observations across cosmic time: recent developments, |
| 342 | Metzger | Kilonovae: NUV/Optical/IR Counterparts of Neutron Star Binary Mergers |
| 244 | *Neiner | Stellar physics with high-resolution spectropolarimetry |
| 309 | Oppenheimer | Imprint of Drivers of Galaxy Formation in the Circumgalactic Medium |
| 450 | Pellegrini | Making the connection between feedback and spatially resolved em-line |
| 409 | Peeples | Understanding the circumgalactic medium is critical for understanding |
| 628 | Pisano | Completing the hydrogen census in the CGM at z~1 |
| 193 | *Rahmani | Quasar absorption lines as astrophysical probes of fundamental physics… |
| 60 | Roederer | The Potential of Ultraviolet Spectroscopy to Open New Frontiers to Study |
| 152 | Roederer | The astrophysical r-process and the origin of the heaviest elements |
| 183 | Roederer | The First Stars and the Origin of the Elements |
| 421 | Tumlinson | The Baryon Cycle, Resolved: A New Discovery Space for UV Spectros- |

*European Principal Author

# 11. REFERENCES

Abbott, B. P. et al., 2017, PhRevL 118, 1101A
Abohalima & Frebel 2018, ApJS 238, 36
Anderson & Sunyaev 2018, A&A 617, 123
Baes, M. et al. 2011, ApJS 196, 22
Baes, M. et al. 2019, MNRAS 484, 4069
Bianchi, L. 2011, Astrophys. Space Sci. 335, 51 (Fig. 3)
Bottema, M . and Woodruff, R., 1971, Appl . Opt., 10, No . 2, 300
Bowen, D. 2018, https://www.astro.princeton.edu/~dvb/pairshome.html
Brauneck, U., Woodruff, R., Heap, S., et al., 2018, Proc . SPIE 10699-120
Butler, J., et al., 2010, Proc. SPIE 7862
Brown, Kochanak, et al., 2018, MNRAS 473, 1130
Cenko, Cucchiara, et al., 2016, ApJ 818, 32
Cochrane, R., Hayward, C. et al. 2019, MNRAS 488, 1779
Des Marais, Harwit, et al., 2002, AsBio 2, 153
Domagal-Goldman, 2015, AAS 22540707
Driver, S., Wright, A. et al. 2016, MNRAS 455, 3911
Ertley, S., Siegmund, O., et al., 2015, Proc. SPIE 96010S
Fleming, B., Quijada, M., et al., 2017, App. Optics, Vol. 56, Issue 36, 9941
Hagopian, J., et al., 2010, Proc. SPIE 7761,
Hayes, M. et al., 2013, ApJ 765, 27
Hayes, M. et al., 2016, ApJ 828, 49
Hayes, M., Östlin, G. et al., 2014, ApJ 782, 6
Harwit, M. 2003, Ast. Soc. Pacific Conf. 291, 9
Heap, S., Danchi, W., et al., 2017, Proc. SPIE 10398
Heap, S., et al., 2016, Proc. SPIE 9905, 990505
Heap, S., Hull, A., et al., 2017, Proc. SPIE 10398
Hodges-Kluck, E. 2016, ApJ 833, 58
Hull, A., Heap, S., et al., 2018, Proc. SPIE 10699
Jouvel, Kneib, et al., 2011, A&A 532, 25
Keeney, Stocke, et al., 2017, ApJS 230, 6
Kendrick, S., Woodruff, R, et al., 2019, AAS 157 .37
Kendrick, S, Woodruff, R., et al., 2018, Proc. SPIE 10699
Kendrick, S, Woodruff, R., et al., 2018, AAS 140 .08
Kendrick, S., Woodruff, R., et al., 2017, Proc. SPIE 10401
Korsch, D., 1991, Academic Press, Inc . San Diego, CA, p. 217-218
Laigle, McCracken, et al., 2016, ApJS 224, 24
Lehnert, Heckman et al., 1999, ApJ 523, 575
Loyd, France, Youngblood, 2018, ApJ 867, 71
Lloyd & France, 2014, ApJS 211, 9
LUVOIR Interim Report, 2018, arXiv:1809 .09668
MacKenty, J., 2016, SPIE 990533
Madau, P. & Dickinson, M., 2014, ARA&A 52, 415
Maréchal, A., 1947, Rev. d'Optique, 26, 257
Martin, C., Dijkstra,M. et al. 2015, ApJ 803, 6

McCandliss, S., et al., 2016, AJ 152, 65
Ménard, B., 2010, MNRAS 405, 1025
Metzger, P. et al. 2015, MNRAS 446, 1115
Meurer, G., Heckman, T., Calzetti, D. 1999, ApJ 521, 64
*Milliquas Catalog*, 2019, heasarc.gsfc.nasa.gov/W3Browse/all/milliquas.html
Moffett, A., et al., 2016, MNRAS 462, 4336
*New Worlds New Horizons* (NWNH), 2010, Nat. Academy Press, Washington D.C.
*NWNH-Panel Reports*, 2010, Nat. Academy Press, Washington D.C.
Nakar & Sari, 2010, ApJ 725, 904
Östlin, Hayes et al., 2014, ApJ 797,11
Pettini, Steidel, Adelberger, 2000, ApJ 528, 96
*Powering Science*, 2017, Nat. Acad. Press., Washington D.C.
Purves, L., 2017, Proc. SPIE 10401
Quijada, M., Del Hoyo, J., Rice, S., 2014, Proc. SPIE 9144, 91444G
Quijada, M. 2019 SPIE conference paper on hot deposition
Robotham, A. et al. 2018, MNRAS 476, 3137 (PROFOUND)
Robotham, A. et al. 2017, MNRAS 466, 1513 (PROFIT)
Roederer, Mateo, Bailey, 2016, AJ 151, 82
Segura, Walkowicz, Meadows, 2010. AsBio 10, 751
Siegmund et al., 2015, Proc. AMOS Conf.
Sullivan, Winn, et al., 2015, ApJ 809, 77
Terrazas, Bell, et al., 2017, ApJ 844, 170
Terrazas, 2019 AAS 2331060
Tian, France, et al., 2014, E&PSL 385, 22
Tumlinson, Peeples, Werk, 2017, ARA&A 55, 389
UVOT, https://www.swift.psu.edu
Vallerga, et al., 2014, Proc. SPIE 91443J
Verhamme, 2018, in *Escape of Lyman Rad. From Galactic Labyrinths*, Sep 11-14, 2018
Wisotzki, Bacon, et al., 2018, Nature 562, 229
Woodruff, R. et al., 2019, JATIS 18084
Woodruff, R. et al., 2019, AAS 443.02
Woodruff, R. et al., 2018, AAS 140 .15
Woodruff, R. et al., 2017, Proc. SPIE 10401

## 12. ACRONYM LIST

| | |
|---|---|
| 4MOST | 4-meter Multi-Object Spectrograph Telescope |
| ACS | Advanced Camera for Surveys |
| ACS | Attitude Control Subsystem |
| AGN | Active Galactic Nucleus |
| AI&T | Alignment, Integration, and Test |
| ALD | Atomic Layer Deposition |
| AOS | Arizona Optical Systems |
| ATLO | Assembly, Test, and Launch Operations |
| BCCG | Brightest Cluster Galaxies |
| BSG | Blue Supergiant |
| CCD | Charged-Coupled Devices |
| CETUS | Cosmic Evolution Through UV Spectroscopy |
| C&DH | Command & Data Handling |
| CDM | Cold Dark Matter |
| CFRP | Carbon Fiber Reinforced Polymer |
| CGH | Computer-generated Hologram |
| CGM | Circumgalactic Medium |
| CIM | Calibration Insert Mirror |
| CMM | Coordinate Measuring Machine |
| CNES | Centre national d'etudes spatiales |
| Conops | Concept of Operations |
| COS | Cosmic Origins Spectrograph |
| CTE | Coefficient of Thermal Expansion |
| DDF | Deep Drilling Field |
| DE | Dynamic Envelope |
| DoF | Degree of Freedom |
| DSN | Deep Space Network |
| EM | Electromagnetic |
| EMCCD | Electron multiplying charged couple detector |
| EOL | End of Life |
| EPS | Electrical Power System |
| EUV | Extreme Ultraviolet |
| FEE | Front End Electronics |
| FGS | Fine Guidance Sensor |
| FoR | Field of Regard |
| FoV | Field of View |
| FPA | Focal Plane Assembly |
| FSW | Flight Software |
| FUSE | Far UV Spectroscopic Explorer |
| FUV | Far Ultraviolet |
| GALEX | Galaxy Evolution Explorer |
| GEO | Geosynchronous Orbit |
| GHRS | Goddard High Resolution Spectrograph |
| GIANT | Goddard Image Analysis Navigation Tool |
| GO | General Observer |
| GRB | Gamma-ray Burst |
| GSFC | Goddard Space Flight Center |
| GW | Gravitational Wave |
| HDST | High-Definition Space Telescope |
| HGA | High Gain Antenna |
| HSF | High Spatial Frequency |
| HST | Hubble Space Telescope |
| HZ | Habitable Zone |
| ICD | Interface Control Document |
| IEM | Integrated Electronics Module |
| IGM | Inter-Galactic Medium |
| IR | Infrared |
| ISM | Inter-Stellar Medium |
| IRSO | Indian Space Research Organization |
| IUE | International Ultraviolet Explorer |
| JATIS | Journal of Astronomical Telescopes, Instruments, and Systems |
| JHU | Johns Hopkins University |
| JWST | James Webb Space Telescope |
| km | kilometer |
| LAE | Lyman alpha emitting |
| LAMOST | Large Sky Area Multi-Object Fibre Spectroscopic Telescope |
| LEO | Low Earth Orbit |
| LGA | Low Gain Antenna |
| LMT | Large Millimeter Telescope |
| LOS | Line of Sight |
| LSF | Line Spread Function |
| LSF | Low Spatial Frequency |
| LSST | Large Synoptic Survey Telescope |
| LUV | Long Ultraviolet |
| LUVOIR | Large UV/Optical/IR Surveyor |
| LV | Launch Vehicle |
| mas | milli-arcsecond |
| MCP | Micro-Channel Plates |
| MDL | Mission Design Lab |
| MEB | Main Electronics Box |
| mm | millimeter |
| MOS | Multi-object Spectrograph |
| MSA | Micro Shutter Array |
| MSF | Mid Spatial Frequency |
| NASA | National Aeronautics and Space Administration |
| NEN | Near Earth Network |
| NG MSA | Next Generation Micro Shutter Array |
| nm | nanometers |
| NIR | Near Infrared |
| NGIS | Northrop Grumman Innovation Systems |
| NUV | Near Ultraviolet |
| NWNH | New World New Horizons |
| OAO | Orbiting Astronomical Observatory |
| OTA | Optical Telescope Assembly |
| PAF | Payload Adapter Fitting |
| PDU | Power Distribution Unit |
| PFS | Prime Focus Spectrograph |
| PI | Principal Investigator |
| PIE | Payload Interface Electronics |
| PL | Payload |
| PM | Primary Mirror |
| PSS | Point/Slit Spectrometer |
| QSO | Quasi-stellar Object |
| OSSA | Outer Sun Shield Assembly |
| RSG | Red Supergiant |
| RTOS | Real Time Operating System |
| RWA | Reaction Wheel Assembly |
| SA | Solar Array |
| SAT | Strategic Astrophysics Technologies |
| SC | Spacecraft |
| SDSS | Sloan Digital Sky Survey |
| SED | Spectral Energy Distribution |
| SI | Scientific Instrument |
| SKA | Square Kilometer Array |
| SM | Secondary Mirror |
| SSL | Space Science Labs |
| STIS | Space Telescope Imaging Spectrograph |
| SUIT | Solar UV Imaging Telescope |
| SW | Software |
| TBD | To Be Determined |
| TBR | To Be Resolved |
| TDE | Tidal Disruption Event |
| TEC | Thermoelectric Cooler |
| TESS | Transiting Exoplanet Survey Satellite |
| TM | Tertiary Mirror |
| TMA | Three Mirror Anasigmatic |
| TMT | Thirty Mirror Telescope |
| ToO | Targets of Opportunity |
| TRL | Technology Readiness Level |
| UV | Ultraviolet |
| UVIT | Ultraviolet Imaging Telescope |
| WCS | Wavelength Calibration System |
| WFIRST | Wide-Field Infrared Survey Telescope |
| WFOV | Wide Field of View |
| WFS | Wavefront Sensor |
| WHIM | Warm Hot Interstellar Medium |
| WMAP | Wilkinson Microwave Anisotropy Probe |
| WR | Wolf-Rayet |

# 13. RESUMÉS

## Sara Heap, CETUS Study Scientist

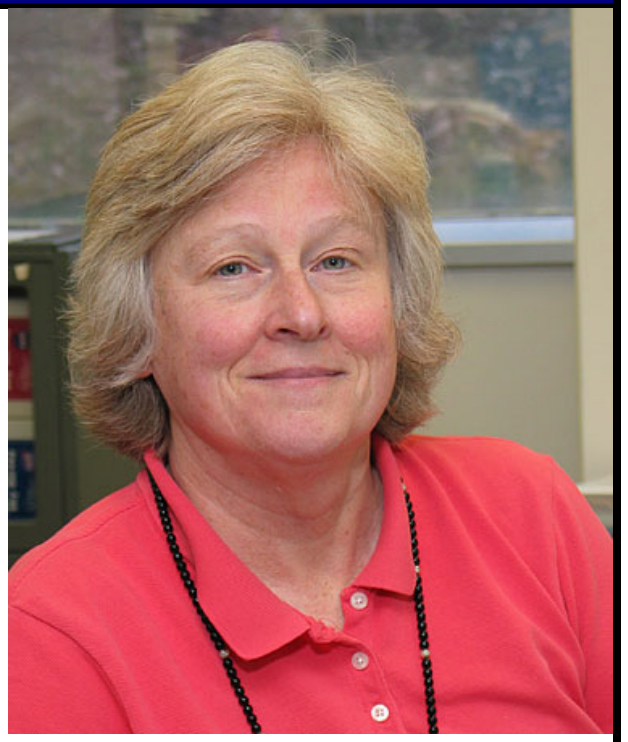

**Education:** BA, Wellesley College (1964); Ph. D., UCLA (1970)

**Professional Experience:**

Adjunct Professor of Astrophysics, University of Maryland (2018-)

Astrophysicist, NASA/GSFC (1969-2015); Emeritus Scientist (2015-)

**Relevant Experience**

International Ultraviolet Explorer (1972-1986)

- Participated in development, integration. and test of IUE
- Led the IUE scientific commissioning for hot stars
- Founded and directed the IUE regional data-analysis facility

Hubble's Goddard High-Resolution Spectrograph (GHRS) (1976-1997) (Heap=Co-PI)

- Led successful proposal team for GHRS
- Worked as scientific lead in development of GHRS (prime contractor, Ball Aerospace)
- Led development for GHRS science data management software

Hubble's Space Telescope Imaging Spectrograph, STIS (1985- ) (Heap = Co-I)

- Led development of STIS science data management software
- Led two STIS Key Projects: coronography of beta Pic; He II Gunn-Peterson absorption

Hubble's Cosmic Origins Spectrograph (COS; PI=Jim Green) as co-I (1997- )

Galaxy Evolution Spectroscopic Explorer (GESE) mission concept development (2012 - )

**Selected Publications**

Heap, S. et al., "IUE observations of hot stars", Nature 275, 385 (1978)

Heap, S. et al., "First results from the Goddard High-Resolution Spectrograph – Spectroscopic determination of stellar parameters of Melnick 42", ApJ 377, L29 (1991)

Heap, S. et al., "The GHRS: in-orbit performance", PASP 107, 871 (1995)

De Koter, A, Heap, S. et al., "On the evolutionary phase and mass loss of the Wolf-Rayet-like stars in R136a", ApJ 477, 792 (1997)

Heap, S. et al., "STIS observations of He II Gunn-Peterson absorption toward Q0302-003", ApJ 534, 69 (2000)

Heap, S., Lanz, T., Hubeny, I., "Fundamental properties of O-type stars", ApJ 638, 409 (2006)

Lebouteiller, V., Heap, S. et al., "Chemical enrichment and physical conditions in I Zw 18", A&A 553, 1

Lebouteiller, V. et al., "Neutral gas heating by X-rays in primitive galaxies", A&A 602, 45 (2017)

Heap, S., Hubeny, I., Lanz, T., "Stars and Stellar Black Holes in the Low-Metallicity Galaxy, I Zw 18, ASPC 519, 267 (2019)

**Jonathan W. Arenberg**
**Chief Engineer, Space Science Missions**
**Northrop Grumman Aerospace Systems, Redondo Beach CA**
**jon.arenberg@ngc.com**

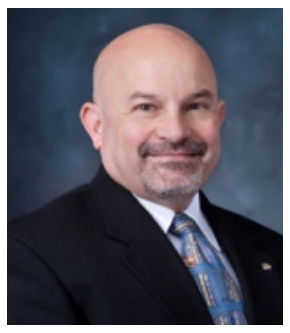

**Current and Previous Positions:**

1989 – present: Northrop Grumman Aerospace Systems
1982-1989: Hughes Aircraft Company, Electro-optics & Data Systems Group

**Relevant Experience :**

Dr. Arenberg has over 29 years of experience working on astronomical programs such as the Chandra X-ray Observatory, development of the starshade, NASA's James Webb Space Telescope and other mission concepts. He held several positions on Webb, system design leader, systems engineering manager and finally chief engineer. While on JWST he led efforts to model contamination and understand its impact on system performance. His current role involves mission development and analysis and technology development. He is conducting research into advanced methods of design to cost and schedule and their optimization.

In addition to his work on astronomical systems, he has contributed to major high-energy and tactical laser systems, laser component engineering, metrology, optical inspection and technology development projects.

Dr Arenberg is recognized for his systems approach to complex problems. He is the co-author of a recent SPIE book on systems engineering and teaches classes on the subject with his co-author.

**PATENTS**

13 US and European Patents

**PROFESSIONAL AWARDS (Selected)**

NASA Technology Development Award, TRW Chairman's Award for Innovation 1991,4 NASA Group Achievement Awards (Chandra X-ray Observatory), 3 NASA Group Achievement Awards (JWST), Goddard Group Award (2006). NGST(TRW) Tim Hanneman Quality Award Finalist (2005), NGST Directed Energy Systems Directorate Safety Champion (2006), SPIE Fellow, Arthur Guenther Best Poster Award-SPIE Laser Damage Conference (2018)

**Recent Relevant Publications**

Arenberg, J., Heap, S., Kendrick, S., et al., "Exploring the UV universe," Proc. SPIE 11116-52, (2019).

Apai, D. et al, "A Thousand Earths: A Very Large Aperture, Ultralight Space Telescope Array for Atmospheric Biosignature Surveys", The Astronomical Journal, 2019; 158 (2)

Jonathan W. Arenberg, "Formulation of the production time and cost of the Lynx x-ray mirror assembly based on queuing theory," J. Astron. Telesc. Instrum. Syst. 5(2), 021016 (2019)

P. A. Lightsey and J. W. Arenberg, *Systems Engineering for Astronomical Telescopes* ,ISBN 9781510616547, published by SPIE press 2018.

J.W. Arenberg, "Statistics of Laser Damage Threshold Measurements", , Chapter in Laser-Induced Damage in Optical Materials, D. Ristau Editor, Taylor & Francis, 2014, ISBN 9781439872161.

J. W. Arenberg, M. Macias, R.C. Lara,"A semi-empirical method for the prediction of molecular contaminant film accumulation (Conference Presentation)", Proc. SPIE 9952, 995202 (7 November 2016)

**Tony Hull; CETUS Architect and Co-Investigator**
Adjunct Professor of Physics and Astronomy, University of New Mexico;
Consultant to SCHOTT AG, KAC, Lockheed, etc. Workshop instructor for ESA.
**tonyhull@unm.edu**

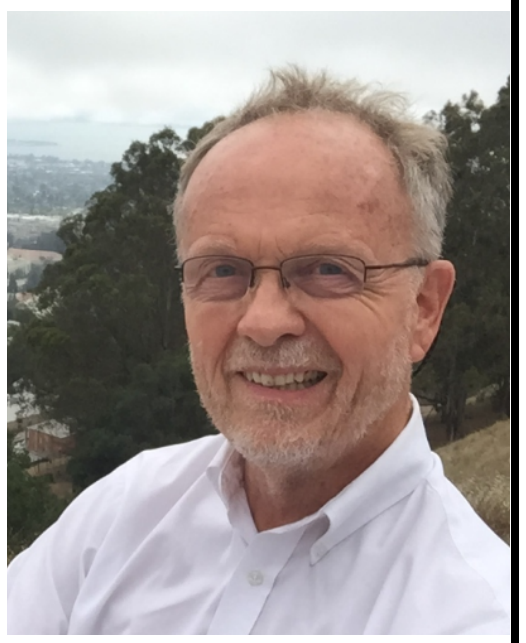

## Previous Positions

| | |
|---|---|
| 2004-2009 | JWST Program Manager (all Optical Fabrication), Director of Large Optics L-3 IOS |
| 1998-2004 | JPL Principal Engineer, NASA's TPF-C Technologist, Founding Manager HCIT, Main Lead of JPL's Team-I |
| 1989-1998 | Optical Corporation of America, VP, Chief Scientist & Director Engineering, Director Program Management |
| 1980-1989 | Director of Program Manager; Beryllium Technology Lead, Perkin-Elmer Applied Optics Division |

## Relevant Experience

Tony Hull has 39 years of experience in space, having held technical development and managerial cost-schedule-performance cognizance responsibilities for companies making a number of optical and electro-optical systems. He presently is consulting at a 75% full time basis, especially in exploring cost sensitive paradigms for materials, mirrors and optomechanics for spaceborne instruments, emphasizing optimum management of dimensional instability. He serves as Adjunct Professor of Physics and Astronomy at the University of New Mexico. His training in Mechanical Engineering and Astrophysics, both at University of Pennsylvania. He serves as Chair of SPIE's Biennial Conference Astronomical Optics…

He has been keynote lecturer at ESA's Space Optics Summer School at Poltu Quatu, and taught there on several space optics topics. Over the last several years, he authored a series of papers with SCHOTT on cost-effective lightweighting of ZERODUR® mirrors, and now consults for SCHOTT supporting trades for developing space missions. The 1.2m lightweight ZERODUR built by SCHOTT is based on his studies, and he has demonstrated its high performance at MSFC's XRCF facility. He has also authored many technical papers, including several papers on CETUS, and its earlier concept, GESE.

He served as Director and Program Manager for optical finishing of the entire JWST suite of mirrors (L-3 Integrated Optical Systems (IOS)), included facility development, establishing and qualifying a trained team, as well as managing the optical fabrication and test efforts in a high-visibility environment.

At JPL, he served as Principal Engineer of Observational Systems, optical lead for the 3.5m diameter Far InfraRed Space Telescope ((FIRST) (an advanced all composite approach conducted at NGIS San Diego), Main Lead of JPL's Concurrent Instrument Design Team (Team-I, now folded into Team-X), Architect for the Eclipse Coronagraph Mission and numerous others, Founding Manager for NASA's High Contrast Imaging Testbed (HCIT), and NASA's Technologist for Terrestrial Planet Finder Coronagraph.

At Perkin-Elmer Applied Optics (became Optical Corporation of America) he Directed both Program Management and Engineering Departments, and was Vice President and Chief Scientist. There he led a state-of-the-art beryllium technology, and defined many of the optical approaches implemented for NASA and SDIO missions. He engaged and contributed technically (often including architecture, definition, stability analysis, error budgets and even radiation testing) to NASA projects ranging from Galileo to the Mars Observer Camera to International Space Station video camera suite, and to a product line of wide-field star sensors.

**Robert A. Woodruff, CETUS Optical Design**
**Principal of Independent Optical Consultant, CAGE Code: 7MXC2**
**rawoodruff@live.com**

## Current and Previous Positions

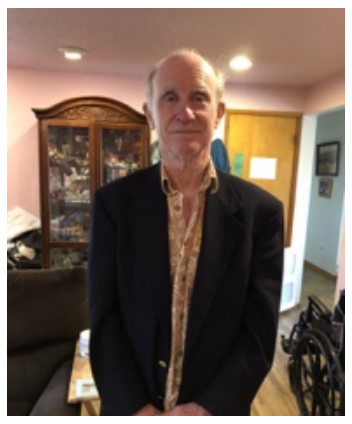

2010 – present: Independent optical design consultant
2000 – present: Associate of Center for Astrophysics and Space Astronomy (CASA), University of Colorado, Boulder,
2002-2010 Lockheed Martin Civil Space, Chief Scientist for Optical Syst
2001-2002 Boeing-SVS, Inc Chief Scientist for Optical Systems
1967-2000 Ball Aerospace & Technologies Corporation: Staff Consultant, Optical Design Numerous Space Missions
1979-1981 Santa Barbara Research Center: Senior Member of Technical Staff,

## Relevant Experience

Mr. Woodruff has over 50 years' experience designing optical systems for United States space program missions. He has made significant contributions to projects ranging from Skylab, Nimbus, Apollo-Soyuz, Galileo, SIRTF/Spitzer, microgravity science, the Hubble Space Telescope (HST), and James Webb Space Telescope (JWST), Terrestrial Planet Finder (TPF), Beyond Einstein, Exo-planet detection, Kepler, as well as others. He has wide and varied experience in the definition and design of optical space-borne telescopes and instruments. His technical specialties are optical physics, optics design, and optical system engineering. He has served in various technical roles in optical design, system engineering, system test, and system calibration in the development of more than 20 flight hardware instruments. Among his accomplishments, two activities standout: involvement in "fixing" the Hubble Space Telescope spherical aberration flaw, designing several HST scientific instruments, and conceiving and generating the optical concept and design for the Kepler mission. His HST experience is directly applicable to CETUS.

## PATENTS

U.S. Patent # 5,898,529 dated 4/27/99. "Deployable Space-Based Telescope"
U.S. Patent # 5,420,681 dated 5/30/95. "Modular Multiple Spectral Imager & Spectral Imager".
U.S. Patent # 4,391,525 dated 7/5/83. "Interferometer". A Michelson Interferometer that is unchirped and inherently insensitive to mechanical perturbations.

## PROFESSIONAL AWARDS (Selected)

2019: Senior Member SPIE, 2015: Senior Member OSA; 2006: Lockheed Martin Fellow
1999: GSFC Group Achievement Award National Resource Science Leadership Group
1997-1999: Four GSFC Group Achievement Awards related to HST ACS, HST STIS, and NGST

## Sample Publications

Bottema, M. and Woodruff, R.A.; "Third-Order Aberrations in Cassegrain-Type Telescopes and Coma Correction in Servo-Stabilized Images" Appl. Opt., 10, 1300 (1971).

Woodruff, R.A., Woodgate, B.E., and Ludtke, C.W.; "Optical Changes to the Space Telescope Imaging Spectrograph (STIS) in Correcting the Hubble Space Telescope (HST) Aberration" Presented at the 177th meeting of the American Astronomical Society, Philadelphia, PA (January 16, 1991).

Woodruff, R. A., et al, "Optical design for CETUS: a wide-field 1.5m aperture UV payload being studied for a NASA probe class mission study." JATIS, Apr-Jun 2019, p. 024006-1.

**Stephen E. Kendrick, CETUS Manager**
**Chief Technology Officer, Kendrick Aerospace Consulting LLC, Lafayette, CO**
**skconsulting@comcast.net**

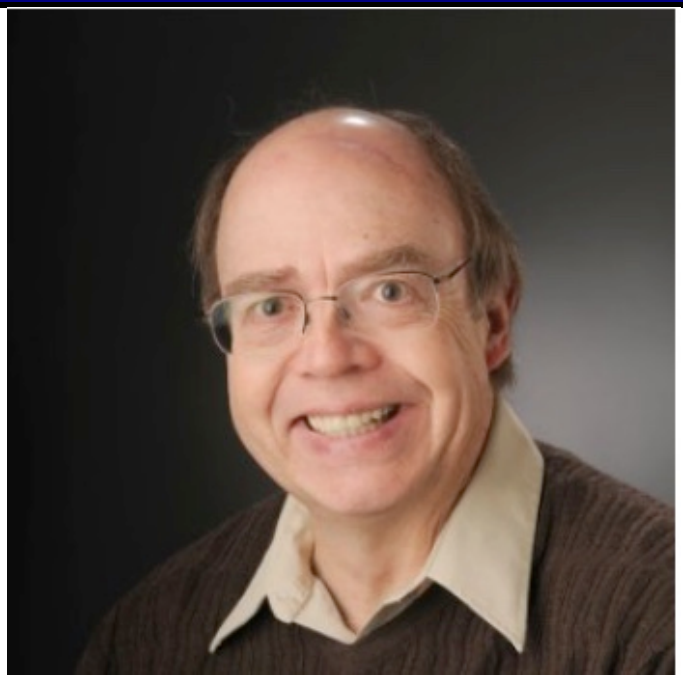

## Previous Positions

| | |
|---|---|
| 1996–2014 | Senior Project Engineer/ Program Manager II/ Principal Systems Engineer, Ball Aerospace & Technologies Corp., Boulder, CO |
| 1984-1995 | Senior Program Manager, Itek Optical Systems, Lexington, MA |
| 1977-1984 | Senior Engineer, Perkin-Elmer Corporation, Danbury, CT |
| 1973-1977 | Engineer, Bendix Corporation, Guidance Systems Division, Teterboro, NJ |

## Relevant Experience

Mr. Kendrick has 46 years of experience in optical and electro-optical systems and 35 years of experience in technical and program management. He was the Study Manager for the CETUS probe study and has led several other technology studies and hardware demonstrations as well as proposals for several new business pursuits. He was the Technology Area Lead for Space Telescopes for Ball Aerospace.

Mr. Kendrick's broad experience includes star sensors/guiders, adaptive optics, spacecraft buses, and space-borne optics from the Hubble Space Telescope to JWST.

His continuing focus has been on program management and technologies for space-borne telescopes including: development of light-weighted mirrors, e.g. the 1.4-m beryllium Advanced Mirror System Demonstrator including actuators, flexures, and reaction structure. At Ball he was the Principal Investigator (PI) for optical coatings and actuator IR&Ds for space telescopes and a star/planet simulator for a high-contrast imaging testbed.

He performed systems analyses to establish error budgets/ alignment requirements for the NASA Spitzer Space Telescope cryogenic telescope assembly and error budgets for the Hubble telescope's wavefront and pointing. He has also performed systems-error budgets, pointing analyses, and SPOT-5 compatibility/feasibility studies for the NASA SAGE III space-borne environmental sensor.

## Professional Societies/Awards

SPIE Senior Member, Optical Society of America (OSA) Senior Member, American Astronomical Society, Sigma Pi Sigma (physics honor society), Beta Gamma Epsilon (business honor society), Past President of Colorado chapter of OSA, Itek Excellence Award for project management

## Recent Relevant Publications

Kendrick, S., Woodruff, R., et al., "Science capabilities enabled by the CETUS NUV multi-object spectrograph and NUV/FUV camera and the driving technologies," Proc. SPIE 11115-3, (2019).

Arenberg, J., Heap, S., Kendrick, S., et al., "Exploring the UV universe," Proc. SPIE 11116-52, (2019).

Kendrick, S., Woodruff, R., et al., "CETUS science capabilities enabled by the CETUS NUV multi-object spectrometer and NUV/FUV camera," AAS 157.37, (Jan. 2019).

Kendrick, S., Woodruff, R., et al., "UV capabilities of the CETUS multi-object spectrometer (MOS) and NUV/FUV camera," Proc. SPIE 10699-116, (2018).

Kendrick, S, Woodruff, R., et al., "UV spectroscopy with the CETUS Ultraviolet multi-object spectrometer (MOS)," AAS 140.08, (January 9, 2018).

Kendrick, S., et al., "Multiplexing in Astrophysics with a UV multi-object spectrometer on CETUS, a Probe-class mission study," Proc. SPIE 10401, (2017).